# The TripleA Principle: Making Knowledge Actionable, Applicable, and Auditable in Post-FAIR Infrastructures

Vogt, Lars[1] orcid.org/0000-0002-8280-0487

Frühstückl, Robert[2] 0009-0005-3773-1039

Alamenciak, Tim[3] 0000-0002-1296-2528

[1]*Leibniz Institute for the Analysis of Biodiversity Change (LIB), Museum of Nature Hamburg, Martin-8 Luther-King Platz 3, 20146 Hamburg, Germany*

[2]*Johannes Kepler University Linz, Altenbergerstraße 69, 4040 Linz, Austria*

[3]*Department of Biology, Carleton University, Ottawa, Canada*

# Abstract

Ecological restoration, species distribution modelling, and invasive species management share a difficulty: knowledge that is findable and reusable carries no explicit account of the conditions under which it can be validly applied, or of the evidence grounding them. Applying it correctly is therefore demanding and expert-dependent, and misapplication usually goes unrecorded. The FAIR and CLEAR principles improved the findability, accessibility, interoperability, reusability, and human-interpretability of knowledge, but these address properties of representation, and reliable action requires more. Bridging the knowledge–action gap requires characterizing knowledge in terms of the operations it supports. Analysing what an operation needs, we derive three capabilities a knowledge representation must support. Actionability is the capacity to supply the knowledge and objective an operation executes. Applicability is the capacity to assess whether it can be reliably performed, through explicit conditions evaluated against context. Auditability is the capacity to assess the empirical grounding for that reliability, through documented success and failure. These form the three criteria of the TripleA Principle, an implementation-indipendent guide for next-generation knowledge infrastructures, jointly sufficient for the representational preconditions of reliably grounded action though not for its justification. Building on the Semantic Units Framework, we realize the principle as *action units*, typed components in which the knowledge an operation executes, the conditions under which it may validly be applied, and its documented successes and failures are addressable and evaluable. Action units form a nested hierarchy in which documented failure refines the conditions of valid use, letting knowledge graphs act as context-sensitive, evidentially accountable decision-support systems. Using biodiversity restoration as the illustrative domain, we introduce ecological fingerprints of ecosystem state, from which geographic knowledge–action-gap maps show where an operation is defined, applicable, and evidentially grounded.

# 1. Introduction: The Knowledge—Action Gap

*Imagine you are a restoration practitioner who has worked with a particular mangrove ecosystem for decades. Your knowledge graph database contains a rich and structured profile of this ecosystem, with detailed measurements of hydrological connectivity, sediment accretion rates, salinity profiles, species composition, and land tenure history, documented as structured, queryable statements. Your knowledge graph is connected to a broader infrastructure that enables you to ask, given this ecosystem profile, what can be done here, by whom, and with what confidence? The system returns a structured response, based on a selected set of priority actions for which the system was developed. For a citizen volunteer with no specialist training, it identifies the species identification and survey techniques that can be performed with the available diagnostic guides. For a data scientist, it identifies the transformation pipelines that can derive biodiversity indicators from the available occurrence records. For you, as an expert restoration ecology practitioner, it identifies relevant restoration procedures at three levels: (i) those that are theoretically appropriate in general; (ii) within these the ones for which your mangrove ecosystem profile meets the specified applicability conditions; and (iii) within these a smaller subset for which prior applications in comparable ecosystems have been documented, including both the sites where the procedure succeeded and the sites where it failed. For one procedure in particular, the system flags that the hydrological conditions at your site fall just outside the range of documented successful applications. The infrastructure does not tell you not to proceed, but it makes the gap between your context and the evidence base explicit, evaluable, and recorded. This is what the framework proposed in this paper is designed to make possible: knowledge that tells you not only what is the case, but what can be done, whether it can be reliably done here, and how well that judgment is grounded in prior experience—what we call* ***actionable knowledge****.*

In the face of accelerating biodiversity loss, ecosystem degradation, and increasingly urgent demands for evidence-informed decision-making (1), the scientific community has never had access to more data, metadata, and formally structured knowledge. Global biodiversity monitoring initiatives, museum collection databases, ecological observatories, remote sensing platforms, and citizen science programs continuously generate vast and rapidly expanding information resources. Yet despite this unprecedented abundance, translating knowledge into effective conservation and restoration action remains a persistent challenge (2,3).

This challenge has been extensively studied. The theory-practice gap, i.e., the disconnect between knowledge generated through scientific research and the needs and practices of knowledge users, is well established across the conservation and restoration community, and its causes span institutional, social, communicative, and political dimensions (4). Empirical work involving broad expert consultation has found that in many cases a lack of information is not the major barrier to biodiversity conservation but mechanisms to translate information into action are. Indeed, some of the information identified as a priority by conservation practitioners likely already exists, but is not accessible, synthesized at an appropriate scale, or in a form that is useful for biodiversity planning and management (5). Evidence-based approaches such as systematic reviews have been proposed as one pathway for strengthening this translation in ecological restoration (6).

This paper does not address the institutional, political, or social dimensions of the knowledge-action gap, which are real, well-documented, and often primary. Effective bridging of this gap requires not merely the transfer of knowledge from science to practice but bidirectional communication and knowledge exchange in which scientific and practical knowledge mutually inform and transform each other, which is a process that is fundamentally social and institutional in character and that lies beyond the scope of the present framework. The framework proposed here is designed to support both directions of this exchange. By making the conditions of valid knowledge application explicit, it enables practitioners to assess whether scientific knowledge applies in their specific context, and it enables scientists to understand under what conditions their knowledge holds or fails in practice, which is information that is currently lost when failure cases go undocumented. In this sense, the representational layer the framework provides is a necessary infrastructure for the kind of bidirectional, mutually informing knowledge exchange that effective science-practice relationships require.

Even when political will exists and practitioners are motivated to act on the best available evidence, the inability to access, interpret, and use that knowledge in specific contexts remains a key contributor to the gap (7). This paper addresses a specific representational dimension of this usability problem, as knowledge cannot reliably guide context-sensitive action unless its operational implications, contextual validity, and procedural requirements are made explicit in the knowledge representation itself. Current knowledge infrastructures do not provide this. The information mobilisation agenda set out by Buxton et al. (5) and the evidence-based restoration agenda advocated by Cooke et al. (6) both call for evidence to be synthesised and used more effectively, and a representation of this kind would help meet those calls. It is the gap the present framework addresses.

Although this paper uses biodiversity science and conservation as its primary application domain, the representational problem it addresses is not domain-specific. In manufacturing and materials science, critical procedural knowledge, including tacit operator knowledge of the form *“when machine X behaves in way Y, do Z”*, is rarely formally documented, making it inaccessible for systematic reuse, transfer, or machine-supported decision-making (8,9). In medicine, knowledge proclaimed to be universally applicable is frequently validated only on narrow demographic populations—historically white adult males—yet applied without explicit representation of these demographic constraints as applicability conditions, with well-documented consequences for patient outcomes across underrepresented groups (10,11). In all these domains, the core problem is the same, with knowledge being available but not being represented in a form that makes its conditions of valid applicability explicit, evaluable, and communicable to both human and machine agents.

The problem is not solely the absence of data or decision support tools, as conservation science has developed a range of frameworks for systematic decision-making (12), but rather that formal knowledge systems rarely represent knowledge in a form that makes its conditions of valid application in specific contexts explicit and evaluable. This may be one reason why practitioners frequently bypass them in favour of word-of-mouth and peer networks to obtain the situationally grounded, procedurally specific guidance they need (13). This structural deficit in current knowledge infrastructures—the absence of a representational layer that makes actionable knowledge as context-sensitive and procedurally grounded as the informal knowledge transmitted through peer networks, but systematically, transparently, and at scale—is what we refer to as the **knowledge-action gap** (Fig. 1). It is not unique to biodiversity research but extends across any domain in which general knowledge must be reliably translated into context-sensitive action.

We do not claim that any infrastructure can close this gap entirely. As we discuss in section 4.4, human assessment of whether general knowledge applies in a particular situation draws on tacit and experiential competence that may resist complete formalisation. The here proposed framework can narrow the gap and make more of it explicit and evaluable, but a residue of non-formalisable judgment may remain, and the aim is to reduce that residue systematically rather than eliminate it.

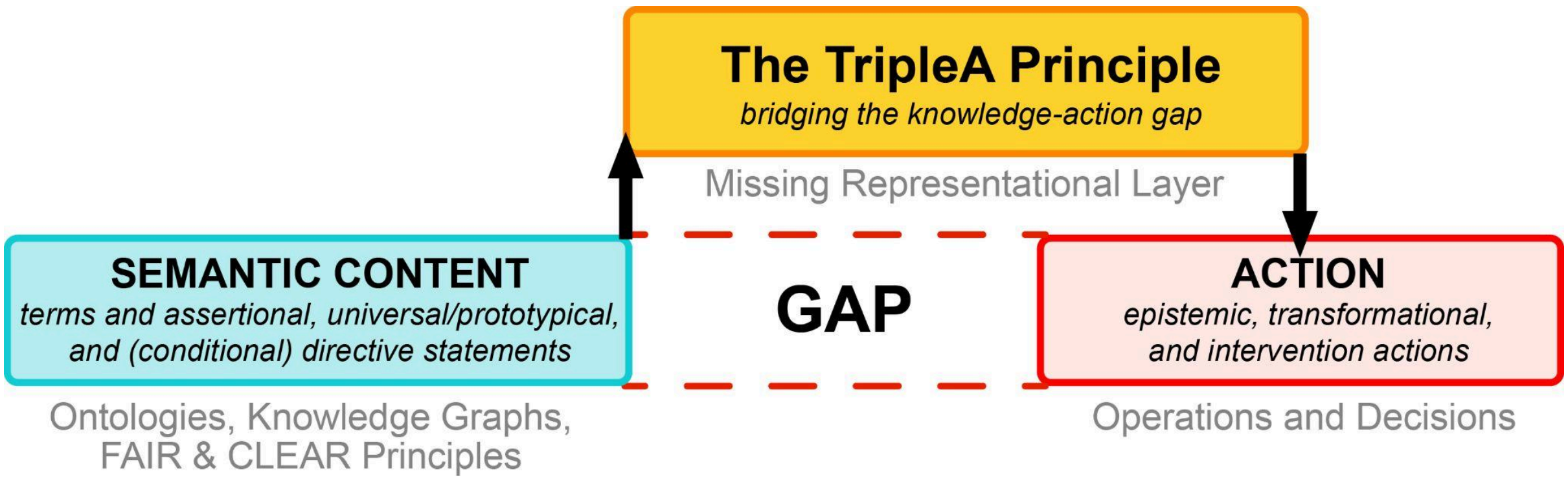

**Figure 1: From semantic content to action—bridging the knowledge-action gap.** The missing layer in FAIR and CLEAR knowledge infrastructures.

Over the past decade, major advances have been made in the representation and accessibility of knowledge and data. The FAIR Principles (14) have advanced the machine **F**indability, **A**ccessibility, **I**nteroperability, and **R**eusability of data, metadata, and knowledge, while more recent additions such as the CLEAR Principle (15) emphasize the importance of **C**ognitively interoperable, semantically **L**inked, contextually **E**xplorable, intuitively **A**ccessible, and human-**R**eadable and -interpretable data and knowledge. In parallel, ontologies and semantic knowledge graphs, grounded in foundational upper-level frameworks such as BFO and DOLCE (16,17), have enabled formal, concept-centric representation of knowledge by modelling entities and their relationships (18,19). Together, these approaches support increasingly sophisticated forms of machine-interoperability and human-interpretability. They remain, however, primarily focused on representing and organizing data and propositional knowledge rather than on enabling their direct use for context-sensitive action. They excel at answering *"What is X?"* or *"How is X related to Y?"*, but provide limited support for determining how terms and knowledge can be applied in practice and what should be done in a specific situation to achieve a specific goal. A critical gap therefore remains between knowledge representation and knowledge-informed action.

This limitation reflects a deeper conceptual issue. Existing approaches do not clearly distinguish three complementary capabilities that a knowledge representation must support if it is to guide action reliably. We call these capabilities: (i) **actionability**, the capacity to support the execution of operations; (ii) **applicability**, the capacity to support assessment of whether knowledge can be reliably used in a specific context; and (iii) **auditability**, the capacity to support assessment of the empirical grounding of prior applications through traceable documentation of where and why they have succeeded or failed—a dimension the conservation community has begun to address through taxonomies of conservation failure (20). Current knowledge infrastructure designs conflate or omit these distinctions, which are essential for ensuring that action is not only executable but contextually grounded and evidentially accountable. Together, they form the basis of the **TripleA Principle** of Actionability, Applicability, and Auditability, which we develop in Section 3 as a technology-agnostic guiding principle for the design of next-generation knowledge infrastructures, extending and complementing FAIR and CLEAR.

The need for these capabilities is particularly critical in ecology and other complex systems, where context-dependence, scale effects, and system variability frequently limit the transferability of general knowledge (21). In such settings the availability of data and formalized knowledge alone is insufficient. Practitioners routinely evaluate whether knowledge is appropriate for the situation at hand, yet current infrastructures provide limited formal support for making this evaluation explicit, consistent, and auditable across agents, contexts, and scales.

To realize the TripleA Principle in a concrete representational form, we build on the **Semantic Units Framework** (22–25), adopting a statement-centred perspective that treats statements rather than isolated terms or triples as the primary units of meaning, and introduce **Action Units** as a class of formal knowledge structures that make the procedural knowledge required for conducting specific operations, the conditions of their valid application, and the empirical evidence supporting it explicit. Action Units are offered as one realization of the TripleA Principle, with the principle itself being framework-agnostic, and they additionally allow knowledge graphs to be extended into graph-native decision-support and workflow-orchestration systems, marking a transition toward **post-FAIR knowledge infrastructures** in which knowledge is not only findable and interpretable but operationally integrated and context-sensitive.

The remainder of the paper develops this argument in stages. Section 2 analyses why current concept- and data-centric infrastructures cannot make knowledge actionable. It characterizes the three classes of operation—epistemic, transformational, and intervention—that actionable knowledge must support, and frames the deficit as three challenges, each motivating one of the three capabilities. Section 3 elevates these capabilities into the TripleA Principle, a framework-agnostic requirement that knowledge representations be actionable, applicable, and auditable. Section 4 argues that the three criteria are not only necessary but jointly sufficient, showing that the operation classes over which they are defined are combinatorially and philosophically closed and closed under composition. Section 5 gives the principle a representational realization, introducing action units within the Semantic Units Framework as compound units that make applicability conditions and empirical evidence first-class, typed components. Section 6 develops the consequences, including (i) the three-level hierarchy of action units, applicable action units, and audited applicable action units, (ii) the typed operational layer that links units to operations, (iii) conditional action units and the transition toward post-FAIR decision systems, and (iv), using biodiversity science and ecosystem restoration as the illustrative domain, the reinterpretation of recurring failures and the construction of ecological fingerprints and knowledge–action-gap maps. Section 7 concludes.

*This paper is a conceptual-framework contribution; implementation and empirical evaluation are deferred to subsequent work.*

# 2. Operations and the Three Challenges of Actionable Knowledge

If knowledge is actionable, it can in principle be used to do something, i.e., to perform a specific operation. Actionable knowledge and the operations it supports are thus two sides of a single structure. An operation is individuated in part by the knowledge it requires, and knowledge is actionable in part by virtue of the operations it can enter into. This gives a starting point from which the requirements on actionable knowledge can be derived. Not by asking abstractly what knowledge *is*, but by examining this structure from both directions: what a knowledge representation must provide for an operation to be performable, and what it takes for the knowledge it carries to be correctly and reliably applied. This section introduces the three classes of operation that actionable knowledge supports, and then derives, as three successive challenges, the capabilities a knowledge representation must provide for knowledge to be performable, reliably applicable, and trustworthy in its application. Each challenge exposes something a bare operation lacks. Together, they motivate the three criteria developed in Section 3.

## 2.1 Actionable knowledge supports operations

Improving the representation, accessibility, and interpretability of knowledge, as achieved by ontologies (16,17) and the FAIR (14,26) and CLEAR (15) principles is a necessary but insufficient condition for enabling action. The transition from interpretability to actionability requires an additional layer, which is the explicit representation of the **operations** through which knowledge is used. Actionability is therefore not a property of representations alone, but emerges from the interaction between representations and the processes that operate on them. We define operations as processes that use knowledge representations to produce outputs, transform representations, or effect changes in real-world systems. In this sense, operations constitute the functional interface between knowledge and action. Without explicit representation of the operations it supports, knowledge remains structurally accessible but operationally inert.

What "*actionable knowledge*" covers is best seen through the range of cases it must include. Consider first a general term such as *cell nucleus*. An ontology may supply its intensional definition, such as the Foundational Model of Anatomy, defining it as "*Organelle which has as its direct parts a nuclear membrane and nuclear matrix*" (http://purl.org/sig/ont/fma/fma63840). This tells us what a cell nucleus *is* but not how to *recognise* one in practice. To be usable, the term additionally requires method-dependent recognition criteria that specify how instances are identified, such as when applying bright-field, Nomarski contrast microscopy, the nucleus appears as a restricted compartment with a granular inside, whereas in transmission electron microscopy and stained with heavy metals, the nucleus appears as a membrane-bound compartment scattered irregularly with electron-dense clusters of varying size to complete electron-dense staining (27). For theoretical terms such as *phylogenetic relatedness* or *invasion success* the gap is sharper still, since there is no naive observation of the referent at all and usability depends entirely on specified criteria. Assertional statements such as "*Salinity at this site exceeds the establishment threshold*" are usable when they can be recognised as applying to a situation, or when used to describe a given situation. Universal statements, including causal hypotheses such as "*Increased salinity inhibits mangrove establishment*," carry generalisable knowledge, but whether they can be *used* in a given case depends on whether they apply there, which their general form does not settle. And a management task, such as dealing with a degraded site and attempting to change it towards a target state, asks for actionable knowledge in the form of procedure specifications for achieving the desired change. Each of these involve actionable knowledge, and in each the knowledge that makes it usable is something the representation does not carry on its face: recognition criteria for the term, conditions of application for the general statement, a procedure for the task.

These cases differ not incidentally but structurally, in *what the use of the knowledge does*. It relates a representation to the world (recognising a nucleus, judging that a statement fits a situation), derives one representation from another (turning observations into an indicator), or changes the world to match a desired state (restoring a site). This is the distinction the three classes of operation capture, and it is why one category will not suffice.

At its most basic, an operation is a process in which an **agent** — human or machine — takes some **input** and produces some **output** (Fig. 2). What the input and output *are* distinguishes three classes of operation that together form a minimal operational basis for actionable knowledge, and that recur throughout this paper.

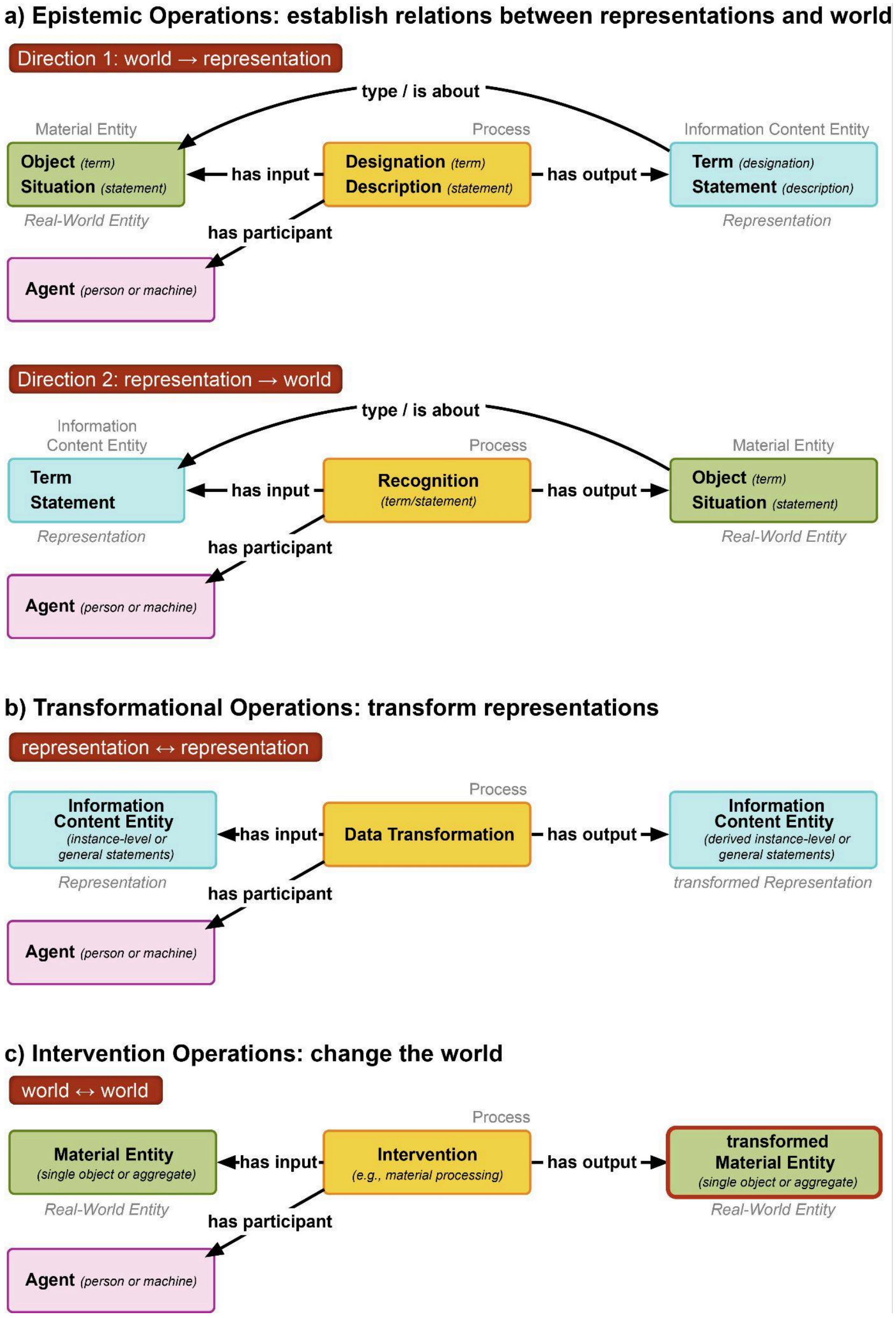

**Figure 2: Three classes of operation in bare form.** Every operation is a process (orange) in which an agent (‘*has participant*’) takes an input and produces an output (‘*has input*’, ‘*has output*’). The classes differ in what the input and output are. **(a)** Epistemic operations relate representations and the world in both directions: *designation* and *description* take an object or situation as input and produce a term or statement (world → representation), while *recognition* takes a term or statement as input and identifies the corresponding object or situation in the world (representation → world). The ‘*type*’/‘*is about*’ arc marks the resulting reference relation between representation (term, statement) and real-world entity (object, situation). **(b)** Transformational operations take an information content entity (instance-level or general statements) as input and produce a derived information content entity as output (representation ↔ representation). **(c)** Intervention operations take a material entity as input and produce a transformed material entity as output (world ↔ world).

**Epistemic operations** establish the relationship between knowledge representations and real-world situations. They operate in both directions (Fig. 2a), as an operation may take a term or statement as input and identify the corresponding object or situation in the world (recognition), or take an object or situation as input and produce a term or statement that describes it (designation and description). Epistemic operations are the grounding layer of actionable knowledge, as they answer *"What is the case in this situation, and does this knowledge apply here?"*

**Transformational operations** operate entirely within the space of representations, taking an information content entity — instance-level or general statements — as input and producing a modified or derived information content entity as output, through processing, aggregation, analysis, filtering, inference, or reformatting (Fig. 2b). They prepare and derive the representations that interpretation and intervention require, answering "*How can available representations be processed so that they support interpretation or action?*" Unlike epistemic operations, transformational operations are agnostic to the epistemic status of the statements they process. They manipulate representations according to formally specified rules without evaluating whether a statement is true, applicable, or relevant, which is what distinguishes the processing of representations from their interpretation and application. Transformational operations vary in inferential character. Some are deductive or computational, applying formally specified rules that preserve truth and can be executed deterministically, such as unit conversions, schema mappings, or logical entailment. Others are ampliative, extending knowledge to new cases without preserving truth, as in statistical generalisation and analogical or abductive inference of the form *"as X relates to Y in one situation, so W relates to Z in another."* Ampliative transformations are not reducible to formal rules and resist complete formalisation, yet they are central to actionable knowledge, since it is partly through such non-deductive transformations that general knowledge is brought to bear on particular situations. A transformational operation therefore may or may not be fully formalisable, and what the class guarantees is only the input and output types, not that the transformation is deductive or automatable.

**Intervention operations** act on the world itself, taking a material entity in a given state as input and producing a transformed material entity as output, to achieve a desired change (Fig. 2c). They are the goal-directed endpoint of actionable knowledge, answering *"What can be done in this situation to achieve a desired outcome?"*

Introduced by Hacking (28), the distinction between *representing* the world and *intervening* upon it has deep roots in the philosophy of science, and the three classes can be seen as a systematic operationalization of that distinction, with epistemic operations corresponding to representing, intervention operations to intervening, and transformational operations to an intermediate class operating wholly within the representational domain. The three are interdependent and ordered, with intervention presupposing epistemic grounding, which transformational operations help prepare, defining a progression from situational understanding, through representational preparation, to system-level action. Their interdependence and composition are examined in more detail in Section 4. This progression is the structure that any infrastructure for actionable knowledge must support.

But specifying that an operation takes an input and yields an output does not yet say what knowledge makes it **performable**, what makes that knowledge **reliably applicable** in a given situation, or what makes its reliability **trustworthy**. Each is something a bare operation lacks, and each motivates one of the three challenges that follow.

## 2.2 The actionability challenge: performing operations requires a plan specification providing directives and a defined objective

A bare operation, as introduced above, specifies only that some agent takes an input and produces an output. But that specification does not yet make the operation *performable*. To apply a concept, recognize a given situation, transform a dataset, or carry out a goal–directed intervention, an agent must possess the knowledge of *how* to conduct the operation and must be directed at an objective the operation is meant to achieve. To know how to identify a cell nucleus, an agent must have the diagnostic knowledge in the form of method-dependent recognition criteria. The first challenge is that current infrastructures, optimised for representing *what is the case*, do not represent these components and, consequently, leave the operations they ostensibly support unperformable. The Foundational Model of Anatomy ontology only provides the intensional definition of *cell nucleus*, specifying what it is ontologically, but not the method-dependent recognition criteria for practically applying the term in designation and recognition epistemic operations. We refer to the capability of a representation to supply what an operation needs to be performed as its **actionability**, and the components it requires are shown in Fig. 3: each operation '*executes*' a **plan specification** containing directive knowledge and specifying an **objective** the operation '*achieves*', the whole being used by an agent that may be human or machine. The objective is a part of the plan specification.

Beyond supplying these components, a further difficulty is that current approaches do not distinguish the different forms actionability takes across representation levels (terms versus statements) and agent types (human versus machine), which is a conflation that obscures what making knowledge actionable actually requires (Box 1).

For **machine agents**, actionability is defined by the availability of executable operations. Machine-actionable knowledge is therefore tied to workflows, APIs, scripts, and reasoning procedures that process formally structured representations (29), aligning with emerging visions of FAIR ecosystems (23,26) that include terminology, schema, operations, and workflow services, enabling automated reasoning and (agentic AI driven) execution. For **human agents**, actionability requires more than formal structure. Human-actionability depends on interpretation (knowing-that), contextualization (judgment), and procedural guidance (knowing-how). Interpretation is necessary but not sufficient for actionability. Knowledge must also guide how to recognize relevant situations and how to act within them, tying human-actionability closely to involving epistemic and intervention operations.

At the level of **terms**, it is useful to distinguish inferential from referential lexical competence (30) (Box 1). Inferential lexical competence concerns the **intensional meaning** of a term—how it is defined, interpreted, and related to other concepts—and is supported by the ontological and intensional definitions that answer *"What is it?"* and *"How is it understood?"* (26). Referential lexical competence on the other hand enables actionable use, supporting **recognition** (given a term, identify a corresponding instance) and **designation** (given an instance, identify the appropriate term), which requires **diagnostic knowledge** of how instances are identified in practice, applying method-dependent recognition criteria that provide the **extensional meaning of a term**, such as measurement protocols, identification keys, assays, diagnostics, decision rules, or observational procedures (27) (Box 1). Ontology-based frameworks typically provide only ontological and intensional definitions (see '*cell nucleus*' in the Foundational Model of Anatomy), supporting inferential but not referential lexical competence, which limits the operational utility of terms. For **machine agents**, Description Logics reasoning over OWL class axioms defined in ontologies supports recognition and designation by classifying instances against necessary and sufficient properties, but requires that lower level representations are already available to conduct the reasoning upon, thus representing a transformational operation.

Crucially, even when terms are made actionable in this sense, they do not by themselves enable action, because meaning in scientific discourse is conveyed primarily through **statements**, and thus propositions about relationships, observations, hypotheses, or processes. Statements can be categorized into three fundamentally different types with distinct roles in actionable knowledge: **instance-level assertional statements**, and two forms of **general knowledge** in the form of **universal and prototypical statements** and **directive and conditional directive statements**. Assertional statements describe relationships among individual entities and function as contextual information and empirical evidence. Their human-actionability relies on epistemic operations such as **recognizing** whether a statement applies to a given situation or **describing** a situation using an appropriate statement, while their machine-actionability depends on the interoperability and the availability of **computational functions** that can process the corresponding statement types, typically requiring standardized representations such as SHACL-constrained data models. **Universal and prototypical statements** express generalizable knowledge, including causal relationships and statistical correlations, that is not tied to specific situations. They provide the basis for explanation and intervention but are not directly actionable without further structure. Procedural knowledge representing methods, algorithms, plan specifications, and practices instructions takes the form of **directive statements** giving step-by-step instructions toward a specific result, frequently including if-then **conditional directive statements**. It is these type-level directive forms that supply the directive content of the specification an operation executes (Fig. 3).

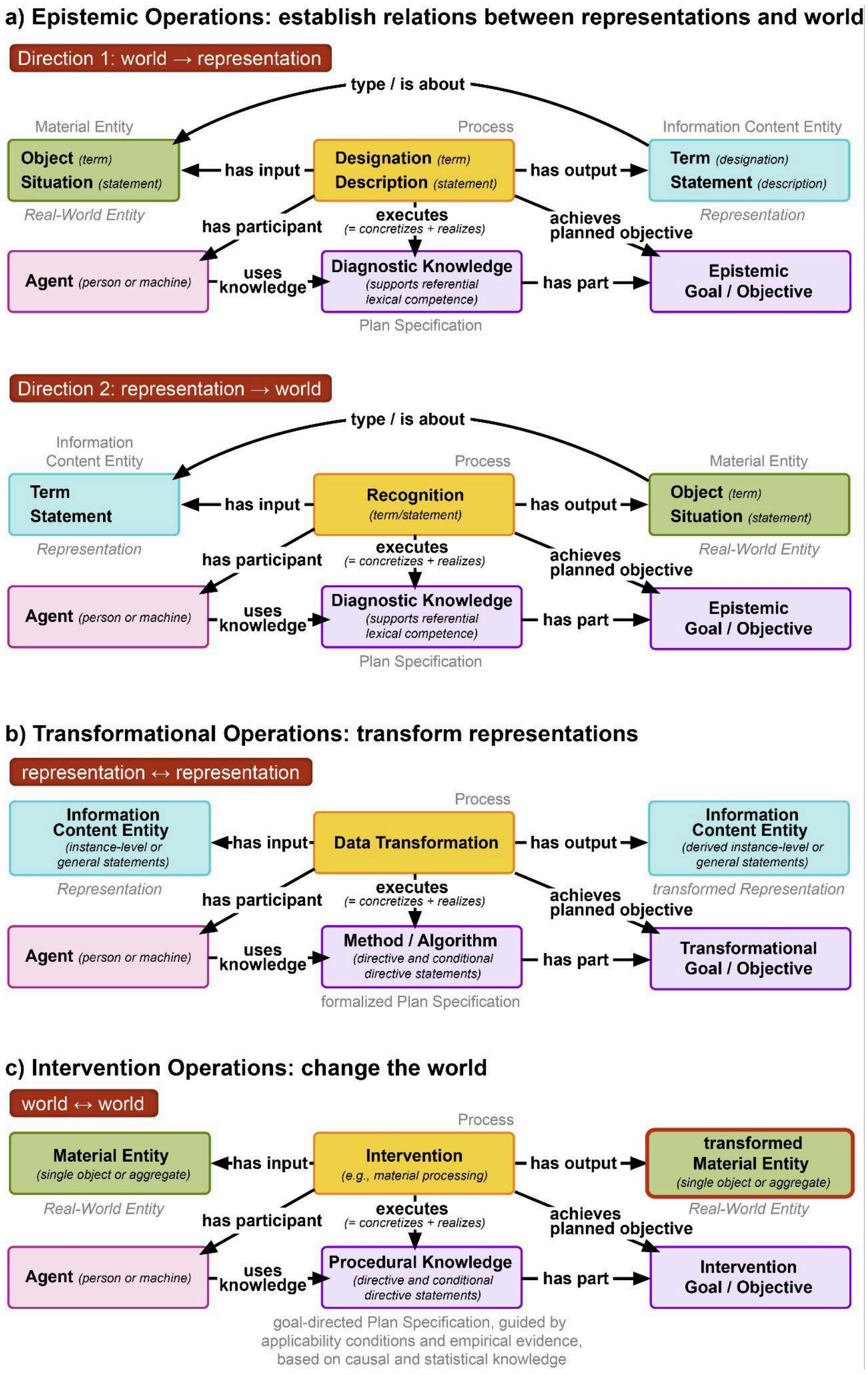

**Figure 3: Actionability requires plan specification and objective.** Extending Fig. 2, each operation now '*executes*' a plan specification, i.e., the directive knowledge it carries out, composed of directive and conditional directive statements (**a**: diagnostic knowledge for epistemic operations, supporting referential lexical competence; **b**: a method or algorithm for transformational operations; **c**: procedural knowledge for intervention operations). The plan specification contains the planned objective that the operation '*achieves*'. The '*executes*' relation is glossed as concretizes + realizes. These are the components a bare operation requires to be performable.

This three-way taxonomy rests on a more fundamental two-class distinction in the philosophy of language. Assertional, universal, and prototypical statements share a **word-to-world direction of fit** in Searle's sense (31,32), aiming to describe how the world is, either at the level of specific situations or at the level of general regularities and dispositions. Directive and conditional directive statements, by contrast, share a **world-to-word direction of fit**, specifying how the world should be changed to match a desired state. The finer distinctions follow Jansen & Schulz (33), and the Semantic Units Framework (25) arrives at the same descriptive/prescriptive structure independently through a representational analysis, indicating a difference not merely in representation but in the epistemic function of statements that represent reality and those that prescribe action. The relevant point here is that current infrastructures do not represent this distinction, and so cannot mark which statements describe and which prescribe, which is a prerequisite for knowing what can be done with them.

The deficit, across all these distinctions, is the same. Current infrastructures support understanding *what is the case*, representing terms intensionally and statements assertionally and universally, but do not represent the directive content an operation executes, the objective that content is directed at, or the diagnostic knowledge that gives terms referential reach. They make knowledge interpretable without making it performable. This is not a peculiarity of any one domain. In manufacturing and materials science, critical procedural knowledge, including tacit operator knowledge of the form "*when machine X behaves in way Y, do Z*", is rarely formally documented, and so cannot be reused, transferred, or executed by machine agents, even where the knowledge plainly exists (8). In biodiversity conservation, broad expert consultation has found that a lack of information is frequently not the principal barrier to action. Much of the information practitioners need already exists but is not available in a form usable for planning and management (3,6). In each case the knowledge is present but not *actionable*, as the directive content required to perform the relevant operations, and the objectives those operations serve, are not represented as a plan specification an agent can execute. Supplying them by adding to each operation the plan specification and its objective as shown in Fig. 3 is what actionability requires. But actionability is not yet sufficient, since a performable operation may still be performed where it does not **validly apply**. That is the second challenge.

**Box 1 Readability, Interpretability, and Actionability of Terms and Statements and Their Human and Machine Dimensions**

| **Human-Centred Actionability** |
|---|
| **Human-readable term** (26)<br>A term is human-readable if it consists of a sequence of characters belonging to a script or writing system known to the reader, such that the term can be identified as a discrete unit. Readability does not require that the term carries meaning, as a string such as “RDF” may still be readable to someone who does not know its meaning. Readability thus requires sender and receiver to use the same set of characters and agree on a reading direction. |
| **Human-interpretable term** (26)<br>A term is human-interpretable if it is readable and can be assigned a cognitive representation of its intensional meaning. This requires a shared **inferential lexical competence** (30), i.e., knowledge of how the term is defined and related to other concepts. Ontological and intensional definitions provide this knowledge, and ontologies provide this knowledge for structured data. |
| **Human-actionable term** (26)<br>Readability, interpretability, and actionability correspond to points on a spectrum of increasingly demanding operations. In this paper, **actionability** specifically refers to the operations of **recognition** (given a term, identify a corresponding instance) and **designation** (given an instance, identify the appropriate term), which require **referential lexical competence** (30), i.e., diagnostic knowledge of how instances can be identified in practice, such as method-dependent |
| recognition criteria, measurement protocols, or classification rules. A competent speaker may read and interpret a term, and use it in communication, without possessing this referential competence. Human-actionability is therefore not a property of the term but of the relationship between a term and a specific agent's referential capacities. Logical reasoners support designation tasks by automatically classifying instances based on their properties and on necessary and sufficient properties specified in an ontology. |
| **Human-readable statement** (26)<br>A statement is human-readable if it consists of human-readable terms organized according to rules that delimit terms and define statement boundaries. A data structure is human-readable if it can be represented in this form. |
| **Human-interpretable statement** (26)<br>A statement is human-interpretable if (i) it is human-readable, (ii) its constituent terms are interpretable, and (iii) it can be mapped onto a syntactic structure (i.e., a syntax tree) with syntactic positions that are associated with specific semantic roles. This requires sender and receiver to share a terminology and a set of common statement structures. Approaches such as **Rosetta Statements** (23,24,34) provide formalized templates that support such interpretation and act as semantic anchors in knowledge infrastructures. |
| **Human-actionable statement** (26)<br>A statement is human-actionable if it is interpretable and supports the operations of **recognition** (given a statement, identify the situation to which it applies) and **description** (given a situation, identify a statement that accurately describes it). Actionable statements can be combined into coherent narratives, provided that entities can be tracked consistently across statements. For general statements (universal and prototypical as well as directive and conditional directive statements), human-actionability additionally depends on assessing whether they can be reliably applied in the situation at hand (see applicability challenge). |
| **Machine-Centred Actionability** |
| **Machine-readable term** (26)<br>“[E]*lements in bit-sequences that are clearly defined by structural specifications*”[(p. 3 in (29)). In the Resource Description Framework (RDF), these are resources in the form of IRIs and typed literals. |
| **Machine-interpretable term** (26)<br>“[E]*lements* [in bit-sequences] *that are **machine-readable** and can be related with semantic artefacts in a given context and therefore have a defined purpose*” (p. 3 in (29), emphasis added). In RDF/OWL, this is achieved through the use of ontology terms with formal semantics. |
| **Machine-actionable term** (26)<br>“[E]*lements in bit-sequences that are **machine-interpretable** and belong to a type* [of element] *for which **operations** have been specified in symbolic grammar*” (p. 3 in (29), emphasis added). In OWL, such operations are provided by logical reasoning based on Description Logics. |
| **Machine-readable statement** (26)<br>A statement composed of machine-readable terms. In RDF, this corresponds to triples following the *Subject–Predicate–Object* structure. Since RDF triples represent binary relationships, there is no one-to-one correspondence between human-readable statements and RDF triples. Within the Semantic Units Framework, independent of the underlying n-arity, all statements are represented as statement units. |
| **Machine-interpretable statement** (26)<br>A machine-readable statement encoded in a formal language (e.g., OWL) that supports semantic interpretation. |
| **Machine-actionable statement** (26)<br>A machine-interpretable statement for which operations are available that take the statement as input. These operations may include logical reasoning, data transformation, statistical analysis, or other computational procedures. Interoperability is a key requirement; formal constraints such as SHACL shapes support the development of reusable operations for specific statement types. |

## 2.3 The applicability challenge: reliable performance requires explicit applicability conditions

Supplying an operation with a plan specification and an objective makes it performable, but not yet *reliably* performable in a given situation. A restoration procedure, sound in general, may fail at a particular site. A causal hypothesis that holds in one ecosystem may not transfer to another. The second challenge is that current infrastructures that do represent the general knowledge required for actionability usually do not represent the conditions under which that knowledge can be validly applied to the situation at hand. We refer to the capability of a representation to support assessment of whether an operation can be reliably performed in a specific context as its **applicability**, and the component it requires is shown in Fig. 4: each operation takes, alongside its plan specification, a set of **applicability conditions** evaluated against contextual information describing the situation.

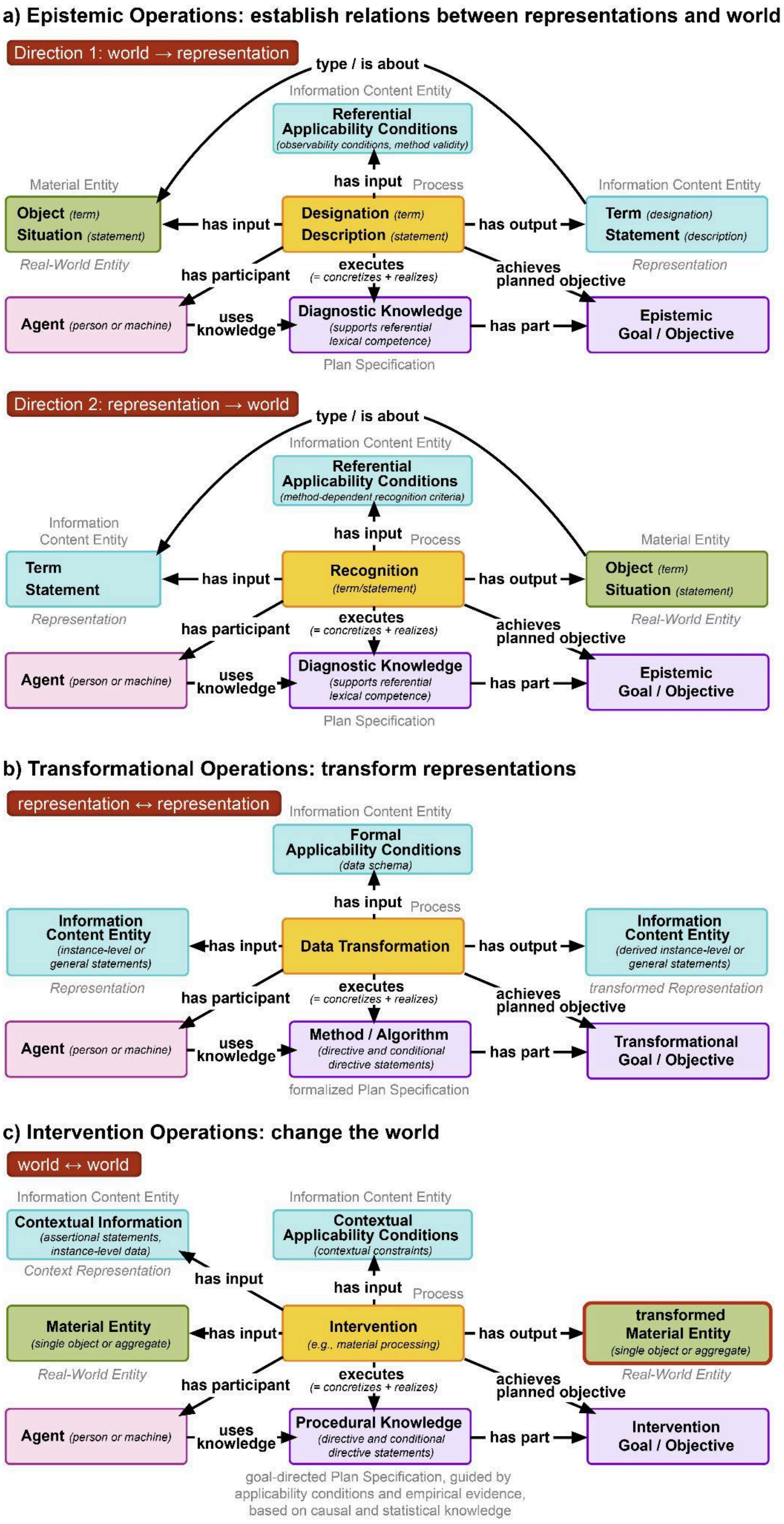

**Figure 4: Applicability requires applicability conditions.** Epistemic, transformational, and intervention operations differ in their input–output structure and in the types of knowledge they execute. **a)** Epistemic operations relate information content entities (ICE) to material entities or processes (here not shown), using 'rdf:type' for terms and 'iao:is about' for statements, and execute diagnostic knowledge, which is the basis for establishing referential lexical competence. **b)** Transformational operations operate on representations, taking ICE as both input and output, and execute algorithms that realize computational functions. **c)** Intervention operations act on material entities, transforming them from an initial to a target state by executing procedural knowledge and realizing goal-directed capabilities. Together, these operation types form the operational backbone of actionable knowledge, linking knowledge representation, data transformation, and real-world action. The properties used in this figure are in line with upper-level ontologies such as the Basic Formal Ontology (BFO) with its OBO Relations Ontology (RO) and extensions such as the Statistical Methods Ontology (STATO), the Ontology for Biomedical Investigations (OBI), and the Information Artifact Ontology (IAO) (see refs (16,35)). For instance the relation 'stato:executes' is understood as a derived relation (a chained property combining ro:realizes and ro:concretizes) linking a process to a plan specification via the intermediate realization of some plan that concretizes the specification, whereas 'iao:is about' is a representation relation between an ICE and the entity this information is about.

Addressing this challenge requires distinguishing three related but conceptually distinct notions: applicability conditions, contextual information, and the applicability relation between them.

**Applicability conditions** define the constraints under which a statement or procedure can be reliably applied in a specific situation. Three kinds are distinguished, corresponding to the three operation classes: (i) **referential** applicability conditions specify method-dependent recognition criteria, observability conditions, and method validity for epistemic operations; (ii) **formal** applicability conditions specify the input data schema for transformational operations; and (iii) **contextual** applicability conditions define the relationship between general knowledge and specific situations for intervention operations, and can be evaluated against a given context to determine whether the required conditions are satisfied. Applicability conditions are related to, but distinct from, the truth conditions of a statement. Where truth conditions concern when a statement is true, applicability conditions are epistemic and practical rather than purely semantic, specifying whether a particular real-world situation satisfies the general conditions of a statement sufficiently to warrant using it to guide action. A statement may have well-defined truth conditions while its applicability in a specific situation remains uncertain, incomplete, or context-dependent. Applicability conditions thus function as the operational bridge between general procedural knowledge and situational decision-making, enabling not only the identification of applicable actions but also the detection of *unmet* conditions, and thereby supporting decision-making under uncertainty and guiding data acquisition or system modification.

Contextual conditions divide in a way that matters for how a statement's scope is refined. The division is not a classification of factors into two kinds but a distinction in the role a condition plays relative to a given causal relation, and the same factor may play different roles relative to different relations. Some conditions are *enabling*. They govern whether the causal relation an intervention exploits obtains, or whether the operation exploiting it can be performed at all, without bearing on the identity of what that relation relates. They stand alongside the relation as separate conditions on it. The availability of mowing crews conditions whether a management regime can be applied, and ongoing physical disturbance at a site conditions whether a planting succeeds, but neither alters what a cut does to a plant population or what a change in salinity does to a propagule. Others are *determining*. Rather than standing alongside the relation, they enter into it, specifying the identity of the changing quality or of its bearer, and so become part of what the general knowledge asserts rather than merely delimiting where it may be used. The mark of a determining condition is that it is absorbed into the causal variable and ceases to be a separate condition at all. Where the tidal regime of a coastal system determines which salinity a propagule is in fact exposed to, the tidal regime does not stand alongside the causal variable but partly constitutes it, and what had been a background condition becomes part of the relatum, interstitial pore-water salinity rather than salinity as such. The distinction is not given in advance and cannot be settled by inspection, since whether a condition merely enables or in fact determines is an empirical question about the system. Recording a condition as one or the other is therefore a modelling decision, revisable in the light of documented outcomes, and Section 6.1 sets out how such a revision proceeds and what it produces.

Referential conditions carry a further burden that the other two do not. Many of the qualities they constrain are not directly observable and are approximated through indicators, and whether a referential condition transfers across indicators depends on how the quality has been represented. Where it is represented as a single underlying property that different indicators estimate with differing reliability, a condition established through one indicator applies, with appropriate uncertainty, to a value obtained through another. Where the indicator is instead taken to partly determine the quality, a condition established through one does not apply to a value obtained through another at all. The distinction matters because it determines what a mismatch between the two *is*, with the former being a loss of precision and the latter a failure of reference, and only the former can be absorbed by widening the uncertainty attached to an assessment. The framework does not adjudicate between the two treatments, which are both defensible and suit different qualities (36). It requires that the choice be recorded, since a referential condition is evaluable only relative to it. Section 6.6 develops the consequences for applicability assessment at scale.

**Contextual information** refers to instance-level descriptions of a specific situation, including any situational factor relevant to determining whether an operation can be reliably executed. Crucially, this encompasses not only biophysical parameters such as environmental conditions, observations, and system states, but also socio-political, economic, institutional, legal, and cultural factors such as land tenure arrangements, community consent, resource availability, governance structures, and political feasibility, which shape what actions are possible and legitimate in a given situation. Contextual information provides the situational input required for assessing the applicability of a given knowledge representation against its applicability conditions, and may be retrieved from a database or supplied by a human or computational agent at the point of use.

**Applicability** itself should not be understood as a type of knowledge, but as a **relation between knowledge and context**. It characterizes whether, and to what extent, a given knowledge representation can be reliably used in a specific situation. Like actionability, applicability is not Boolean but graded, ranging from unknown, through plausible and supported, to empirically validated, depending on the availability of evidence and contextual information. This gradation reflects the inherent uncertainty of real-world systems, particularly in ecology, where context-dependence, scale effects, and system variability frequently limit the transferability of general knowledge (21).

The process that relates knowledge to a specific situation, by combining contextual information with applicability conditions, is **contextualization**. Evaluating that relation, determining whether, and to what degree, the contextual information satisfies the applicability conditions, is **applicability assessment**. For **assertional statements** properly grounded through prior epistemic operations, meaning those whose subject terms have been correctly designated and linked to specific material entities or system states, contextualization is largely implicit, since they already refer to specific situations and so themselves provide contextual evidence. For **universal and prototypical statements**, and for **directive and conditional directive statements**, contextualization is essential and must be performed explicitly, as the validity of these general statements cannot be assumed but depends on whether the conditions under which they hold are satisfied in the situation at hand.

Much of the knowledge relevant for decision-making in biodiversity science and many other domains is expressed as **causal hypotheses** specifying how changes in one part of a system influence others, such as environmental drivers of biodiversity change, ecosystem responses to interventions, interactions between species and their environment. The representation of causal hypotheses draws on interventionist concepts of causation (37), particularly the characterization of causal relationships by their invariance under interventions and the conditions under which they hold and thus concepts directly relevant to specifying applicability conditions, without implying that interventionist accounts provide a complete theory of causation.

This characterization can be made precise, and doing so bears directly on where the applicability conditions of a causal hypothesis attach. A quantity does not by itself cause anything, since a number of species, a salinity level, a biomass is not the kind of thing that brings something about. Rather the *change* in a quantity can function as a cause. In the account we adopt, a causal relationship holds between processes of change of some quality, disposition, or relationship, such that a change in one may trigger a causal reaction expressed as a change in another (36). The universal statement *increased salinity causes decreased mangrove establishment rates* accordingly relates an increase in the salinity of a site to a decrease in the establishment rates of mangroves there, with the causal relation holding between the two changes and the direction of each carried by the relatum rather than by the predicate. Two consequences follow. First, the applicability conditions of a causal hypothesis are conditions on the *bearers* of the changing qualities, i.e., on the sediment regime, hydrology, and species pool of the system in which the change occurs, and not on the causal relation itself, which is why they are evaluated against a description of the system rather than against the hypothesis. Second, an intervention is itself an induced change of some quality, so the objective of an intervention action unit and the relata of the causal hypothesis that justifies it are entities of the same kind. This is what allows a causal hypothesis to license an intervention operation directly rather than by analogy: the change the intervention induces is the change the hypothesis relates, and the invariance under intervention that the interventionist account appeals to is invariance of precisely this relation.

Robust causal claims typically depend on a plurality of evidence types, what Illari and Russo call the *causal mosaic* (38). Causal hypotheses provide the scientific justification for the effectiveness of intervention operations, but their usefulness depends critically on whether they are applicable in a specific context. Their validity depends on environmental conditions, spatial and temporal scale, system composition, methodological assumptions, and boundary conditions. Contextual applicability conditions make this dependency explicit.

Applicability requirements are not uniform across operations but depend on the class of operation. **Transformational operations**, such as unit conversions, are typically **intrinsically applicable** as they require only that their input, typically assertional statements, satisfy specific data-schema and format constraints, so applicability is determined by schema compatibility through formal applicability conditions, and the operation is typically effectively deterministic. **Epistemic and intervention operations**, such as applying the concept of an ecological niche or a restoration method, are by contrast **contextually applicable**. Consequently, they require evaluation of referential and contextual applicability conditions against the situation, and often involve uncertainty. This distinction shows that suggesting operations grounded in general knowledge is not sufficient for reliable action, since a knowledge system may contain fully FAIR general knowledge yet, lacking mechanisms to evaluate applicability, be unable to determine whether executing that knowledge is valid in a given situation. General knowledge is in this sense not directly actionable. Universal, prototypical, and directive statements become operational only when embedded in a structure that specifies how they can be used to achieve a goal under defined conditions, which is what linking them to contextual information through applicability conditions provides.

A clarification of terminology is warranted, because our sense of valid application is deliberately bounded. When we describe an operation as *validly performed* or knowledge as *validly applied*, we mean that the conditions for its reliable execution and the expected achievement of its specified goal and thus its success conditions are satisfied on current knowledge. This is distinct from whether an action is justified in a fuller normative sense as ethically appropriate, socially legitimate, or consistent with the values and rights of all affected parties. Satisfying the applicability conditions is a necessary but not sufficient condition for justified action.

This bounding matters especially for the socio-political and institutional factors just admitted as contextual applicability conditions. Representing community consent, land tenure, or governance arrangements as conditions on the reliable and legitimate performance of an operation is deliberate, since an intervention that is ecologically sound but legally or socially impermissible is not one that can be validly performed. But representing such a condition is not the same as satisfying it, and evaluating it as satisfied is not the same as establishing that an action is just. The conditions under which knowledge may be applied to a community's land, and by whom, are articulated in the **CARE Principles for Indigenous Data Governance** (**C**ollective benefit, **A**uthority to control, **R**esponsibility, **E**thics) (39), which were formulated in explicit response to the governance-neutrality of the FAIR Principles (40). Our framework can represent the conditions that CARE requires and make their satisfaction evaluable. It cannot supply the governance through which they are decided. We return to this in Section 6.7.

Here we address the representational preconditions for reliable action. Questions of full normative justification involve additional ethical, social, and institutional dimensions that lie beyond the scope of the present work but are no less important. This bounded understanding resonates with the *adequacy-for-purpose* view in the philosophy of science (41), on which representations are evaluated not against an abstract standard of correctness but in terms of their fitness for particular purposes in specific contexts of use. Applicability conditions make that purpose- and context-relative validity explicit and evaluable rather than leaving it implicit in modelling conventions or tacit expert judgment.

The cost of leaving applicability conditions unrepresented is visible in recurring failure patterns in conservation practice. Large-scale mangrove restoration has repeatedly failed where projects relied on the well-supported general principle that planting seedlings yields ecosystem recovery, but applied it as if contextually valid without evaluating whether the conditions under which it holds were satisfied at the target site (42). Reported causes include inappropriate hydrological conditions, unsuitable sediment regimes, excessive wave energy, salinity outside species tolerances, and inadequate consideration of land-use history (20,42). In several cases seedlings were planted where mangroves could not colonize under any conditions. The general causal knowledge was structurally actionable via a procedure linked to an ecological goal, but the contextual applicability conditions defining when the procedure can be reliably applied were not represented as explicit, evaluable components, even where technical guidelines specifying such conditions existed (42). In the transfer of species distribution models (SDM) across spatial and temporal contexts, the causal and statistical relationships an SDM encodes constitute universal or prototypical statements whose transferability to new regions, time periods, or environmental regimes is consistently identified as a major challenge (43,44): models calibrated in one region frequently produce unreliable predictions when applied beyond the environmental range of their training data, which is a well-documented source of prediction failure, yet projecting models trained on current conditions to future climates is among the most common applications in conservation planning. In both cases the knowledge was structurally present but applicability-incomplete, and the failure is not one of biological knowledge but of representational infrastructure.

Representing applicability conditions by adding to each operation the typed conditions and the contextual information they are evaluated against, as shown in Fig. 4, is what applicability requires. Applicability assessment, however, remains an epistemic process that may require evidence, inference, or judgment, and the conditions themselves are rarely known completely in advance, but are discovered and refined only through **documented experience** of where an operation has held and where it has failed. This residual, i.e., the gap between conditions assessed as satisfied and outcomes observed in practice, is what the third challenge addresses.

## 2.4 The auditability challenge: trustworthy reliability requires documented empirical grounding

Representing an operation's applicability conditions allows their satisfaction to be assessed before action, but assessing a condition as satisfied is not the same as knowing that acting on it will succeed. Applicability conditions are rarely known completely in advance. In complex systems they are discovered, and refined, precisely through documented experience of where an operation has held and where it has failed. The third challenge is that current infrastructures represent neither this evidence nor the feedback it provides. If they record what is claimed to hold, and the conditions under which it is claimed to hold, they do not record where prior applications under those conditions actually succeeded or failed. We refer to the capability of a representation to support assessment of the empirical grounding for successfully applying an operation, through traceable documentation of where and why prior applications have succeeded or failed, as its **auditability**, and the component it requires is shown in Fig. 5.

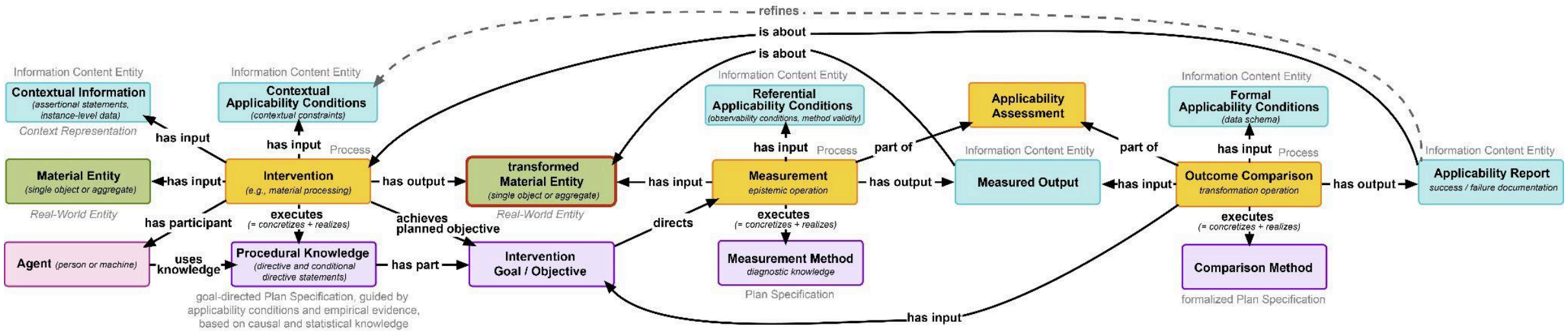

**Figure 5: Auditability requires documented empirical grounding via applicability reports.** Extending Fig. 4, the realized output of an operation is evaluated against its objective to document whether it succeeded, producing the empirical evidence on which auditability rests. This retrospective applicability assessment is a composite operation reusing the operation classes already introduced. An epistemic 'Measurement' operation takes the realized output (here the transformed material entity produced by an intervention) and produces a 'Measured Output', directed by the objective as to what to measure. A transformational 'Outcome Comparison' operation then compares the 'Measured Output' against the objective specification, executing a comparison method, and produces an 'Applicability Report' documenting success or failure. Documented failure can feed back (dashed '*refines*' arc) to refine the contextual applicability conditions against which future use is assessed — the self-correcting loop that distinguishes auditability from mere record-keeping. The measurement and comparison steps are themselves governed by referential and formal applicability conditions respectively, which is where the evidential regress terminates. The figure shows the intervention case, in which the material output requires measurement. For a transformational source operation the output is already an information content entity and feeds the comparison directly, and an epistemic operation is evaluated by the inverse epistemic operation.

The key to representing this evidence is that it is not a new kind of component appended to the operation, but the **record of an operation that actually ran and was evaluated against its objective**. The same applicability assessment introduced in the previous section as a prospective check (*does this situation satisfy the conditions for reliable use?*) reappears here in its retrospective form (*did the realized output bear those conditions out?*). Applicability assessment is thus a single epistemic capability that operates in both directions of the knowledge–action cycle, prospectively before action and retrospectively after it, and it is the retrospective pass that generates the evidence on which auditability rests.

This retrospective assessment can itself be a composite operation, built from the operation classes already introduced (Fig. 5). In case of intervention operations, its first step is an epistemic 'Measurement' that takes the intervention's realized output as its input and produces a measured representation of it, directed by the objective as to *what* to measure. Its second step is a transformational 'Outcome Comparison' that compares this measurement against the objective specification and produces an 'Applicability Report' documenting success or failure.

Whether the first step is required depends on the output type of the operation being evaluated. An intervention produces a material entity, which must be measured before it can be compared, so Fig. 5 shows this case in full. A transformation produces an information content entity, which may feed the comparison directly without a transformation step. And an epistemic operation is evaluated by the inverse epistemic operation, with a designation or description checked by recognition, and conversely, with success understood as the reliable recovery of the world–representation relation. That the measurement and comparison steps carry their own referential and formal applicability conditions is what keeps the evidence itself auditable and terminates the evidential regress.

The component that completes the figure is the **feedback from evidence to applicability conditions**. A failure documented in the applicability report, i.e., an outcome that diverged from the objective despite ostensibly satisfied conditions, shows those conditions to have been incomplete or mis-specified, and so have to be *refined* (Fig. 5, dashed arc). Auditability is therefore not merely layered on top of applicability but continually revises it. The applicability conditions against which future use is assessed are themselves the product of accumulated evidence about where prior use has succeeded and failed. Negative evidence is for this reason often as informative as positive evidence in determining the boundaries of valid application, and the systematic accumulation of failure evidence also supports the progressive construction of what Illari and Russo call the *causal mosaic* (38): where a difference-making relationship fails to hold despite ostensibly similar conditions, the pattern of failures identifies the contextual constraints that point toward the underlying mechanism, enabling a transition from difference-making toward mechanistic knowledge that documented success alone could not provide.

This refinement is, however, less direct than it appears, because a documented failure does not by itself say what has failed. A restoration procedure may fail at a site because the causal hypothesis that underwrites it is false, or because that hypothesis is true and the site falls outside the conditions under which it holds. The two call for different responses. While the first is a reason to revise the general knowledge the plan specification executes, the second is a reason to narrow the applicability conditions under which that knowledge may be used, leaving the knowledge itself intact. Auditability records the failure and the context in which it occurred. It does not adjudicate between these readings, and we do not claim that a representational framework can. What it provides is the structure in which the adjudication becomes possible, since outcomes are documented together with the conditions under which they were obtained. Outcomes that show structure in the context space, with successes concentrated in some regions and failures in others, are evidence that the conditions were drawn too widely, and it is these that the refinement loop of Fig. 5 acts on. Outcomes that show no such structure, with failures scattered across conditions the operation was expected to hold, bear instead on the general knowledge itself, and refining the conditions in response to them would preserve a hypothesis that the evidence tells against. The distinction matters for the mosaic just described, since it is the clustered failures that delimit where a difference-making relationship holds and thereby point toward the mechanism, while the scattered ones put the relationship itself in question. Auditability accumulates the evidence that supports this judgement without replacing the judgement, which remains a scientific one.

From the three capabilities (actionability, applicability, auditability), auditability is the one most clearly absent from current knowledge infrastructure design, and its necessity can be argued independently of any particular representational solution. The conservation and restoration community has recognized the need from the practice side, developing explicit taxonomies of conservation failure (20) and repeatedly calling for systematic, evidence-based learning from both success and failure (3,6), on the grounds that the field cannot improve its interventions while the conditions under which they fail go unrecorded. Yet, this evidential dimension remains largely absent from the infrastructures that represent the knowledge such learning depends on.

The cost is sharply illustrated by the control of common ragweed (*Ambrosia artemisiifolia*), an invasive annual that has spread rapidly across Europe and contributes substantially to seasonal pollen allergies (45). Roadsides are among its most prevalent habitats and act as primary dispersal corridors, with seeds transported by maintenance machinery and the airflow of passing vehicles (46), so mowing is central to confining the species, and the underlying causal hypothesis, that cutting reduces seed production and thereby population growth, is well-supported. Yet, the prevailing Austrian roadside regime of two cuts, in June and September (47), was shown in a five-year multi-site experiment to perform *worst* of all tested regimes, while alternatives including a cut before female flowering reduced projected population growth dramatically, in many cases below replacement (45,47). The reason is evidential in a way the applicability challenge alone does not capture, as mowing suppresses population growth through three distinct mechanisms, i.e., compensatory-regrowth suppression, height and fecundity reduction, and seed-viability reduction, and which mechanism dominates depends on phenological timing, plant-height distribution, and population density at the specific site. Which causal pathway a given regime triggers, and therefore whether it succeeds, can be established only by documenting outcomes across sites and years. A representation that carried the procedure, the objective, and even candidate applicability conditions, but no record of how the regime had fared across documented sites, would reproduce exactly the regime that persistently underperforms. Not because the biological knowledge is lacking, but because the evidential infrastructure to learn from documented failure is.

A representation can therefore be actionable, and even support applicability assessment, while carrying no record of how its applicability claims have fared in practice. And a field whose infrastructures cannot document failure cannot systematically learn from it. Auditability names the capability whose absence this represents. With it, Figure 5 shows the full set of components an operation requires to be performable, reliably applicable, and trustworthy in its application. How the empirical-evidence component and the progression from assessed to evidentially grounded applicability are realized representationally is developed in Section 5. Auditability is the third of the three challenges this section has set out, and together they motivate the three capabilities a knowledge representation must provide for knowledge to be actionable, reliably applicable, and trustworthy in its application. These three capabilities form the basis of the TripleA Principle introduced in Section 3.

## 2.5 Synthesis: three challenges, three capabilities

The three challenges set out in this section are not independent defects to be patched one by one. Each arose from asking what a bare operation lacks if it is to be performed not merely at all, but reliably, and on a trustworthy basis. And each exposed the absence of a component that current infrastructures, optimised for representing *what is the case*, do not provide.

The first challenge is that performing an operation requires the directive knowledge it executes and the objective it is directed at, which is a plan specification that most current infrastructures, representing terms intensionally and statements assertionally and universally, do not supply. The second is that performing an operation *reliably in a given situation* requires the applicability conditions under which its general knowledge holds, evaluated against contextual information describing the situation and thus conditions whose type depends on the operation class, and which current infrastructures leave implicit. The third is that *trusting* that reliability requires documented evidence of where prior applications have succeeded and failed, fed back to refine the conditions themselves, which is an evidential dimension current infrastructures do not represent at all. Across the three operation classes (epistemic, transformational, intervention) that actionable knowledge requires, the same deficit recurs at each level: support for performing operations, without support for assessing whether and when they may validly be performed, or how they have performed before.

These three challenges correspond to three complementary **capabilities** that a knowledge representation must support if it is to guide action reliably: (i) **actionability**, the capability to supply what an operation needs to be performed; (ii) **applicability**, the capability to support assessment of whether it can be reliably performed in a given context; and (iii) **auditability**, the capability to support assessment of the empirical grounding for that reliability through documented success and failure. The accreting figure that has run through this section (Figs. 2–5) is the visual summary of their cumulative effect, beginning from a bare operation of agent, input, and output, it acquires in turn the plan specification and objective that make it performable, the applicability conditions that make its reliable use assessable, and the documented outcomes that make that reliability auditable. With the feedback from documented outcomes to applicability conditions, it closes into the self-correcting cycle on which learning from practice depends.

The next section develops them as the TripleA Principle. Bridging the knowledge–action gap, on the account given here, begins with recognising that an actionable representation must be not only findable, interoperable, and interpretable, but performable, reliably applicable, and auditable in its application.

# 3. The TripleA Principle

The operations and challenges sections above identified three challenges in how current infrastructures represent knowledge for action. These challenges are not independent shortcomings to be patched one by one. They reflect the absence of three complementary capabilities that any knowledge representation must support if it is to guide action reliably. We name these capabilities **actionability**, **applicability**, and **auditability**, and together they constitute the **TripleA Principle**, which is a guiding principle for the design of next-generation knowledge infrastructures that extends and complements the FAIR (14,26) and CLEAR (15) Principles.

We state the principle as a requirement on representations, independent of any particular representational technology or framework. The Action Units introduced in Section 5.1 are one concrete realization of the principle within the Semantic Units Framework (22–25), but the principle itself is framework-agnostic, as it specifies *what* infrastructures must support, not *how* that support must be implemented, and future designs can be evaluated against it regardless of the formalism they adopt.

## 3.1 Actionability

Actionability is the capability of a knowledge representation to support the execution of operations, and thus the extent to which a given representation can be used by an agent to perform epistemic, transformational, or intervention actions, specifying what can, in principle, be done with it. It depends on interpretability and on the availability of operations that can process the representation as input. It is the most basic of the three capabilities, in that applicability and auditability both presuppose that there is some operation whose validity and grounding can be assessed.

Actionability is not a Boolean property but a multidimensional spectrum, determined by three dimensions: (i) **agent type** (human or machine), reflecting differences in interpretative and operational capabilities; (ii) **meaning-bearing representation level**, distinguishing non-propositional representations (terms) from propositional, meaning-bearing representations (statements); and, for propositional representations, (iii) **statement generality**, distinguishing assertional (instance-level) statements from universal, prototypical, directive, and conditional directive (type-level) statements. These dimensions jointly determine the degree and type of actionability, with statement generality being defined only for propositional representations.

As established in Section 2 earlier, current concept- and data-centric infrastructures support understanding but not the operations through which knowledge is put to use — a representational limitation whose roots are examined in Section 5. Actionability in the sense intended here is therefore largely unmet, since most infrastructures support *what is the case* and *how X relates to Y*, but not the operations through which knowledge is put to use. Actionability names the capability whose absence this represents.

## 3.2 Applicability

Applicability is the capability to support assessment of whether a knowledge representation can be validly applied in a specific context, and thus whether an operation can be reliably performed in a given situation, in accordance with the conditions under which the knowledge has been shown to hold. It depends on the relationship between the contextual information describing the situation and the applicability conditions associated with the representation. It is not a guarantee of success, since outcomes in complex systems are never certain, but a principled assessment of whether the known conditions for reliable application are satisfied in the context at hand. Applicability can be understood as the epistemic characterization of actionability in context. Where actionability describes the structural potential for action, applicability determines whether that action can be reliably realized here.

Like actionability, applicability is graded rather than Boolean, and multidimensional. It is assessed against a structured set of individual conditions, each of which may be satisfied, partially satisfied, or unsatisfied, so that the applicability profile is the structured set of these assessments rather than a single aggregate measure, and whether partial satisfaction warrants action is a reflective judgment weighting the relative importance of the conditions against the costs of action versus inaction. The capability does not eliminate the uncertainty inherent in such judgments but makes the explicitly representable conditions visible and evaluable, while acknowledging that human judgment always involves a tacit residue the schema cannot fully capture (Section 4.4).

Applicability requirements depend on the type of operation. Some operations are intrinsically applicable on the basis of representational compatibility alone, whereas others require context-dependent evaluation of applicability conditions. This is the capability the corresponding challenge section identified as absent, since current infrastructures that support the execution of operations often do not support evaluation of whether those operations can be reliably executed in a given context, and so enable action without ensuring its appropriateness.

## 3.3 Auditability

Auditability is the capability of a knowledge representation to support assessment of the empirical grounding for successfully applying an operation in a specific context, through traceable documentation of where and why prior applications have succeeded or failed. Where actionability describes the structural potential for action and applicability the assessment of whether action can be reliably realized in a given context, auditability concerns the evidential basis on which such applicability claims rest. It can be understood as the evidential characterization of applicability: where applicability assesses whether the known conditions for reliable application are satisfied, auditability assesses the empirical grounding of those conditions through traceable evidence of prior success and failure. Like actionability and applicability, it is graded rather than Boolean, ranging from no documented evidence, through a single application, to extensive documentation across comparable contexts, including failure instances.

Of the three capabilities, auditability is the one most clearly absent from current knowledge infrastructure design, and its necessity can be argued independently of any particular implementation. As Section 2.4 set out, applicability conditions are rarely known completely in advance but are discovered and refined through documented experience of where an operation has held and where it has failed, a pattern that recurs across domains from conservation to medicine. A representation can therefore be actionable, and even support applicability assessment, while carrying no record of how its applicability claims have fared in practice, and a field whose infrastructures cannot document failure cannot systematically learn from it. Auditability names the capability whose absence this represents, and, because documented failure feeds back to refine the conditions it tests, it does not merely sit atop applicability but continually revises it.

This argument stands on the epistemology of application and the documented needs of practice across domains, independent of how auditability is realized representationally and implemented technically. Section 5 shows how the Semantic Units Framework makes this capability concrete through the progression from action units, to applicable action units, to audited applicable action units. With respect to the TripleA Principle, it is sufficient that the capability is necessary (and typically unmet).

## 3.4 The Principle

The three capabilities are nested in their dependencies. Applicability presupposes actionability, as there is no question of whether an operation can be reliably performed unless the representation supports that operation in the first place. Auditability presupposes applicability, because documenting where an operation has succeeded or failed is documentation of where its applicability conditions held or broke down. The three therefore form an ordered progression rather than an unordered list, ranging from structural potential, to contextual assessment, and to evidential grounding. The dependency is not, however, purely one-directional. The evidential grounding that auditability provides feeds back to refine the applicability conditions on which it depends, so that the progression is also a cycle in which documented use continually revises the conditions against which future use is assessed. The capabilities apply both prospectively, when deciding whether and how to act in a specific situation, and retrospectively, when evaluating whether past actions were grounded in knowledge appropriate to the context in which it was applied.

These three capabilities form the basis for the three criteria of the TripleA Principle, that is the requirement that knowledge representations, whether they are concepts or propositions, be actionable, applicable, and auditable wherever possible. Actionability applies to both individual concepts and proposition, though in different senses as terms are actionable at the level of recognition and designation, whereas the full operational actionability that applicability and auditability presuppose is a property of statements, since only statements carry the procedural specifications, objectives, and conditions an operation requires. Where the capabilities are properties a representation can possess to a graded degree, the criteria are the normative requirements a knowledge infrastructure should satisfy. The principle is framework-agnostic and elevates the three capabilities into design criteria against which infrastructures can be evaluated independent of the representational formalism they use to satisfy them (Box 2).

**Box 2 The TripleA Principle.** A guiding principle for the design of next-generation knowledge infrastructures, extending and complementing the FAIR and CLEAR Principles. Grounded in the three capabilities of actionability, applicability, and auditability, the principle comprises three criteria that knowledge representations ideally should meet:

| |
|---|
| **1) Actionable:** explicitly specifying, for a given knowledge representation (concept or proposition), the operations through which it can be used, together with the plan specification those operations execute (diagnostic knowledge, methods, procedural knowledge), the objective they are directed at, and the agents and processes they involve. |
| **2) Applicable:** explicitly specifying the conditions under which a given knowledge representation (concept or proposition) can be validly used in a specific context, and enabling systematic evaluation of those conditions against the contextual information describing the situation before action is taken. |
| **3) Auditable:** making the empirical evidence for an operation explicit and traceable, in the form of documented outcomes of prior applications in comparable contexts, including both successes and failures, and assessable by human and machine agents, so that the applicability conditions and the boundaries of valid application can be refined over time. |

The TripleA Principle is intended to extend and complement, not supersede, the FAIR (14,26) and CLEAR (15) Principles. FAIR advanced the machine findability, accessibility, interoperability, and reusability of data, metadata, and knowledge, while CLEAR added the cognitive interoperability that makes knowledge accessible and interpretable to human agents. Both are concerned principally with how knowledge is *represented, organized, and made accessible*. TripleA addresses the complementary question of how knowledge is *put to use* and thus whether a representation supports operations, whether their valid application in context can be assessed, and whether the empirical grounding of that application is documented and traceable. A representation can be fully findable, interoperable, and cognitively accessible, and thus FAIR and CLEAR, and still provide no basis for determining what can be done with it, whether doing so is valid here, or how that judgment has fared in prior practice.

The three criteria are also what the opening vignette illustrated before they were named. The infrastructure identified what operations were available to each agent (actionability), narrowed them to those whose conditions the specific ecosystem profile satisfied (applicability), and flagged where the documented evidence base did and did not cover the case at hand (auditability). The principle names the three criteria, as general requirements on knowledge infrastructures, the capabilities that scene depended on. Satisfying the criteria marks a transition toward what we term **post-FAIR knowledge infrastructures** and thus infrastructures in which knowledge is not only findable, interoperable, and cognitively accessible, but operationally usable, contextually assessable, and evidentially accountable. The following section introduces Action Units within the Semantic Units Framework as one concrete strategy for attaining the principle, and the section thereafter develops the infrastructure implications and the post-FAIR transition in detail.

# 4. Completeness and Sufficiency of the Principle

Earlier, we derived three capabilities as answers to successive questions about an operation, and in the previous section elevated them into the three criteria of the TripleA Principle. That each criterion is *necessary* follows from the derivation as each addresses a distinct deficit, i.e., unperformable operations, unassessable reliability, ungrounded reliability, that the others leave untouched, so none can be dropped without reopening a challenge. Necessity, however, is the weaker half of what a principle must establish. The stronger claim is *sufficiency*, meaning that these three criteria, and no fourth, capture what a knowledge representation must support for the valid use of actionable knowledge. A principle that is merely a useful list invites the question of what it omits, whereas a principle that can argue its own completeness answers that question in advance. This section makes the sufficiency argument, and bounds it honestly.

The argument has three parts. First, the three classes of operation on which the capabilities are defined are combinatorially exhaustive, so the capabilities are not defined over an open-ended space of operation types (Section 4.1). Second, this combinatorial closure converges with an independent philosophical delimitation of the same space, lending the typology support from two directions rather than one (Section 4.2). Third, the composition of operations, and thus the way real tasks chain operations of different classes, introduces no requirement the three capabilities do not already meet, so compositional complexity does not call for a fourth criterion (Section 4.3). Together these bound a sufficiency claim that is deliberately restricted to the *valid-use* aspects of actionable knowledge representation, leaving aside both the full normative justification of action and the irreducibly tacit residue of human judgement (Section 4.4).

## 4.1 The operation classes are combinatorially exhaustive

The three capabilities are not defined in the abstract but as requirements on operations, and the operations themselves fall into three classes (epistemic, transformational, intervention) distinguished in Section 2.1 by what their input and output are. This grounding is what makes a completeness argument possible. If the space of operation classes is closed, then capabilities defined as requirements on operations inherit a bounded domain, and there is no unaccounted class of operation smuggling in a need for a fourth capability.

That the three classes are exhaustive can be shown combinatorially. At the level at which operations relate the world to its representation, an operation's input and output are each of one of two kinds: a **material entity** (a real-world object or system) or an **information content entity** (a representation). The two kinds, taken as input and as output, yield four possible input→output combinations, and every operation falls into one:

- **information content entity → material entity**: a representation is taken as input and the corresponding entity or situation is identified in the world, which are the components of *epistemic operations* in the recognition direction;
- **material entity → information content entity**: an entity or situation is taken as input and a representation describing it is produced, representing the components of *epistemic operations* in the designation and description direction;
- **information content entity → information content entity**: a representation is taken as input and a derived representation produced, wholly within the representational domain and thus representing *transformational operations*;
- **material entity → material entity**: a system in one state is taken as input and a modified system produced, as it is the case in *intervention operations*.

These four cases exhaust the possibilities, and the three classes cover them. Epistemic operations occupy the two boundary-crossing cases (the recognition and designation/description directions), transformational operations the representation-internal case, and intervention operations the world-internal case. Given that an operation's input and output are each either a material entity or an information content entity, there is no further case and hence no fourth class of operation. This makes the three classes exhaustive rather than merely useful. However, the claim must be bounded in two ways.

First, the closure is relative to the two-category distinction between material entities and information content entities and depends on this ontological commitment, which is BFO-aligned and adopted throughout this work, not a claim that no other categorisation of inputs and outputs is possible. Second, the exhaustiveness is of operation *classes*, not of the contextual factors any particular operation must weigh. The set of potentially relevant applicability conditions in a real-world system is in principle unbounded, and the typing of conditions into referential, formal, and contextual kinds is pragmatic rather than exhaustive (Section 4.4). The combinatorial argument closes the space of operation *types* and does not claim to close the space of *conditions* those operations must assess.

## 4.2 The combinatorial closure converges with an independent philosophical delimitation

The combinatorial argument shows that the three operation classes exhaust the input→output combinations available given two categories of entity. On its own, this might be read as an artefact of the chosen categorisation. It carries more weight because an independent line of argument, owing nothing to the combinatorics, delimits the same space and partitions it the same way. Where two arguments proceeding from different starting points converge on one structure, that structure is less likely to be an accident of either.

The independent delimitation comes from the philosophy of language and the philosophy of science, and it operates at two levels: the statements an operation executes, and the operations themselves.

At the level of statements, the relevant delimitation is the one introduced in Section 2.2: Searle's two directions of fit (31,32). It is exhaustive. A representation either answers to the world or the world answers to it, and the finer statement distinctions of Jansen and Schulz (33), as developed within the Semantic Units Framework (25), fall under one or the other without remainder. The statement types an operation can execute are therefore bounded by a two-valued distinction, not an open list.

At the level of operations, the corresponding delimitation is Hacking's between **representing and intervening** (28). Hacking argued that representing the world and intervening upon it are irreducible modes of scientific engagement, neither reducible to the other, with intervention having a life independent of representation. The three operation classes map onto this distinction exactly, with the epistemic operations aligning with Hacking's representing, intervention operations with his intervening, and transformational operations the limiting case in which the activity remains wholly within the representational domain with representing without world-involvement on either side. The two directions of fit and the representing/intervening distinction are the same boundary seen from two sides. World-to-word statements are what intervention operations execute, word-to-world grounding is what epistemic operations establish, and the boundary between representation and world that Hacking's distinction marks is the boundary that the combinatorial argument's two entity categories define. The philosophical and combinatorial delimitations are thus not merely consistent but coextensive, as they pick out the same partition because they are tracking the same underlying asymmetry between representing the world and acting upon it.

A third convergence appears within the epistemic class, where the combinatorial argument already distinguished two directions, i.e., recognition (representation → world) and designation/description (world → representation). These instantiate Kant's distinction between **determinative and reflective judgement**. Recognition, in which a representation is given and the task is to identify the instance that falls under it, is determinative judgement, as the universal is given and the particular is subsumed. Designation and description, in which a particular is given and an adequate representation must be found or constructed for it, is reflective judgement, as here the particular is given and the universal must be found. The two epistemic directions identified combinatorially are therefore not an arbitrary subdivision but correspond to the two forms judgement can take, thus providing a third independent delimitation landing on the same structure. The distinction also marks the point at which machine support for epistemic operations reaches its limit. Determinative operations apply a given concept or statement to a case, and are in principle amenable to machine execution over explicitly specified recognition criteria. Reflective operations require the agent to perceive and articulate what is diagnostically significant about a particular for which no concept is yet given, a form of engagement with the particular that resists complete prior specification. The reflective direction is therefore where the tacit residue discussed in Section 4.4 concentrates, and where human judgement remains least replaceable.

These three delimitations (combinatorial, drawn from the categories of entity; semantic-pragmatic, drawn from directions of fit; epistemological, drawn from the forms of judgement) converge on one partition and represents the strongest available evidence that the partition is real. The three operation classes are not three conveniences but the joint fixed point of three convergent ways of carving the space, and the capabilities defined over them inherit that closure.

## 4.3 Composition introduces no fourth criterion

The combinatorial and philosophical arguments bound the *classes* of operation. But real tasks rarely consist of a single operation of a single class but, instead, chain operations together, often across classes. A natural worry is that composition, meaning the assembling of operations into higher-order operations, is a further capability in its own right, requiring representation beyond what the three criteria provide, and so a fourth criterion. This section shows this is not the case, as composition is handled by the part–whole structure the framework already carries, and a composite operation makes no demand that its component operations, each already covered by the three capabilities, do not.

Operations compose in two ways. The looser is *sequencing*, with the output of one operation becoming the input of the next, as when an epistemic operation creates data that a transformational operation then transforms and analyses. The tighter, and more revealing, is *constitutive composition*, in which an operation of one class requires, as a necessary internal step, an operation of a different class. A recognition operation may require, as its first step, an intervention that prepares the material entity to be recognised. In histological tissue composition analysis, the tissue must first be dissected and chemically stained (an intervention operation) before its composition can be identified by microscopy (an epistemic recognition operation) (Fig. 6). Also recall the method-dependent recognition criteria for '*cell nucleus*' discussed earlier, which requires an intervention operation before the epistemic operation. The intervention is not a separate operation that merely precedes the recognition but is constitutive component of it, because without the staining the referential applicability conditions for compositional recognition cannot be satisfied. The coarse-grained recognition has both the intervention and the finer recognition as parts, takes the untreated tissue as its overall input, and produces the compositional description as its overall output.

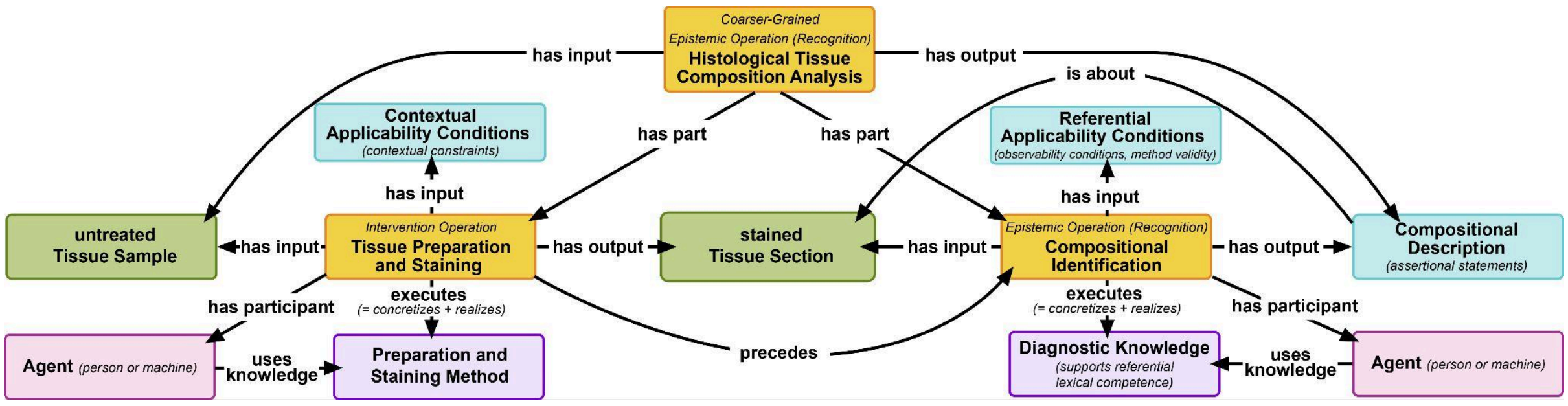

**Figure 6: A granular composite operation spanning multiple operation classes.** A coarser-grained epistemic recognition operation, *Histological Tissue Composition Analysis*, decomposes into two ordered steps of different classes connected by a *precedes* relation. Step 1 is an intervention operation (*Tissue Preparation and Staining*) that takes the untreated tissue sample as input and produces a stained tissue section, governed by contextual applicability conditions and a preparation and staining method. Step 2 is an epistemic recognition operation (*Compositional Identification*) that takes the stained tissue section as material input, together with diagnostic knowledge and referential applicability conditions, and produces a compositional description. The composite operation has both steps as parts, takes the untreated tissue sample as its overall input, and produces the compositional description as its overall output. Step 1 is a constitutive precondition for step 2, as without prior staining, the referential applicability conditions for compositional recognition cannot be satisfied. For clarity, the *executes* relations and agent participants of the parent operation are not shown, since these are inherited from the component operations through the *has part* relation.

Two features of this example bear on sufficiency. First, the boundary between representing and intervening that Section 4.2 drew is not, at the level of composite operations, a clean sequential boundary, since a higher-order operation of one class may constitutively require a step of another, so the classes, though exhaustive and mutually exclusive *as classes*, interleave freely *in composition*. This is not a breach of the typology but a consequence of it. Composition can only chain operations that the typology has already individuated. Second, and decisively for the present argument, the composite operation introduces no representational requirement that its components do not. Each component operation is itself epistemic, transformational, or intervention, each therefore has its own plan specification, objective, applicability conditions, and, once executed, evaluable outcome, and the composite operation inherits these through the part–whole relation rather than introducing components of its own (Fig. 6). Its plan specification, applicability conditions, and objective are those of its parts, arranged. The composite's actionability is the coordinated actionability of its parts, its applicability is the conjunction of theirs, together with the ordering constraints among them, and its auditability is the auditability of its parts plus the record of the composite's overall outcome. Nothing in the composite exceeds the three capabilities applied to its components and to their arrangement.

The arrangement itself, i.e., the ordering of component operations within a composite, is carried by the same part–whole logic that governs compound representations generally, with the ordering constraints held in the plan specification of the higher-order operation. Composition is therefore a structural relation among operations already covered by the three criteria, not a new kind of demand on representation. A framework that can represent each operation, and the part–whole relations among operations, can represent any composite of them. It requires no fourth capability to do so. The three criteria are closed under composition. Applied to the parts and to their arrangement, they suffice for the whole.

## 4.4 The bounded sufficiency claim

The three arguments so far, i.e., combinatorial closure, philosophical convergence, and closure under composition, support the sufficiency claim that the three criteria capture what a knowledge representation must support for actionable knowledge to be validly used. But the claim must be stated at the right strength, because two things it might be taken to assert, it does not. Naming them is not a retreat from the claim but a specification of it as the criteria are sufficient for a bounded target, and the boundary is principled.

The first boundary is **normative**. The criteria concern whether an operation can be *reliably* performed, i.e., whether the conditions for its successful execution and the achievement of its objective are satisfied on current knowledge. This is distinct from whether performing it is *justified* in a fuller sense, including whether it is ethically appropriate, socially legitimate, or consistent with the rights and values of those affected. Satisfying the applicability conditions is necessary but not sufficient for justified action. The framework addresses the representational preconditions for *reliable* action. Questions of full normative justification, however, involve ethical, social, and institutional dimensions that lie beyond its scope but are no less important.

Where such constraints can themselves be represented as applicability conditions (community consent, tenure rights, regulatory compliance), the framework accommodates them as evaluable components rather than placing them wholly outside its reach. This is not, however, to absorb the normative question into the representational one. The conditions under which knowledge may be applied to a community's land, and by whom, are articulated in the **CARE Principles for Indigenous**

**Data Governance** (39) and it might be asked why these do not constitute a fourth criterion alongside the three. They do not, because they do not operate on the same axis. Actionability, applicability, and auditability are criteria on what a representation must supply for an operation to be *reliably grounded*, whereas CARE's criteria constrain whether an operation *ought to be performed at all*, and by whom. A representation can fully satisfy the three criteria and be used to do something that ought not to be done. Adding a governance criterion to the three would not extend the closure argument but change its subject. The sufficiency claim is therefore for the *valid-use* aspects of actionable knowledge and thus reliability of application, not for the full justification of action, and our exposure along the second axis is developed in Section 6.7. This bounded target is the one the **adequacy-for-purpose view** in the philosophy of science already recommends (41), with knowledge being evaluated for fitness to a purpose in a context, not against an abstract standard of correctness, and the criteria make that purpose-relative fitness explicit and evaluable.

The second boundary is **epistemic**, and it has two layers. The shallower is that the *typing* of applicability conditions into referential, formal, and contextual kinds does not exhaust them, and need not. The number of potentially relevant contextual factors in a real-world system is in principle unbounded, and the framework's claim is not that a finite set of typed conditions can capture all of them, but that making the most important known conditions explicit provides a structured scaffold that supports judgement rather than replacing it. The three *operation classes* are exhaustive (Section 4.1) whereas the *conditions* an operation must weigh are not, and the criteria do not pretend otherwise.

The deeper layer concerns what resists explicit representation altogether. Skilled human assessment of applicability draws on tacit knowledge and embodied expertise and thus what Polanyi called the **tacit dimension** (48) that grounds judgement in ways that cannot be fully articulated propositionally. A practitioner may assess whether a restoration procedure fits a site through experiential pattern recognition that has never been, and perhaps cannot be, codified. This maps onto the distinction drawn in Section 4.2 that the framework supports the *determinative* components of applicability assessment, in which explicit conditions are applied to a case, while the *reflective* components, i.e., finding the right characterisation of a novel situation when no rule fully specifies it, depend on judgement that resists complete specification.

The same residue appears within the transformational class, in its ampliative operations. Judging that a new situation is relevantly similar to a documented one, which is the analogical inference on which much applicability assessment rests, is not a formal deduction but an inductive or abductive judgment that accumulated cases can inform but not determine. The same limit has long been recognised in legal theory, where subsuming a particular case under a general rule has been argued to involve an ineliminable element of judgment that codification and accumulated precedent can guide but never fully determine (49,50). The parallel is close, since assessing whether a situation falls under a general applicability condition is structurally the same problem as deciding whether a case falls under a general rule, and in both, documented prior cases narrow the space left to judgment without closing it.

The framework does not claim to replace this judgement but makes the explicitly representable components visible and evaluable, and, through progressive articulation, extends the reach of what has been made explicit over time, converting previously tacit insight into stated conditions as evidence accumulates, without supposing that full articulation is ever reached.

Beneath even this lies a limit intrinsic to any explicit approach. Every specification of applicability conditions rests on background assumptions that are themselves left unstated . This refers to what Wittgenstein called **hinge propositions**, so foundational to a practice that doubting them would unravel it (51). The reliability of instruments, the stability of natural regularities, the uniformity of the systems under study function as the unquestioned background against which any applicability assessment proceeds. These cannot be made fully explicit without regress, as stating them would invoke further assumptions that themselves go unstated. The framework inherits its hinges from the human practices that design and populate it, and acknowledges this inheritance as a limit of any representational approach rather than a defect of this one. It bears on the sufficiency claim only to fix its register, as the criteria are sufficient for what a representation can *make explicit and evaluable* about the valid use of knowledge, against a background of practice they do not themselves constitute.

Within these boundaries the claim stands. For the valid-use aspects of actionable knowledge, i.e., whether an operation can be performed, whether it can be reliably applied in a context, and whether that reliability is empirically grounded, actionability, applicability, and auditability are jointly sufficient and individually necessary. They are sufficient because the operations over which they are defined are combinatorially and philosophically closed, and closed under composition. They are necessary because each addresses a deficit the others leave standing. The three criteria are therefore not a useful list of desiderata but a complete specification of what a knowledge representation must support to make the valid use of knowledge explicit, evaluable, and accountable, which is what a principle, as distinct from a checklist, must be able to claim of itself.

# 5. From Principle to Representation: Action Units and the Semantic Units Framework

The TripleA Principle specifies what a knowledge representation must support: the operations through which knowledge is used, the conditions under which their use is valid, and the evidence grounding those conditions. It does not specify *how* these are to be represented. Section 4 argued, at exactly that level of abstraction, that the three capabilities are necessary and jointly sufficient, as a requirement on representations, not a representation. This section makes the transition from that result to one concrete way of meeting it, showing how the Semantic Units Framework provides the capabilities the principle requires. It proceeds in three steps. First, a specification of the representational components an operation requires, in the form of **action units** (Section 5.1). Second, the observation that these components cannot be represented by single resources, posing a representational challenge the principle's abstraction left open (Section 5.2). And third, the framework that resolves it, in which statements are treated as first-class units of meaning, i.e., the **Semantic Units Framework** (Section 5.3).

## 5.1 Action units as a specification of representational components

The operation figures of Section 2 (Figs. 2–5) describe operations as processes, with an agent taking an input, executing a plan specification directed at an objective, under applicability conditions, and, once run, producing an output that can be evaluated. But a process is something that *happens*. For knowledge to be actionable in advance of any particular execution, what must be represented is not a completed process but the *specification* of everything an operation of that kind requires. That specification is what an action unit provides.

An action unit is a structured representation that names the components an operation needs, drawn together as a single referenceable object (Fig. 7). It extends the notion of a **plan specification**, i.e., in the standard sense of an information content entity specifying how a process is to be carried out, by making the components the TripleA Principle requires explicit, typed, and individually addressable. Its mandatory components comprise: (i) **input and output specifications**, defining the types and roles of the entities the operation takes and produces; (ii) a **plan specification** proper, carrying the diagnostic, methodological, or procedural knowledge the operation executes; and (iii) an **objective**, characterising the intended outcome. These are the components an operation needs to be performable, providing the representational counterpart of actionability.

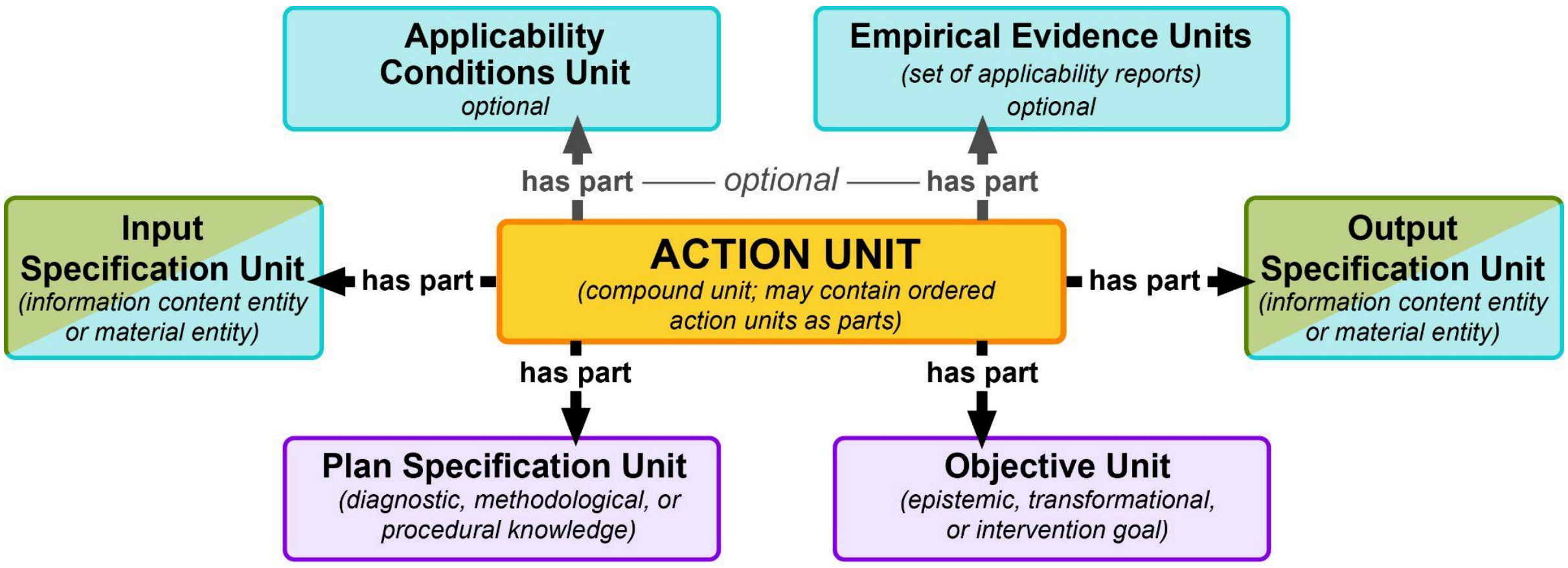

**Figure 7: Action unit.** A structured object whose typed components are related to it by '*has part*'. The input/output specification, plan specification, and objective components are mandatory for every action unit and represent actionability. The applicability conditions component (optional at the base level, mandatory for applicable action units) represents applicability. The empirical evidence component (mandatory only for audited applicable action units) includes the corresponding applicability reports and represents auditability. Action units may be granular, a coarser-grained action unit having finer-grained action units as ordered parts (cf. Fig. 6). The component labelled *plan specification unit* carries the procedural, methodological, or diagnostic knowledge the operation executes. This is distinct from the sense in which the action unit as a whole extends the notion of a plan specification, since the action unit specifies the operation's full structure, including conditions, evidence, and objective, while the plan specification component carries only the executed knowledge.

Two further components are optional at this base level but become mandatory as the principle's higher criteria are met: an **applicability conditions** component, specifying the referential, formal, or contextual constraints under which the operation can be reliably applied, and an **empirical evidence** component, documenting prior success and failure instances in comparable contexts in the form of applicability reports. The result is a three-level structure that mirrors the three capabilities exactly. An **action unit** with its three mandatory components provides all information necessary to perform the corresponding operation in principle, embodying actionability. An **applicable action unit** adds the applicability conditions, and so additionally supports assessment of whether, on current knowledge, the operation can be reliably applied in a given context, embodying applicability. An **audited applicable action unit** adds documented empirical evidence, and so additionally supports assessment of the grounding for that reliability, embodying auditability. The three levels are strictly nested, with every audited applicable action unit being an applicable action unit, and every applicable action unit being an action unit, each level enriching the last with the components the next criterion requires. The action-unit hierarchy and the three capabilities are thus two views of one structure. The hierarchy naming the representational levels, the capabilities naming what each level supports (the levels are developed further, with their gradation and infrastructural consequences, in Section 6).

Because the components an operation requires depend on its class, action units specialise into three types corresponding to the three operation classes. **Epistemic action units** carry diagnostic knowledge and referential applicability conditions, and represent operations that ground representations in the world through recognition, designation, and description. **Transformational action units** carry methods or algorithms and formal applicability conditions, and represent operations that derive representations from representations. **Intervention action units** carry procedural knowledge and contextual applicability conditions, and represent operations that effect goal-directed change in material systems. Being the goal-oriented endpoint, they are the most integrative, coordinating contextual information, applicability conditions, and procedural knowledge toward a specified objective.

Action units may also be composed, with a higher-order action unit having lower-order action units as ordered parts, carrying the compositional structure of Section 4.3 to the representational level, with the ordering constraints held in the plan specification of the higher-order unit.

What matters for the argument that follows is the status of this specification. An action unit, as Fig. 7 presents it, is a compound object whose parts are labelled by *role*. These labels name what knowledge each component must carry, but they do not say how that knowledge is itself represented. The action unit specifies the roles. It does not, on its own, discharge the obligation to represent what fills them, and that obligation is the representational challenge of the next section.

## 5.2 The representational challenge: action unit components are not single resources

The action unit specifies what an operation requires, but it specifies it as a set of *roles*: a plan specification carrying diagnostic, methodological, or procedural knowledge the operation carries out, an applicability conditions component carrying evaluable constraints, an objective component specifying the goal of the operation, an input and an output specification component declaring what the operation requires as its input and what it produces as its output, and an empirical evidence component connecting the operation to evidence about its successful and failed applications in the form of one or more applicability reports. Fig. 7 draws each component as a single node. The representational challenge is what happens when one asks what fills those nodes, because each component cannot be adequately represented by a single resource without losing relevant semantic content.

Consider what a plan specification component must carry to be practically usable: a procedure of ordered, often conditional steps of the form "*if the site's tidal inundation falls within range X, proceed with natural regeneration; else initiate active replanting.*" That is not a term to be looked up but a structured set of directive statements. An applicability conditions component must carry a set of constraints, each evaluable against the situation at hand, which, again, are represented as statements and not as individual resources. An objective is a specification of a target state, the input and output specifications declare the types and roles of the entities the operation takes and produces, the empirical evidence component, as a collection of applicability reports, records prior outcomes together with the conditions under which they occurred. Every component of an action unit, on inspection, resolves into *statements* and thus propositions relating entities, specifying conditions, prescribing steps, and often into many of them, bound together into a coherent whole. Each component node in Fig. 7 is a placeholder for a statement-structured body of knowledge, not for a single addressable thing that can be adequately represented by a single resource.

This is the representational challenge, and it has two parts. First, to represent an operation's components is to represent statements as **first-class, referenceable units**, i.e., units that can be pointed to, typed, associated with conditions and provenance, and reasoned about as wholes. Second, the components require not statements of a single kind but the **distinct statement types** the operation's structure demands. Plan specifications are represented via universal, directive, and conditional directive statements, applicability conditions via conditional constraints, empirical evidence via assertional statements, and the objective specifies a target state — representing the same descriptive/prescriptive division introduced in Section 2.2, now as a representational requirement rather than a conceptual distinction. Representing action units therefore requires a framework in which statements are both first-class and typed by kind, including the directive and conditional directive types that carry prescriptive, world-to-word content. Whether current knowledge infrastructures can do this is the question the action unit forces, and the answer, across the paradigms that shape those infrastructures, is that most of them do not. Not, because statements are absent from them, but because most of them neither treat statements as first-class units, nor provide the typed statement forms, in particular the prescriptive ones, that actionable knowledge requires.

Four paradigms shape current infrastructures, and each improves some dimension of representation while leaving this one unmet. **Ontologies** provide formal, concept-centric representations, answering "*what is X?*" through intensional definition, but generally lack the diagnostic and procedural knowledge required to recognise instances or guide action (26,27), supporting conceptual clarity, not operational use. **RDF/OWL-based knowledge graphs**, the primary technical infrastructure for implementing ontologies (18,52), encode information as binary *Subject–Predicate–Object* triples (53). This model is well-founded for binary relations but inherently so. Representing an n-ary statement, such as a context-dependent claim or a procedural specification, requires decomposing it across multiple triples, through reification or intermediate nodes (54), which fragments a semantically coherent statement across the graph and makes it difficult to treat as a unit, to associate with provenance and applicability conditions, or to use directly in context-sensitive reasoning. **The FAIR Principles** have transformed the findability, accessibility, interoperability, and reusability of data (14), but as a data-centric framework FAIR concerns discovery and integration, not how knowledge is translated into context-sensitive action. **The CLEAR Principle** complements FAIR by emphasising cognitive interpretability and contextual exploration, and points implicitly toward statements as meaning-carrying units (15), but provides no explicit mechanism for linking knowledge to procedures, goals, or interventions. Together the four paradigms improve access, structure, and interpretability, but none supplies what the action unit's components require, which is the representation of statements as first-class, typed, actionable units of knowledge.

That statements, not terms, are the primary carriers of meaning is the premise this challenge rests on, and it is worth making explicit, because it is easy to grant in principle and lose in practice. A term such as '*salinity*' in isolation makes no claim and conveys nothing about any situation. Embedded in a statement such as "*Salinity at site X exceeds the threshold for mangrove establishment*," it makes a claim that can be assessed, acted upon, or refuted. Meaning in scientific discourse is primarily carried by propositions about relationships, observations, and processes, not by isolated terms, and the recognition that these propositions must themselves be first-class has recurred across communities. The nanopublication model was proposed precisely to make core scientific statements, rather than raw triples, findable, citable, and provenance-bearing (55), and foundational work in Linked Data emphasised that reusable data requires coherent, self-contained statements with clear semantic roles rather than decontextualised triples (56). Even technical extensions that attach metadata to individual triples, such as RDF-star (57), do not resolve the deeper problem that a statement requiring several triples to be modelled cannot be referenced or reasoned about as a unit without an explicit representational commitment to the statement level. And building such units alone is not enough. The prescriptive statement types the action unit's components depend on, the directive and conditional directive forms, have received far less first-class representational treatment than descriptive assertions, so that even where statements are made addressable, the types that specify *how to act* remain the least well-supported. That twofold commitment, i.e., statements as first-class units, and the full range of statement types, prescriptive as well as descriptive, as distinct typed forms, is what the action unit's components require and what current infrastructures withhold, and it is what the framework of the next section supplies.

## 5.3 The Semantic Units Framework

The representational challenge sets two requirements: statements must be representable as first-class units, and the distinct kinds of statement the action unit's components demand must be representable as distinct typed forms. The **Semantic Units Framework** meets both. A **semantic unit** is a modular, self-contained, referenceable unit of meaning that can be interpreted independently within a knowledge graph (22–25,34). Where the triple is the technical unit of the underlying graph, the semantic unit is the unit of *meaning* laid over it. As such, an individual semantic unit may correspond to many triples, but is addressable, represented in the graph with its own typed resource, and reasoned about as a single whole, which is precisely the first-class status the representational challenge requires, provided independently of the underlying n-arity of its semantic content.

The framework distinguishes two categories of semantic unit. A **statement unit** represents an individual proposition and a **compound unit** represents a semantically coherent collection of statements — and, recursively, of other compound units — corresponding to structured knowledge objects such as causal hypotheses, workflows, or system descriptions. This modular, potentially nested organisation lets knowledge be decomposed into units that retain semantic integrity while remaining linkable and reusable across contexts, so that a statement fragmented across many triples at the graph level is reassembled into a single referenceable unit at the level of meaning.

Crucially for the action unit, statement units are **typed by kind**. The descriptive and prescriptive statement types distinguished in Section 2.2 (i.e., assertional, existential, universal, prototypical, directive, and conditional directive) are represented as distinct types of statement unit, each with its own structure and admissible roles (25). This is what discharges the second half of the representational challenge, as the framework does not merely make statements first-class, it gives the prescriptive types, and thus the directive and conditional directive forms that carry world-to-word content and that current formalisms support least well, the same first-class, typed representational status as descriptive assertions.

Typing statements by kind requires giving each type a formal structure, and the framework does this through SHACL shapes supporting machine-interpretability, and formalized natural language statement schemata (26,34) supporting human-interpretability. A formalized natural language statement schema, implemented in the Semantic Units Framework as '*dynamic label*' schema (15,22,25), specifies three things for a given type of statement: (i) its structural organisation as a predicate–argument form familiar from linguistics, built over the subject–predicate–object syntax of the underlying graph; (ii) the semantic role each syntactic position carries; and (iii) the constraints on the values admissible in each position. These three components allow the alignment of a given formalized natural language statement schema with its corresponding SHACL shape. Making these explicit is what lets statements of a given type be represented, interpreted, and validated consistently across datasets and domains, and it is what supports statement-level epistemic operations in both directions, recognition from representation to world and description from world to representation. The schemata also carry the framework's claim to serve human and machine agents at once, as the CLEAR Principle (15) recommends. For a human agent, a statement is interpreted through its predicate–argument structure and the semantic roles of its positions. For a machine agent, the same statement is processed through schema-constrained matching, querying, and validation against its corresponding SHACL shape. The pairing of a formalized natural language schema with its corresponding SHACL shape is what both read, which is how the framework makes a statement cognitively interpretable and computationally processable at the same time.

The schemata also clarify how general knowledge relates to the situations it governs. When represented through formalized natural language schemata, assertional statements become epistemically actionable directly, since they describe concrete situations and so provide the contextual grounding against which applicability is assessed. Universal and prototypical statements are not epistemically actionable in isolation. The universal statement "*increased salinity inhibits mangrove establishment*" describes a general causal relationship, but on its own it says nothing about any particular site. An assertional statement instantiates it only by asserting that this relationship holds in a concrete case, for example "*at site X, elevated salinity has inhibited mangrove establishment*," which asserts not merely a salinity level but the causal consequence the universal predicts. Applying the universal statement to site *X* therefore requires an epistemic operation that relates it to the situation through the evaluation of applicability conditions, a relation analogous to the instance–type relation at the term level. The framework represents this relation explicitly: the assertional statement unit *satisfies* the universal statement unit it instantiates (36). Establishing that it does requires explicit epistemic grounding, since it involves asserting that the predicted consequence obtains and not merely that a triggering condition is present, rather than following from the representation alone. Naming the relation gives the components of Section 5.1 their attachment points. Applicability conditions are stated on the universal, whose scope they delimit by constraining the bearers of the qualities whose changes it relates (Section 2.3). Contextual information is carried by the assertional statements describing the situation. And an applicability report, which records the outcome of an operation performed in a particular case, hangs from the assertional statement that satisfies the universal, carrying with it the contextual information under which that outcome occurred. This is why applicability conditions and contextual information, though evaluated against one another, are distinct in kind, as the one belongs to the general statement, the other to the case that instantiates it, and applicability assessment is the relation between them. At the level of representation, this is what Section 2.3's observation amounts to, that general knowledge becomes actionable only when linked to a situation through applicability conditions.

Two properties of the framework matter for what follows. First, it is **technology-agnostic**, meaning that semantic units can be implemented over RDF/OWL knowledge graphs, property graphs, or relational systems, since the contribution lies in the explicit representation of meaning through modular, referenceable, typed units, not in any particular storage technology (25). This is how the framework resolves the triple-fragmentation limitation of Section 5.2 rather than escaping it, by supplying the statement-level commitment those infrastructures lack, over whatever substrate they use. Second, the framework admits **progressive formalisation**. A component's knowledge need not be fully formalised at the outset. It can begin as natural-language text, such as a field protocol or an identification key, and be formalised toward typed statement units incrementally, as modelling effort and ontological resources allow, following the **Semantic Ladder** approach (24). The Semantic Ladder describes a sequence of levels of increasing formalisation, from free natural language text, through controlled and templated forms, to semantically fully formalised typed statement units grounded in ontology terms. Each level is a valid resting point, and machine-actionability increases as a component moves up the ladder while the authoring effort required rises with it, so a component is formalised only as far up the ladder as its intended use requires. Rosetta Statement units (34) occupy an intermediate rung on this ladder. They provide templated, human-interpretable RDF-based statement structures that are to certain degree machine-interpretable, acting as semantic anchors that keep a statement CLEAR-compliant even before it is semantically fully formalised, and they give authors a practical entry point from which further formalisation can proceed. An action unit whose plan specification is represented at a lower level of formalisation remains structurally valid but corresponds to a lower degree of machine-actionability and FAIRness, providing human-readable operational guidance without yet-automated applicability assessment. This graded character means the representational challenge can be met incrementally, rather than requiring full formalisation as a precondition for any operational use.

With the framework in place, the action unit can now be given its representational identity within the Semantic Units Framework. An action unit *is* a compound unit, i.e., a structured collection of typed semantic units, comprising input and output specification units, a plan specification unit carrying typed directive and conditional directive statements, an objective unit, and, optionally, an applicability conditions unit and empirical evidence units, all bound into a single referenceable whole. In representing an operation's components as typed semantic units, the action unit extends the notion of a plan specification with exactly the components the TripleA Principle requires, and does so in a form that is first-class, typed, addressable, and progressively formalisable. The specification of Section 5.1 and the challenge of Section 5.2 are thereby met together, as the roles the action unit named are filled by typed semantic units, and the statements those units carry — descriptive and prescriptive alike — are represented as units in their own right.

# 6. The Action-Unit Hierarchy and Infrastructure Implications

Section 5 established the action unit as the representational realization of the TripleA Principle and identified its three levels, action unit, applicable action unit, and audited applicable action unit, as the reification of the three capabilities. This section develops that hierarchy into its consequences. It sets out the graded structure of actionable knowledge (Section 6.1) and the two directions in which knowledge is used (Section 6.2), then turns to mechanism: how conditional action units let a knowledge graph act as a decision system (Section 6.3), and how a typed operational layer links units to operations and makes workflows discoverable (Section 6.4). It then draws the implications for infrastructure design, including how representing objects makes action units discoverable and reframes the knowledge-action gap as an action-knowledge gap that turns practice needs into research priorities, and how action units serve as a shared interface among scientists, practitioners, and communities (Section 6.5). Finally, using biodiversity science and ecosystem restoration as the illustrative domain, it shows how ecosystems are described as fingerprints and how the action-knowledge gap is made geographically explicit (Section 6.6). The section closes by turning the framework's own resource on itself, examining where its strengths create exposures that a representational framework cannot neutralise, in particular around governance, the ambiguity of documented absence, and the ontological commitments of its substrate (Section 6.7). At several points, the mangrove, species-distribution-model, and ragweed failures that motivated the three challenges in Section 2 return, no longer as symptoms but as cases the framework can diagnose.

## 6.1 The hierarchy of actionable knowledge

The three levels of action unit form a **nested hierarchy**. Every audited applicable action unit is an applicable action unit, and every applicable action unit is an action unit. The sets are thus strictly nested, and each level does not replace the one below but enriches it with the components the next capability requires. An action unit, with its input and output specifications, plan specification, and objective, encodes what can be done in principle, representing structural actionability. An applicable action unit adds an applicability conditions component and so links the operation to the contexts in which its conditions are satisfied. An audited applicable action unit adds documented empirical evidence and so grounds those conditions in the record of prior success and failure. The transition from each level to the next is qualitative, depending on the presence or absence of applicability conditions, then of empirical evidence, and marks the point at which the corresponding capability becomes supported.

Within each level, however, the gradation is **continuous**. Among applicable action units, some carry a rich set of environmental, methodological, and institutional conditions while others carry only the minimal conditions current knowledge affords. Among audited applicable action units, some rest on extensive documentation across many comparable contexts, including failures, while others rest on a single recorded application. The threshold between levels, however, is qualitative, whereas the degree of specification and evidential support within a level is continuous. Actionable knowledge is therefore not a Boolean property a representation either has or lacks, but a position on a continuum running from structurally specified but contextually unassessed, through theoretically applicable on current knowledge, to empirically validated through documented evidence, with knowledge representations moving along this continuum as conditions are refined and evidence accumulates. This mapping between the levels of the hierarchy and their representational form is summarized in Table 1.

**Table 1: The action-unit hierarchy and its representational form.**

| Epistemic status of knowledge | Representational form | Capability supported |
|---|---|---|
| Structurally actionable | Action unit | Actionability |
| Contextually applicable | Applicable action unit | Applicability |
| Empirically grounded | Audited applicable action unit | Auditability |

The operation architecture is inherently iterative. The outcomes of intervention operations generate new observations and system states, re-enter the graph as assertional statements, and feed back into epistemic operations that refine contextual understanding and reassess applicability. The upward movement along this continuum is the representational counterpart of that iteration, and it is not merely additive but self-correcting, as Section 2.4 anticipated. Documented failures at the audited level reveal where an operation's applicability conditions were incomplete or mis-specified, and so feed back to refine the conditions carried at the applicable level. The evidence accumulated at the top of the hierarchy grounds and corrects the judgements encoded lower down. Over time this supports the progressive construction of the causal mosaic (38), as patterns in where a difference-making relationship holds and fails can help to identify additional contextual constraints on the hypothesized relationship and may even contribute to mechanism discovery, so that the hierarchy can provide representational infrastructure for the combination of difference-making and mechanistic knowledge, rather than presupposing that we already have both types of knowledge. Failure documentation in this sense, far from merely being a record of things that went wrong, can be a source of evidence and a tool for finding the boundaries of valid application, which is why its systematic absence from current infrastructures is so consequential (3,6).

The self-correction operates only on conditions that enter the record. A modulating factor held constant across an entire literature never varies, casts no shadow in the pattern of success and failure, and cannot be surfaced by the audit loop, since a genuinely unknown condition is invisible until a case reveals it. Because outcomes re-enter the graph under the condition vocabulary that currently exists, an outcome partly produced by an unrepresented factor is attributed to the represented ones, and iteration can entrench rather than expose the omission. The framework does not escape this, but what it changes is the threshold at which such a factor becomes visible and the surface on which it can be corrected. The enabling-to-determining migration converts the first documented violation of a taken-for-granted constant into structural revision rather than dissipated anomaly, and the requirement that failures re-enter alongside successes creates the contrast that a success-only record cannot. The residue is the same that section 4.4 names. No explicit representation can guarantee it has captured every relevant condition, because every specification rests on a background it does not itself state. The claim is not that the architecture is immune to systematic blind spots but that, unlike the tacit practice it augments, it makes them nameable and revisable once suspected.

Refinement of this kind has a structural consequence. The hierarchy described so far is vertical, tracking levels of grounding for a single operation. But documented outcomes are rarely uniformly distributed across it. Successes and failures are of course disjoint at any given point. What is not guaranteed is that they are spatially organised in contrast to being more randomly distributed. Where they are, so that successes cluster in some regions of the context space and failures in others rather than interleaving throughout, the pattern does not merely show that the applicability conditions were drawn too widely. It shows where the boundaries lie, and it shows that what had been treated as a single operation or causal hypothesis was in fact several, distinguished by conditions the original specification did not record. The result of responding to such a pattern is therefore not a revised action unit but a *specialisation* of the original into two or more distinct units, each executing the same plan specification toward the same objective under a distinct set of conditions. Consider an action unit whose objective is to establish mangrove propagules, executing the general knowledge that increased salinity inhibits establishment. Documented outcomes need not fail uniformly. Suppose applications succeed across microtidal sites but fail repeatedly across macrotidal ones, the failures concentrating where tidal range is large. The pattern does not show the unit simply wrong. It shows that tidal regime, recorded at best as background, marks a boundary the original conditions did not, and it specialises the unit into two: one for microtidal settings and one for macrotidal, each carrying the tidal condition explicitly and each grounded in the outcomes from its own region. The parent, asserting the salinity relation the two share, remains what they have in common. The parent is not discarded, since it may remain applicable elsewhere, and it remains the statement of what the children have in common. Repeated across contexts, this yields a **family of action units**, with a general parent whose conditions are broad and weakly grounded, and children whose conditions are narrower and mutually distinct, each grounded in the evidence from the region it covers. The relation between parent and child is checkable rather than curated, since one unit specialises another exactly when its applicability conditions entail the other's. Siblings need not stand in any such relation to one another, and their conditions may be disjoint. Where the general knowledge an action unit executes is a causal hypothesis, this is the discovery that what had been asserted as one causal hypothesis is in fact several. Where the separating conditions are merely enabling, the siblings differ only in the range of systems for which one and the same relation holds. Where those conditions prove determining, they differ in the relation itself, one sibling asserting a more determinate causal quality than another in the sense set out above, as interstitial pore-water salinity is more determinate than salinity. The parent does not become false. It becomes underspecified, and the family is the specification, in scope and, where the conditions are determining, in content. An infrastructure therefore grows in two directions at once. Vertically, as evidence accumulates and units rise through the levels. And horizontally, as the structure of documented outcomes reveals divisions the original specification did not record. This is what the incremental expansion of Section 6.5 consists in, and it is why a site may satisfy a parent's conditions and fail every child's, a distinction the maps of Section 6.6 depend on.

Not every condition specialises in the same way, and only some produce children at all. Referential conditions do not, and among contextual conditions the distinction drawn in Section 2.3 between enabling and determining conditions determines what a child asserts. Where a quality is a theoretical construct approximated through indicators, as ecological fitness or phylogenetic relatedness are, refining the criteria by which it is recognised often alters neither what a hypothesis asserts nor the range of systems it covers. It alters how reliably the assertion can be assessed and how confidently evidence can be attributed to it, which is movement along the vertical axis rather than the horizontal one. Sharper recognition criteria can raise a unit through the levels of the hierarchy, and can equally lower it, since a criterion that reveals two studies to have measured different things withdraws from a hypothesis the evidence it was previously credited with. What they cannot do is divide it. Enabling contextual conditions can, as a child restricted to sites free of ongoing disturbance asserts exactly the parent's causal relation, and asserts it of fewer systems. Determining contextual conditions do more. Where the tidal regime of a coastal system determines which salinity a propagule is in fact exposed to, a child restricted to microtidal systems does not merely apply the parent's claim less widely, but instead reveals that the quality actually doing the causal work is a more determinate one, namely interstitial pore-water salinity rather than salinity as such, and the condition is thereby absorbed into the causal variable. The child then stands to the parent as a determinate stands to its determinable, and asserts something the parent could not. The three kinds of condition therefore divide along a single line. Vertical movement changes what is known about a hypothesis, whereas horizontal movement changes which hypothesis it is. Referential conditions govern the first and contextual conditions the second, with enabling ones narrowing the range of systems a hypothesis covers and determining ones specifying what its causal variables are.

Whether a contextual condition is enabling or determining cannot be settled in advance, and this is what the audit loop supplies. A condition initially recorded as merely enabling may, as outcomes accumulate, prove determining, and its migration into the causal variable is not just a bookkeeping adjustment but also reflects objective features of the system. The one apparent exception confirms the rule. Where a refinement of recognition criteria establishes not that a quality has been measured better but that two methods were never measuring the same quality, the discovery is not referential but determining, since it concerns the identity of the relata. What is presented as one hypothesis assessed in two ways is then two hypotheses about two qualities, and the split is horizontal. A referential finding has occasioned it, but a determining fact has produced it.

The distinction between conditions that determine a causal relation and conditions that merely enable it is an old one in the philosophy of causation, and the framework adopts it rather than resolving it. Mill held that the cause of an event is its whole antecedent, and that singling out one condition as *the* cause is a matter of subjective, practical interest rather than an objective fact about the causal relation itself (58). Mackie made the difficulty precise, arguing that a cause is an insufficient but necessary part of an unnecessary but sufficient condition, which makes the striking of a match and the presence of oxygen structurally alike, and he located the difference between them not in the world but in the *causal field*, the background the enquirer holds fixed in posing the question (59,60). To the fire investigator the oxygen is background and the spark the cause, but to the engineer specifying an inert store the assignment reverses. If the distinction is field-relative in this way, so is ours, and which conditions a practitioner holds fixed in posing the question reflects the purpose of the inquiry rather than the system alone. This much we grant. However, we disagree with the view that selection is entirely interest relative and unsystematic (61,62). Field-relativity governs what is held fixed, not what follows once it is. Given a field drawn one way, whether a backgrounded condition in fact modulates the relation is settled by the system and not by the enquirer's interest, and it is here that migration is a discovery rather than a re-framing. The fire investigator who treats oxygen as background has fixed a field by interest. That oxygen concentration modulates combustion within the range their cases span, and so will not stay in the background, is not a further choice but something to outcomes force. What is purpose-relative is the frame and what migration reports is the reference system's verdict on a condition the frame had set aside. The framework changes the standing of both. In existing infrastructures the causal field is nowhere represented, residing in the interests of the enquirer and the tacit practice of a discipline. Contextual applicability conditions represent it, the enabling/determining line is the boundary the field is drawn at, and the audit loop is what allows a field once drawn to be found wrong, since a condition recorded as background that proves to modulate the relation migrates into its relata under the pressure of documented outcomes. The field so drawn is one of the things that fixes the space of conditions against which a situation is described, since an enabling condition becomes a dimension along which sites vary, whereas a determining one does not, having entered the causal claim itself. It is not the only thing.

Referential conditions are fixed by methodological rather than causal considerations, and the conditions governing whether an operation can be performed at all are fixed by institutional and practical ones. The conditions space is therefore only partly causal in origin, and the context space across which documented outcomes are compared is representationally determined by it. A migration across the enabling/determining line does not merely relabel a condition. It redraws a dimension of the space in which the evidence is read. The choice Mill and Mackie identify remains a choice. It ceases to be an unexamined one.

The explicit representation of applicability conditions bears on a long-standing difficulty. Causal generalisations in the special sciences are rarely stated without an implicit **ceteris paribus clause**, and the standard objection is a dilemma: read as strict, such generalisations are false, and hedged by an indefinite clause, they appear to lack testable content, since any failure can be attributed to unmet background conditions rather than to the generalisation itself (63,64). **Completer approaches** respond by proposing that the missing conditions be supplied, whether from a more fundamental science (65) or, closer to the mechanism described here, retrospectively, as independently confirmable factors explaining each counterinstance (66). Such accounts have been criticised as almost vacuous, since a completion adduced case by case, subject only to independent confirmability, licenses generalisations that are causally accidental (67,68). The framework proposed here does not resolve that difficulty, but it changes what a completion must satisfy. Applicability conditions are stated in advance rather than adduced after failure. Their revision is licensed not by any explanation of a counterinstance but only by structure in the documented record, so that failures explained individually and severally, however well confirmed, do not narrow a condition, whereas successes and failures separating across the context space do. And the record itself is inspectable, which is what distinguishes a narrowing that tracks a boundary from one that merely absorbs an inconvenient failure. The distinction between the two is otherwise tacit, residing in a discipline's practices for identifying which disturbing factors are worth listing (69). Auditability makes it explicit, and thereby makes it contestable.

This structure has an ecological precedent. The **hierarchy-of-hypotheses approach** (70) organises ecological theory by nesting an overarching hypothesis into progressively narrower subhypotheses whose lower branches connect directly to empirical results, and it distinguishes *operational* subhypotheses, which narrow the overarching hypothesis by accounting for the particular method or study design used to test it, from *mechanistic* subhypotheses, which narrow it with respect to the underlying causes assumed to be at work. The first corresponds closely to the referential applicability conditions of Section 2.3, since both concern how a quality is recognised and measured. The second does not correspond to contextual applicability conditions, which is instructive, as a mechanistic subhypothesis specifies *which cause* is proposed to operate, whereas a contextual applicability condition specifies *under what circumstances* a proposed cause in fact operates. The approach's stated aim of assessing the range of applicability of a hypothesis is therefore pursued through a structure that does not itself represent the conditions of application. The authors of the approach observe that hierarchical decomposition can reveal that studies combined under one header are in fact addressing very different things. The mechanism described here is the evidential counterpart of that observation, as the structure of documented outcomes can reveal that what was specified as a single operation, and the causal hypothesis it executes, are in fact several.

What the present framework adds is threefold. It makes the ordering checkable rather than curatorial, since one action unit specialises another exactly when its applicability conditions entail the other's, whereas the authors of the approach note that the placement of subhypotheses rests on individual choices and that full reproducibility is unlikely to be reached. It represents the conditions of application as evaluable components rather than as an interpretation of where evidence happens to accumulate. And it supplies a generative mechanism, since clusters of successes separated by failures are what produce children, so that a family is not only a way of organising hypotheses already held but the record of how their boundaries were found. In doing so it takes up two aims the authors of the approach set for its further development, namely a formalised version with enhanced expressiveness of the relationships among subhypotheses, to be pursued through semantic-web methods and ontologies, and an extension capable of representing causal relationships and multicausality, which they note the hierarchy in its present form does not depict (70).

A philosophical commentary on the approach (71) identifies two further limitations, both of which the present framework addresses. Schurz argues that (i) theory-driven hierarchies neglect the effects of combinations of causes, and proposes to remedy this by adding a layer of system conditions, and (ii) that evidence-driven hierarchies, in incorporating the empirical tests by which hypotheses are assessed, raise unresolved problems about the relation between causal hypotheses and the correlation claims that are actually testable. The applicability-conditions component supplies the first, and supplies more than the proposed remedy asks for. An effect of combined causes is what one has when an enabling condition proves to be a modulating co-cause, and the framework admits two representations of it: (i) absorption into a more determinate causal variable, where the condition determines what is changing, and (ii) an additional causal relation in the network, where it alters the strength of a relation whose relata are unchanged. Which applies is a modelling decision the evidence constrains but does not dictate, and what the framework requires is that it be recorded and revisable rather than settled correctly at the outset. A layer of system conditions represents the circumstances under which a hypothesis holds. The distinction between enabling and determining conditions represents, in addition, whether those circumstances belong to the hypothesis's scope or to its content. The second is addressed by representing causal and correlational claims as distinct but related statements, so that the interpretive step between them becomes explicit and auditable rather than left implicit (36).

The hierarchy also gives the failures of Section 2 a precise diagnosis. Each, the mangrove plantings on unsuitable sites, the species distribution models transferred beyond their training domain, and the ragweed regime applied as if universally effective, shares a single structure that can be translated to an action unit that was used *as if* it were an audited applicable action unit, its validity treated as established, when it was at best an action unit with unassessed applicability conditions. Three representational deficiencies recur across the cases. The **contextual information** describing the actual situation was not structured so that it could be evaluated against applicability conditions. The **applicability conditions** themselves were left implicit rather than represented as explicit, evaluable components — the mangrove causal hypothesis presupposed hydrological and sediment conditions that were never formalized and the SDM presupposed environmental stationarity and adequate niche sampling that were never stated as referential conditions. And the **operational grounding** was incomplete. The transition from action unit to applicable action unit requires applicability assessment, and the transition to audited applicable action unit requires documented evidence, yet in each case the transition was assumed rather than performed. The hierarchy converts these from post-hoc diagnoses into pre-operational checks. What failed silently in practice becomes, in an action-unit-structured infrastructure, a question that must be answered before action is taken: is this an action unit, an applicable action unit, or an audited applicable action unit here?

The ragweed case shows what the answer looks like when it is made explicit. The mowing regime, represented as an intervention action unit, comprises an input specification identifying a ragweed population in its roadside context, a plan specification prescribing cuts in June and September with associated cutting heights and machinery (47), an objective specifying suppression of intrinsic population growth below replacement, and an output specification characterising the post-intervention population state. The component that current management workflows leave implicit, and that the framework requires be made explicit, is the contextual applicability conditions unit. For this regime it should include at least four conditions. A **phenological condition** specifying that the June cut precedes the period during which compensatory regrowth and increased size-dependent fecundity outweigh the suppression effect (45,47). A **size-distribution condition** specifying that mean adult height at the site falls within the range over which the height-reduction mechanism dominates over fecundity compensation, drawn empirically from prior demographic studies (45). A **seed-bank condition** specifying that local seed survival rates permit seed-bank depletion within the management horizon, since regimes that drive growth rates marginally below replacement still fail under high seed survival (45). And an **institutional condition** specifying that mowing crews and machinery can be scheduled for the dates the biological conditions require.

The four are not of a kind, and the framework requires the difference to be recorded. The institutional condition is enabling, with the crew availability determining whether the regime can be carried out, but not what a cut does to a ragweed population. The seed-bank condition is likewise enabling, gating whether suppression of growth translates into local extirpation without altering the causal relation between cutting and fecundity. The phenological and size-distribution conditions may prove determining, since if the effect of a cut depends on when in the growth cycle it falls and on the size class it removes, then the quality actually doing the causal work is not cutting as such but the removal of a determinate proportion of reproductive biomass at a determinate phenological stage. Whether they are enabling or determining is not settled by the ecology as currently stated, and it is the accumulated record of where the regime succeeds and fails that would settle it.

That record already exists. The five-year multi-site experiment compared the prevailing regime against three alternatives across sites and years, and found the regimes separating according to which mechanism dominated at each site (45,47). Those outcomes are what an applicability report would carry, and their structure across the context space is what would specialise the parent regime into the family of regimes the experiment in effect discovered, without the infrastructure to record it as such. Made explicit in this form, the regime can be evaluated against site context before execution rather than after, and accumulated outcomes across sites would tighten the boundaries of its valid application and, where successes and failures separate, specialise it into the family of regimes it always was. The demographic literature on ragweed contains most of the evidence needed to populate these conditions. What is missing is the infrastructure within which that evidence can be represented as evaluable applicability conditions rather than as caveats in methods sections.

That the framework makes such assessment *possible* does not, of course, make it *happen*. Whether practitioners engage with an infrastructure of this kind depends on access, training, and the social and institutional relationships through which actionable knowledge is actually acquired and trusted (13), which lie beyond the representational framework proposed here. The framework provides the infrastructure for systematic validity assessment, but its adoption is a separate, and equally necessary, undertaking.

## 6.2 Two directions of use: the forward and reverse perspectives

The hierarchy describes what an action unit *is* at each level of grounding. How it is *used* depends on which of two directions the knowledge–context relation is traversed, and these are not two conveniences but the two directions of fit that Section 4.2 identified at the root of the operation typology, now appearing at the level of use.

In the **forward, decision-oriented** perspective, a specific situation and an objective are given, and the task is to find the knowledge and procedures that apply: *here is this site, this is the state I want it in, what can be done?* This is the perspective of the practitioner and the decision-maker, acting under context-specific constraints, and it instantiates the world-to-word direction of fit with the world being given, and the task being to find the directive knowledge that specifies how to change it to match a desired state. An infrastructure organized around action units answers this query by matching the situation's contextual information against the applicability conditions of candidate action units, returning those whose conditions are satisfied, and, where auditable, those whose prior applications in comparable contexts are documented.

In the **reverse, knowledge-oriented** perspective, general knowledge and an objective are given, and the task is to find the contexts in which that knowledge applies: *here is this causal hypothesis, this restoration method, where, and under what conditions, does it hold?* This is the perspective of the scientist, and it instantiates the word-to-world direction with the knowledge being given, and the task being to determine the situations it correctly describes or to which it reliably applies. The same action units serve this query in reverse. The applicability conditions that a forward query evaluates against a situation are, read the other way, the specification of the situations to which the knowledge applies, and the accumulated empirical evidence is the record of where it has been shown to hold or fail. This perspective supports systematic validation, comparison across contexts, and the analysis of knowledge transferability, and thus the very problem the species-distribution-model failures of Section 2 exhibited, now approached as a first-class representational question rather than left to implicit modelling judgement.

The two perspectives, and the assessment that relates them, correspond to the three forms of inference Peirce (72) distinguished, and the correspondence is not incidental, since the framework's units are the *rule*, the *case*, and the *result* over which his account is stated. Applicability assessment, which applies a general statement to a described case in order to determine what to expect, is deductive. The refinement of applicability conditions from accumulated outcomes, which infers from cases and their results what the rule must be, is inductive, and it is the movement the reverse perspective performs and the one that Section 2.4's clustered failures license. And the forward perspective's characteristic question, *given this situation and this objective, what can be done here*, is abductive, since it seeks the operation whose applicability conditions, if satisfied, would account for the situation admitting the desired change. Three directions of inference are thus realised as three uses of a single representational structure, and the same statement units serve each. Whether this correspondence is structural or merely suggestive we leave open, as with the parallels noted in Section 7.

The two perspectives are not symmetric. As Section 2.1 noted and Section 4.2 grounded, epistemic operations that establish what is the case are a precondition for reliable intervention, whereas intervention does not presuppose prior representation in the same way. Knowledge must be assessed as applicable before it can be reliably used, but the reverse dependency does not hold. This asymmetry is why the forward perspective (i.e., deciding what to do here) rests on epistemic grounding that the reverse perspective (i.e., determining where knowledge holds) supplies. The two are linked through the action unit and its applicability conditions, which make the transition between them explicit and evaluable, as what the scientist establishes about the conditions and evidence for a piece of knowledge is precisely what the practitioner's situation is matched against, and what the practitioner's documented outcomes contribute is precisely what refines the scientist's account of where the knowledge holds. Actionable knowledge is in this sense bidirectional, with knowledge informing action, while contexts constraining and organizing the use of knowledge. And the same representational structure serves both readings, a duality that runs from the direction-of-fit distinction in the philosophy of language through to the design of the infrastructure itself.

## 6.3 Conditional action units and graph-native decision support

An applicable action unit carries the conditions under which its operation can be reliably performed. When those conditions are represented as an executable query rather than a static description, the action unit becomes able to evaluate its own applicability against the current state of a knowledge graph. This is the basis of the **conditional action unit**, which extends the applicable action unit into executable decision logic (Fig. 8).

A conditional action unit is a conditional directive statement, an IF–THEN rule in which the condition is represented as a readily executable query. The IF clause is an executable question unit (25), typically realized as a graph query such as a SPARQL ASK or SELECT, that evaluates whether the contextual information available for a situation satisfies the applicability conditions of the action unit, and so identifies the cases to which the THEN clause applies. The THEN clause is an executable directive unit, and it is here that the operation class varies. Where the conditions identified by the IF clause are satisfied, the directive may be carried out as a further data transformation, which is fully automatable, as the invocation of an external service, or as the initiation of a workflow involving human agents, which is an intervention in the world.

Because the IF clause is executable, the applicability conditions cease to be descriptive constraints that a human must check and become computable predicates that the graph can evaluate. The directive content of the THEN clause likewise becomes an executable procedure rather than a passive record of what should be done. A knowledge graph populated with conditional action units can therefore evaluate context-dependent conditions, identify the action units that are applicable in a given situation, trigger the corresponding operations, and coordinate workflows across internal and external components. It functions as a graph-native decision-support system rather than a passive repository.

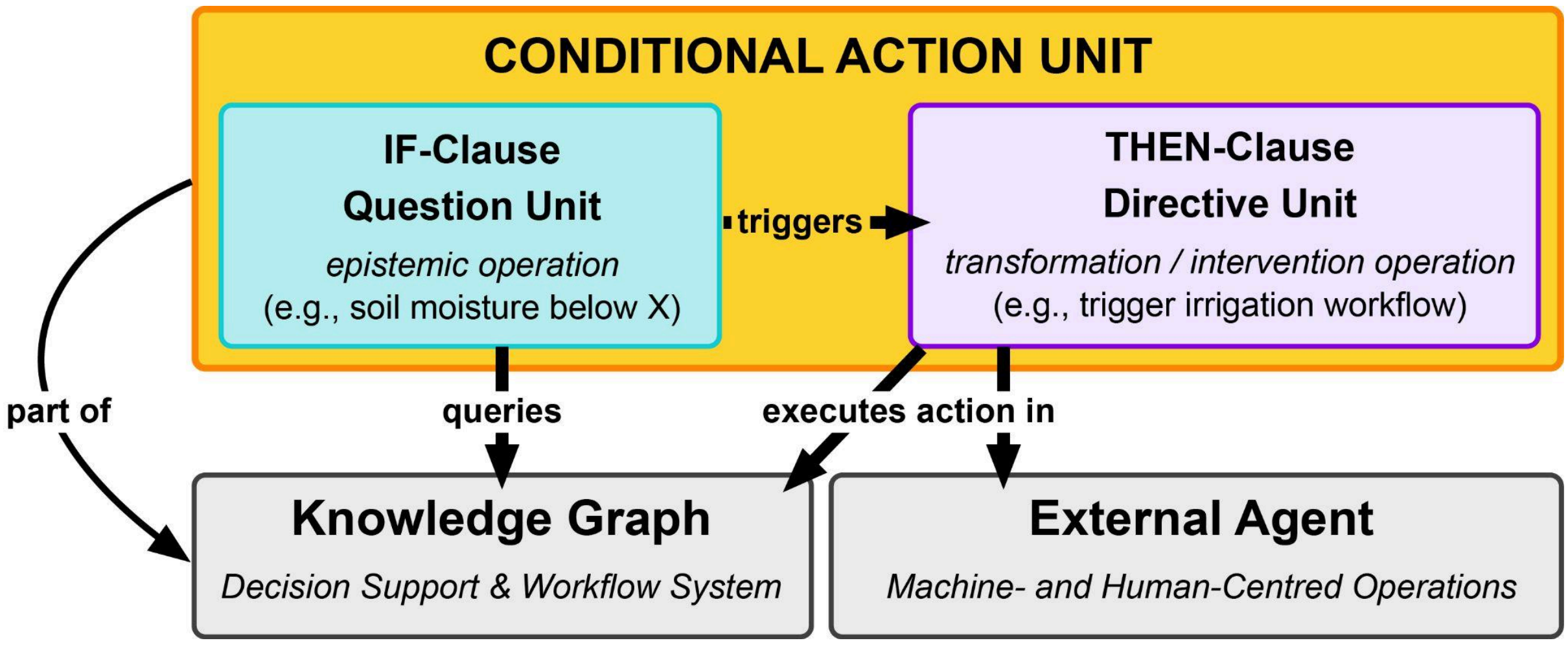

**Figure 8: Conditional action units in knowledge graphs.** The IF clause is represented as an executable question unit that queries whether the contextual information for a situation satisfies the applicability conditions of an action unit, for example whether a measured value falls below a threshold. The THEN clause is represented as an executable directive unit that triggers an associated operation or workflow, such as a data transformation, a service call, or a human-executed intervention. Together they link context evaluation, data preparation, and action execution within a single structure, allowing the knowledge graph to function as a decision-support and workflow-orchestration system.

This capability is particularly useful for connecting causal hypotheses to the evidence that grounds them. A conditional action unit can be defined so that if the measurements relevant to the variables of a causal hypothesis are available for a given situation, then an assertional statement is constructed recording that the hypothesised relation obtains, or fails to obtain, in that situation. It is linked to the universal statement it satisfies, and carries the measurement and the contextual information under which they were obtained. Patterns of this kind allow observational data to be integrated with the hypotheses they bear on, support the automated detection of missing data, and enable counterfactual queries that test how applicability changes as contextual parameters vary. A restoration example shows the same structure guiding action. A conditional action unit for mangrove restoration might work as follows. If the hydrological connectivity recorded for a candidate site satisfies the conditions for natural regeneration, the unit triggers a passive regeneration action unit. If connectivity is instead restorable through engineering, it triggers a hydrological restoration action unit as a precondition for planting. If neither holds, it flags the site as currently unsuitable and schedules a reassessment. The IF clause evaluates the site's contextual information against the applicability conditions, and the THEN clause selects the appropriate intervention, so that the assessment which failed silently in the restoration projects of Section 2 becomes an explicit, recorded step. The same mechanism that identifies applicable action units also identifies their absence, which becomes important for infrastructure design in Section 6.5.

More generally, conditional action units allow closed-loop workflows to be represented within the knowledge graph, in which context evaluation, data transformation, and intervention are coupled rather than carried out separately. This is what distinguishes a post-FAIR infrastructure from a FAIR one. A FAIR infrastructure makes knowledge findable, accessible, interoperable, and reusable. A **post-FAIR knowledge infrastructure** additionally makes it operationally integrated, executable, and context-sensitive, so that the graph does not merely store what is known but evaluates where and how it can be used.

## 6.4 The operational layer: typing, compatibility, and operation discovery

Representing action units as first-class objects establishes an operational layer between semantic representation and procedural execution. This layer connects what knowledge is with what can be done with it, and its formal structure is what allows operations to be matched to knowledge and workflows to be assembled from them. Two ingredients define it: the typing of semantic units by schema, and the typed relations that determine when an operation can attach to a unit.

Semantic units are typed by formal schemata, and operations are defined by their required inputs, outputs, and plan specifications. An operation can attach to a unit only when the unit's type satisfies the operation's input requirements, and the relation that decides this differs by operation class. Epistemic operations depend on **referential compatibility**, determined by the diagnostic criteria that link a representation to a material entity or system state. Transformational operations depend on **formal compatibility**, determined by schema constraints on input and output representations. Intervention operations depend on **contextual compatibility**, determined by applicability conditions evaluated against situational context. These three compatibility relations are the operational counterpart of the three applicability-condition types, and they type not merely the operations but the matching relation that governs their composition.

Referential compatibility is not, however, settled by type identity alone. Where a quality is not directly observable and is approximated through methods that do not yield interchangeable values (Section 6.6), two units may share a type while the operations that produced them do not license their joint use. Structural compatibility then holds while referential compatibility does not, and the mismatch is silent, since nothing in the type declaration records how the value was obtained. The consequence is more serious here than for the maps of Section 6.6, because an agent composing a pipeline by following type compatibility does so at scale and without the opportunity for the human check that a map at least invites. Making the method by which a value was obtained part of what is represented is therefore a precondition of the operational layer, not an addition to it. Where it is represented, the composition can be checked, but where it is not, an operations meta-graph will assemble pipelines whose inputs were never comparable and will report them as valid.

Once operations and units are typed in this way, the operational affordances of a representation become explicit and queryable. For a given semantic unit, it becomes possible to ask which operations are compatible with its type, and to expose the result as an affordance layer showing which recognition procedures, transformations, or interventions are structurally available for that kind of entity. Conceptually, typed operations over typed units form an **operations meta-graph**, a layer over the knowledge graph in which units define available inputs and outputs and operations define the transitions between them. Workflows are then paths through this meta-graph, compositions of operations whose input and output types are compatible, which both human and machine agents can assemble by following typed compatibility rather than by hand. Structural operation discovery, the ability to query which operations a given unit affords, is the primitive on which the objective-driven action discovery and the coverage maps of the following subsections rest.

This operational layer has a concrete realization in emerging FAIR service ecosystems, in which terminology services manage controlled vocabularies, schema services define semantic schemata with formal constraints such as SHACL shapes, operation services expose executable functions over those schemata, and workflow services compose operations into pipelines (23,26). Within such an ecosystem, deductive and computational transformational operations can become fully machine-actionable. Ampliative transformational operations, such as analogical or abductive inference, can be supported and partially automated, for instance through similarity retrieval or case-based reasoning, but retain a non-formalisable element, so full determinism is a property of the deductive subclass rather than of transformational operations as a whole. Where operations are defined with explicit input and output schemata, tested, and registered as reliable services, an agent (including AI agents) given a goal and input data can construct a deterministic or semi-deterministic pipeline over the available operations to reach the desired result, which is a form of **traceable, reproducible agentic execution**. Not all operations reach this level. Epistemic and intervention operations often involve observation, judgement, or physical action on material systems, and sometimes cannot be executed as services. The operational layer accommodates this by representing human-, machine-, and hybrid-executable operations uniformly through the same typed structure, so that an operation that cannot be executed automatically is still represented with its inputs, outputs, procedure, and applicability conditions, and remains discoverable and assessable even where its execution depends on a human agent.

This operational layer is not offered in isolation from existing standards. Several established frameworks link knowledge representations to processes and workflows, and action units are best understood in relation to them. Process ontologies such as the Ontology for Biomedical Investigations (OBI) (35) represent planned processes, their participants, and their outputs, grounded in the Basic Formal Ontology (BFO) (16). The PROV-O provenance ontology and its extension P-Plan represent the provenance of processes and link plan specifications to their execution records (73,74). Workflow frameworks such as RO-Crate and the Common Workflow Language package research artefacts with their metadata and specify portable, reusable computational pipelines (75,76). Like plan specifications in these frameworks, action units specify participants, procedural knowledge, and objectives. They differ in a consequential respect. These frameworks document processes that have been or will be carried out, representing what happened, what was planned, or how a workflow should execute. They do not make contextual applicability conditions explicit as first-class, typed, evaluable components, nor do they distinguish the structural capacity of a representation to support an operation from the epistemic validity of performing that operation in a given situation. Action units address precisely this prior layer. Where the Common Workflow Language specifies how a tool is invoked given conformant inputs, an action unit specifies when the invocation is contextually valid and what contextual information establishes this. Where PROV-O and OBI record what an activity used and generated, an action unit specifies the conditions that must hold for the activity to be epistemically justified. Action units thus provide pre-operational infrastructure that complements provenance and workflow standards rather than replacing them, making the representational and epistemic preconditions of execution explicit and evaluable before a process is run.

## 6.5 Implications for knowledge infrastructure design

The action unit gives knowledge infrastructures a design target beyond storage and retrieval. To support the TripleA Principle, an infrastructure must represent action units as first-class objects, so that operations, their applicability conditions, and their evidence are queryable in the same way as the facts they act on. This means representing not only what is the case but what can be done with it, under which conditions, and with what evidential grounding. An infrastructure that does this supports an explicit epistemic layer in which applicability conditions are evaluated against contextual information, often under uncertainty, rather than leaving that evaluation to the tacit judgement of whoever happens to use the data.

Such an infrastructure should not aim to represent every possible action. Actionability is bounded and graded, and both human and machine agents work under cognitive and computational limits. Representing all possible operations, contexts, and procedures for a domain would demand a level of formalization that is neither achievable nor maintainable. A more realistic design principle is to prioritize a bounded set of high-value action units, chosen by frequency of use, importance for decision-making, and the availability of sufficiently robust knowledge and data. The action unit is modular, so an infrastructure built this way can expand incrementally. New action units are added as data and procedures become available, and existing ones move up the hierarchy from action unit to applicable and audited applicable action unit as conditions are refined and evidence accumulates. Boundedness is a design principle rather than a limitation, because it allows an infrastructure to be useful early and to grow without requiring completeness before it can be used at all. This also accommodates decision-making under time pressure, which is common in conservation practice. The framework does not require full applicability assessment before any action can be taken. It makes explicit the level of grounding an action currently rests on, so that a practitioner who must act on an unassessed action unit does so knowingly rather than on an unexamined assumption of validity.

Representing objectives as first-class components enables a second capability, which is objective-driven action discovery. Given a situation and an objective, an infrastructure can identify which action units are applicable and therefore which operations can validly be performed. Read in reverse, given an action unit, it can identify the situations in which the operation applies and the objectives it can serve. When no applicable or audited applicable action unit exists for a given operation and objective in a given context, the same query that would have found one instead reveals its absence. This turns the knowledge-action gap into a structured statement of what is missing, an action-knowledge gap that specifies what practitioners need science to provide (subject to the qualifications of Section 6.7). Knowledge gaps that were previously tacit become explicit and queryable, and can be expressed in the same vocabulary that scientists use to represent findings, so that the infrastructure supports the translation of practice needs into research priorities as well as the translation of knowledge into action.

Because action units are built from typed components, they also serve as a shared interface among the different agents who contribute to and draw on actionable knowledge. Scientists produce the general knowledge components, the causal and statistical statements, the diagnostic and procedural knowledge, and the applicability conditions under which these hold. Policymakers work mainly with objectives and with audited applicable action units, selecting among operations that have been assessed for a context. Practitioners work mainly with plan specifications and applicability conditions, executing procedures under real-world constraints that are ecological and institutional alike. Citizens and local communities contribute assertional statements through observation and monitoring, enriching the contextual information available for assessment. They also hold procedural, diagnostic, and conditional knowledge of the systems they inhabit, which the framework can in principle represent through the same components, and which it is a limitation of the representational substrate rather than of that knowledge that it is often harder to accommodate. Each agent engages a different part of the same structure, which lets the framework integrate different forms of knowledge without privileging scientific over local knowledge or technical over practical expertise, which is an aspiration whose representational preconditions are not yet fully met, as Section 6.7 sets out.

The framework sits alongside several established traditions rather than replacing them. Decision-theoretic and planning approaches assume that the elements of a decision problem, the states, actions, and objectives, are already formally specified, and they optimize over that specification (77,78). Research on knowledge-intensive processes addresses how knowledge workers navigate procedures that cannot be fully specified in advance (79,80). In both cases the framework addresses the prior question these traditions presuppose, which is how heterogeneous and partly formalized knowledge is to be represented so that it can support action at all. The framework is therefore pre-decisional infrastructure that prepares knowledge for use in planning, optimization, and case-management systems rather than competing with them. It is also complementary to evidence-synthesis initiatives such as the *Conservation Evidence* and the *Collaboration for Environmental Evidence* initiatives, which assess in human-readable form where interventions have succeeded and failed. Those initiatives produce exactly the empirical grounding that audited applicable action units require, and the framework offers a representational structure through which that grounding could be made machine-evaluable for specific contexts rather than read and applied by hand.

Taken together, these implications describe a shift from FAIR to post-FAIR knowledge infrastructure. A FAIR knowledge infrastructure makes knowledge findable, accessible, interoperable, and reusable. A **post-FAIR knowledge infrastructure** additionally makes it actionable, applicable, and auditable, so that knowledge is not only available but usable in context, with its conditions of valid use and its evidential grounding represented as explicitly as the knowledge itself.

## 6.6 Ecological fingerprints and knowledge-action-gap maps

Two consequences of representing knowledge as action units become visible only at the scale of a populated infrastructure, where many action units and many described ecosystems are held together in one graph. Both are stated here at the level of the framework and illustrated with biodiversity, where they take their most concrete form.

The first concerns the description of a specific system. The outputs of *description* epistemic (world→representation directed, involving statements) and transformational operations are typically assertional statements. *Description* epistemic operations generate instance-level records, and transformational operations aggregate and normalise them into derived indicators. Assembled for a single ecosystem at a specific time, these assertional outputs constitute a structured, multidimensional description of its state. In biodiversity science this description is an **ecological fingerprint of an ecosystem**, integrating measurements across dimensions such as biome type, climate class, soil type, landscape form, species composition, functional and ecological trait composition, and ecosystem functions and services, with *Essential Biodiversity Variables* (81) providing widely adopted intermediate indicators between raw observation and fingerprint. The ecological fingerprint captures the bio-physical state of a site, but reliable action often depends on more than biology. The broader **ecosystem profile** extends the fingerprint with the socio-political, legal, institutional, and socio-economic descriptors of a site, such as land tenure and use contracts, governance arrangements, and resource availability, which constitute further contextual information against which the applicability conditions of an operation are evaluated. As section 2.3 noted, applicability conditions span biophysical and institutional factors alike, and the profile is the site description that makes both representable. This broader characterization concerns the feasibility and legitimacy conditions of action and not its full normative justification, which lies beyond the framework's scope. Fingerprints and ecosystem profiles are time-indexed. Because monitoring data are collected repeatedly, a single location holds a series of fingerprints/profiles that together track how its ecosystem changes across cycles, and a sufficiently drastic change, such as a fire, may alter not only the ecosystem's state but its type, so that the fingerprints/profiles at one geographic location need not all describe an ecosystem of the same kind.

A fingerprint/profile is a description, not a specification. In the framework's terms it is contextual information, the instance-level characterisation of a situation against which the applicability conditions of an action unit are evaluated. This is what makes it operationally useful. Once an ecosystem's fingerprint/profile is represented, the framework can evaluate it against the applicability conditions of the available action units and report, for that ecosystem at that time, which operations are available and at what level of the hierarchy: which are actionable, in that a structurally specified action unit exists; which are applicable, in that their conditions are satisfied at this site on current knowledge; and which are audited applicable, in that documented evidence supports them in comparable contexts. The fingerprint/profile is the input to applicability assessment, and the tiered set of available operations is its result. This is the forward, decision-oriented perspective of Section 6.2 at the scale of a whole ecosystem, with the situation given as a fingerprint/profile and the framework returning the operations its description makes available.

This evaluation is not a matter of comparing numbers. Many ecological qualities are not directly observable and are approximated through methods that do not yield interchangeable values. Phylogenetic relatedness estimated as an abundance-weighted mean phylogenetic distance is not the same measurement as phylogenetic relatedness estimated as a nearest-taxon distance, and indices intended to capture the same underlying property may disagree (82). An applicability condition that constrains such a quality was established through some particular method, and a fingerprint records its value as obtained through some particular method, and these need not be the same. Where they are not, matching the fingerprint against the condition returns a verdict the evidence does not support, and returns it silently. This is a referential applicability condition in the sense of Section 2.3, i.e., a condition of method validity, and it applies not only to the epistemic operations that generate a fingerprint but to the fingerprint's subsequent use as contextual information. Two representational strategies address it (36). On a **reflective** reading, the quality is a single property that different methods estimate with differing reliability, with each measurement being an assertional statement carrying, as metadata, the method by which it was obtained, and comparability across methods becomes an explicit question the infrastructure can pose rather than an assumption it makes. On a **formative** reading, the method partly constitutes what is measured, with each method-dependent approximation being a distinct subclass of the general quality, and a condition stated over one subclass simply does not match a fingerprint recording another. The two strategies differ in their ontological commitments and suit different qualities, and the choice between them is a modelling decision that the framework requires to be made explicitly rather than left implicit, and it is this choice that determines whether a method mismatch can be absorbed as uncertainty or must block the assessment (Section 2.3) (36). What matters for the present argument is that in either case the method becomes part of what is represented, and a mismatch between the method behind a condition and the method behind a fingerprint becomes detectable. Without this, a knowledge–action-gap map can report a site as applicable on the strength of a value that was never comparable to the evidence the condition rests on, and the map's authority would then exceed its warrant in exactly the way the framework is designed to prevent, and the same mismatch propagates into the operational layer, where it is composed rather than merely displayed (Section 6.4).

The second consequence is this evaluation carried out across many ecosystems at once, in the reverse, knowledge-oriented direction. It rests on how action units connect to geographic places, which they do only indirectly. An action unit is a type-level specification and carries no geographic extent. What links it to the world is its applicability conditions, and among these can be a condition on ecosystem type, specifying the kind of system to which the operation applies. Places enter only through fingerprints, each of which describes a specific ecosystem, including its type, at a specific location and time. The applicability conditions are therefore evaluated against a fingerprint, not against a location, and a location's ecosystem type is a property of its fingerprint at a given time rather than a fixed attribute of the place. Ecosystem type, holding at a time through a fingerprint, is the join between action unit and geographic location.

For a given type of ecosystem (and its subtypes) and a given type of operation (and its subtypes), a **knowledge–action-gap map** is built by locating the fingerprints/profiles of that type that the graph holds and indicating, for each, the level at which the operation applies to the ecosystem it describes. The operation is present at three levels of grounding, which the map distinguishes: (i) as an action unit, where the operation is structurally defined for the ecosystem; (ii) as an applicable action unit, where the fingerprint satisfies the operation's applicability conditions on current knowledge; and (iii) as an audited applicable action unit, where documented evidence supports the operation in comparable contexts. Because ecosystem type is necessary but not sufficient for applicability, a mangrove site must still satisfy the remaining hydrological, sediment, and salinity conditions. Where these narrower conditions belong to a specialised child of a more general action unit (Section 6.1), the map can distinguish sites at which the parent applies from those at which a child, with its more specific evidence, applies as well. The level a site attains is therefore not assigned by type match but determined by evaluating its fingerprint/profile against the full applicability conditions. Each mapped ecosystem carries, for this operation, the highest level of grounding it attains, and the sites where the operation applies at no level are the genuine knowledge gaps for it. Such maps are readily rendered for a region and inspected directly, and where several operations are relevant to an ecosystem type, one such map is produced for each.

Because fingerprints/profiles are time-indexed, such a map is evaluated for a chosen time, or as a temporal sequence across cycles. Where a site's ecosystem has changed type between cycles, it enters and leaves the maps for different operations accordingly, and the operations available at that location shift with its changing state. A time-indexed map therefore shows not only where an operation is grounded but how the possibilities for action at a site change as the ecosystem itself changes, which is particularly informative after a disturbance, when the operations relevant to a site's new state differ from those relevant to its former one.

A knowledge–action-gap map thus shows the coverage of an operation across sites at three levels, with the gap appearing as the residual where coverage does not reach. Such maps are not static products but query results that update as fingerprints/profiles are added, action units formalised, and evidence documented. They make the knowledge–action gap geographically explicit and turn it into the action-knowledge gap of Section 6.5 in directly usable form, showing scientists and funders where the gaps that most constrain practice actually lie.

For the audited applicable level of an operation, the documented success and failure instances carry spatial and temporal provenance, since each occurred at a site whose fingerprint/profile is represented in the graph. Mapping these yields an evidence map for the operation, showing across sites of the ecosystem type the contexts in which the operation has been shown to work and those in which it has failed. Regions of documented success mark where the operation can be reliably applied, regions of documented failure delimit where it cannot, and the boundary between them is the spatial expression of the applicability conditions themselves, refined with every further application.

## 6.7 Governance and the limits of representational neutrality

The TripleA criteria are governance-neutral. They specify what a representation must support in order to guide action, but they say nothing about who may act, on whose land, or on whose terms. That neutrality is a design fact with consequences rather than an absence of consequences, and the consequences become visible precisely where the framework is strongest.

**Auditability** is the framework's point of maximum exposure. An audited applicable action unit records, with spatial and temporal provenance, where an operation has been applied and where it has failed. A failure record is not an abstract datum, as it is about a place, frequently about a project a community requested, hosted, permitted, or resisted, and sometimes about land whose governance is contested. The evidence maps discussed in the previous section aggregate such records and render them queryable at scale. Under the CARE Principles' criterion of **Authority to control**, the parties with authority over a site have a legitimate interest in whether its failure record is represented, aggregated, and exposed to query at all, representing an interest that FAIR, being concerned with the mobility of data rather than with the terms of its circulation, does not address. Making failure first-class, which is what auditability requires and what distinguishes the framework from infrastructures that record only claimed successes, therefore also makes the framework accountable in a way that representation-only infrastructures are not. This is not an argument against auditability, but an argument that the criterion cannot be adopted without the governance question being adopted along with it.

Auditability also presupposes something it does not itself secure, namely that the outcome of an application can be ascertained. An audited action unit records where an operation succeeded and where it failed, but not every outcome resolves cleanly into one or the other. An intervention may leave the system in a state whose success is genuinely undetermined at the time of record, and an operation reported as failed may have partly taken effect before the report was made. The framework treats outcomes as recordable once known, and says nothing about how reliably knowing is achieved. That is a real limit, but it is prior to the framework rather than internal to it, since it concerns the ascertainment of outcomes in the world and not their representation once ascertained. Where ascertainment is itself uncertain, the audit record inherits that uncertainty, and an infrastructure should mark an outcome as undetermined rather than resolve it by default to success or to failure.

**Absence** on a map is ambiguous, and the ambiguity is not symmetric. A knowledge–action-gap map shows, for an operation and an ecosystem type, the sites at which the operation is grounded at each of the three levels, with the residual appearing as the region where it is grounded at none. We described that residual as the genuine knowledge gap for that operation. Strictly, it is the region where the *infrastructure holds no grounding*, which is not the same thing. It may be a region where the relevant knowledge exists but is held by communities who have not been asked, or have been asked and declined, or whose knowledge is not expressible in the representational forms the graph admits. Another possible scenario is that these are locations where fingerprints have never been collected because monitoring capacity is unevenly distributed or where operations have been performed and their outcomes never documented. Read as a research agenda, which is how Section 6.5 proposed the action-knowledge gap should be read, such a map risks directing effort toward regions that are under-documented rather than under-known, and away from regions where the knowledge exists, but the infrastructure has not been able, or permitted, to hold it. Under CARE's criteria of Responsibility and Ethics, this asymmetry is not a technical artefact to be noted and set aside. A map whose coverage tracks the historical distribution of documentation capacity, and which is then read as a map of what is known, reproduces that distribution as an apparent fact about the world. The framework's own resources make this diagnosable, since every level of grounding is explicit and every applicability report carries provenance. Consequently, a TripleA compliant infrastructure based on the concept of action units can distinguish, and should be required to distinguish, a site at which an operation has been assessed and found inapplicable from one at which no assessment has been made, and a region without evidence from a region without documentation. A related asymmetry is not diagnosable at all. The audit loop can reveal that a condition was mis-assigned only if the condition is a dimension of the conditions space, which is to say only if someone has already thought to state it and the fingerprint records it. A determining condition that has never been represented is invisible to the loop, since failures attributable to it cannot cluster along an axis that does not exist. They will appear scattered, and scattered failure, on the account given in Section 2.4, bears against the general knowledge rather than delimiting it. An unrepresented determining condition therefore presents exactly as a false hypothesis. No amount of auditing corrects this, because the deficiency lies in what the infrastructure is able to describe rather than in what it has recorded. It is a further reason for fingerprints to be rich and standardised, since a condition can be discovered to matter only where it is first recorded as something that varies, and it is a further respect in which the coverage of an infrastructure determines what it is capable of learning.

**Applicability conditions** are also where governance constraints become evaluable. Against the exposures just described stands a corresponding capability, and it is the reason CARE bears on this framework more directly than on FAIR. CARE was formulated because FAIR is silent on governance, as it makes data mobile without specifying the terms of its circulation. TripleA does not make CARE's requirements superfluous, but it does provide, for the first time in this lineage, a structure in which they can be represented as conditions on *action* rather than as metadata about *data*. A contextual applicability condition of the form *this operation may be performed at this site only where documented consent of the relevant knowledge holders obtains*, evaluated against the site's ecosystem profile (not only its fingerprint), is a condition of exactly the kind Section 2.3 defines, and an infrastructure that implements it will not return the operation as applicable where the condition is unmet. Governance constraints thereby enter applicability assessment on the same footing as hydrology and salinity, rather than sitting alongside the representation as a policy the representation cannot see. The same structure supports data visiting rather than data sharing, as semantic units are individually addressable and typed so that access to a component of the graph can be mediated by queries its owners authorise, so that contextual information can inform an applicability assessment without being released. The graded structure of the hierarchy applies here too. Where consent is unrecorded, an operation remains an action unit and does not advance to applicable, and the absence is visible as an unmet condition rather than an unexamined assumption.

**A third limit** the framework cannot represent away is of a different kind. The two exposures above are ones the framework's own resources allow it to make visible, because every level of grounding is explicit and every applicability report carries provenance, so that an infrastructure can be required to distinguish what it has assessed from what it has never held. A third exposure is not of that kind, because it concerns the substrate rather than what is built on it. Action units, like the semantic units that realise them, are built over a subject–predicate–object structure and, in the implementations we envisage, over the categorial distinctions of an upper-level ontology such as BFO, among them the division between *continuants* and *occurrents*. The framework's technology-agnosticism, noted in Section 5.3, concerns the storage substrate, since semantic units can be implemented over RDF, property graphs, or relational systems alike. It does not extend to these ontological commitments, which the framework does make, and it is these, not the choice of storage technology, that not all knowledge systems share. A worldview on which an entity exists only in and through its relations, or on which a river is a bearer of agency and of rights rather than a material entity with qualities, does not map onto that substrate without loss. And the loss is not evenly distributed across the knowledge systems the framework proposes to integrate, since it falls entirely on those whose commitments diverge from the ones the substrate encodes. To claim, as Section 6.5 does, that action units serve as a shared interface among scientists, practitioners, and communities is therefore to claim more than the framework can currently underwrite, and we state the qualification here rather than leaving it to be discovered in use. Interoperability across such divergences is achievable in degree rather than in kind. Terminological interoperability, relating the terms of different ontologies, has an established mechanism in entity mappings and the SSSOM standard (83). Propositional interoperability (26), relating alternative schemata for the same type of statement, would require schema mappings, for which no comparable standard yet exists. That absence is a substantive limitation of the present framework and of the semantic-web infrastructure it builds on, and the cost of relating two systems rises with the ontological distance between them. We do not resolve this here, and we do not think it is the kind of problem a representational framework can resolve on its own. We note only that a framework proposing to make knowledge *actionable* across communities of practice inherits, in a sharper form than a framework proposing only to *represent* it, the question of whose knowledge the representation is able to hold, and that answering it is a precondition of the shared interface Section 6.5 describes, not a refinement of it.

# 7. Conclusion

The growth of data, metadata, and formally structured knowledge has not closed the gap between what is known and what is acted upon. FAIR and CLEAR have made knowledge findable, accessible, interoperable, reusable, and human-interpretable, but these are properties of representation, and reliable action needs more than a well-represented body of knowledge. It needs knowledge whose operational use is explicit, whose conditions of valid application are represented, and whose evidential grounding can be assessed.

This paper has argued that meeting these needs requires treating knowledge in terms of the operations it supports. An operation must be performable, which requires an objective and the diagnostic, methodological, and procedural knowledge it executes. It must be reliably performable in a given situation, which requires the conditions under which its knowledge applies. And that reliability must be trustworthy, which requires documented evidence of where prior applications have succeeded and failed. These three requirements define three capabilities a knowledge representation must support, which we have named **actionability**, **applicability**, and **auditability**, and which together constitute the **TripleA Principle**. We have shown that the three capabilities are necessary, each addressing a deficit the others leave standing, and sufficient for the valid-use aspects of actionable knowledge, since the operations over which they are defined are combinatorially and philosophically closed and also closed under composition. We have given the principle a representational realization in **action units**, which extend the notion of a plan specification by making applicability conditions and empirical evidence first-class, typed components, and which are built on the **Semantic Units Framework** so that the statements their components require are represented as units in their own right.

The framework's reach extends beyond biodiversity science. In any domain where empirical claims are produced through experimental work, a scientific paper can be understood and interpreted in terms of the operation and action-unit structure that the framework provides. As papers are currently written, the materials-and-methods section reports what was done, and so describes the operations that were executed, from which a reader must reconstruct the underlying specification. An ideal paper written for **reproducibility** would state that specification directly. Its materials-and-methods section would provide the action units of the approach, representing the materials-and-methods as plan-specification units and the materials as input and output specification units, together with the applicability conditions under which the approach was expected to hold and the objective it aimed at. The paper’s results section would then provide the operation report, the record of the executed operations and their realized outputs, together with the evidence from evaluating those outputs against the objectives in the form of applicability reports, including where the approach succeeded and where it failed. A paper of this form would, in the framework's terms, contribute to an audited applicable action unit. The difficulty of reproducing results across contexts is, on this view, partly a consequence of materials-and-methods sections that report an executed operation without stating the action unit behind it, and in particular without stating the applicability conditions on which the result depended.

The three-level hierarchy also gives a more precise vocabulary for reproducibility than the binary distinction between reproducible and irreproducible. Some claims lack validation evidence because the conditions that produced them cannot recur, as when a species was observed only before its extinction. For these the action-unit level is not a deficiency but an honest representation of the limits of repeatability. For others, the progression from action unit through applicable to audited applicable action unit describes the accumulation of replication evidence over time.

The framework's reach has limits that its own resources make visible, and we have not left them implicit. Because every level of grounding is explicit, and every applicability report carries provenance, an infrastructure built on these criteria can be required to distinguish what it has assessed from what it has never held, and a region without evidence from a region without documentation. But some limits lie deeper than this. Making failure first-class, which auditability requires, also makes the framework accountable to those on whose land failure is recorded, in ways that data-centric infrastructures are not. A gap on a map may mark absent knowledge or only absent documentation, and reading the second as the first would reproduce the historical distribution of documentation capacity as an apparent fact about the world. And the substrate over which action units are built encodes ontological commitments that not all knowledge systems share, so the claim that action units serve as a shared interface across communities holds in degree rather than in kind (see Section 6.7). We state them here because a framework that proposes to make knowledge actionable across communities of practice inherits, in a sharper form than one that proposes only to represent it, the question of whose knowledge the representation is able to hold, and answering that question is a condition of the contribution rather than a qualification of it.

The framework also suggests a view of *actionable understanding*. Understanding is not a static internal state that precedes action. In many settings it is refined, and often first achieved, through the application of knowledge in concrete cases, and we discover what we do and do not know by trying to act on what we believe we know. This view has a precise parallel in Wittgenstein's later philosophy, where understanding is shown not in an accompanying inner state but in the ability to apply knowledge correctly in new cases, what the secondary literature on the Philosophical Investigations calls the *ability to go on* (84). His remark that an inner process stands in need of outward criteria captures the framework's central motivation. Applicability conditions are the outward, publicly assessable criteria that make the otherwise tacit judgement of whether knowledge applies in a situation explicit, evaluable, and communicable to human and machine agents alike. These outward criteria never wholly exhaust the judgement they support, as Section 4.4 conceded in the language of Polanyi's *tacit dimension* and Wittgenstein's *hinge propositions*. The framework represents this limit rather than removing it. What it externalises is the part of the judgement that can be made explicit, against a background of practice and tacit competence that it presupposes rather than replaces. Understanding of this kind is what the term *actionable understanding* has named throughout this paper (Box 3). Actionable understanding is the specifically human-centred form of actionable knowledge. A machine agent can carry actionable knowledge, execute operations, evaluate applicability conditions, and document outcomes, but understanding in the sense intended here is the capacity of a human agent to interpret that knowledge, judge its applicability, and select and justify an action accordingly. As a conceptual claim about what that capacity must involve, rather than an empirical claim about human cognition, actionable understanding requires two components. It requires epistemic operations, which establish the relationship between a knowledge representation and the situation at hand, for without them there is no assessment of whether the knowledge applies. And it requires intervention-oriented reasoning, which determines an appropriate course of action given that relationship, for without it there is no action guidance. The framework does not merely describe this capacity but helps to constitute it, by making the conditions of valid use explicit and representable.

**Box 3 Actionable Knowledge and Actionable Understanding**

| |
|---|
| **Actionable knowledge**<br>Knowledge that enables one or more classes of operations—epistemic, transformational, or intervention—depending on the agent, representation type, and context. In its full operational sense it concerns propositional representations (statements), since isolated terms do not carry propositional knowledge independently, though terms are themselves actionable at the level of recognition and designation. Actionable knowledge includes both human- and machine-actionable statements, and takes three forms depending on the class of operation it supports: epistemically actionable knowledge, transformationally actionable knowledge, and interventionally actionable knowledge, each defined below. |
| **Epistemically actionable knowledge**<br>A form of actionable knowledge that supports **epistemic operations** such as recognition, designation, description, and the evaluation of whether a statement applies to a given situation. For general statements (universal and prototypical as well as directive and conditional directive statements), epistemic actionability is realized through **applicability assessment**, i.e., determining whether the referential applicability conditions under which the statement holds are satisfied in a specific context. Epistemically actionable knowledge enables the question: *“Does this knowledge apply in this situation?”* |
| **Transformationally actionable knowledge**<br>A form of actionable knowledge that supports **transformational operations** on representations, such as data processing, |
| aggregation, filtering, inference, or reformatting. It enables the transformation of representations into forms that support subsequent epistemic interpretation or intervention operations. Transformational actionability depends on the compatibility between input and output representations (formal applicability conditions), as well as the availability of operations (e.g., workflows, scripts, or graph-based transformations) that can be applied to them.<br>Unlike epistemically actionable knowledge, which relates representations to real-world situations, and interventionally actionable knowledge, which supports changes in real-world systems, transformationally actionable knowledge operates within the space of representations (information content entities) themselves. It enables the question: *"How can this representation be transformed into a form that supports interpretation or action?"* |
| **Interventionally actionable knowledge**<br>A form of actionable knowledge that supports **intervention operations** aimed at modifying a system to achieve a specified goal. It requires the integration of semantic content, **contextual information** (assertional statements describing the actual state of the specific situation at hand), procedural knowledge, explicit goals, and **contextual applicability conditions** (the constraints that the situation must satisfy for the procedure to be reliably applied). Applicability assessment consists of evaluating whether the contextual information satisfies the contextual applicability conditions. Interventional actionability depends on prior applicability assessment. Interventionally actionable knowledge enables the question: *"What can be done in this situation to achieve a desired outcome?"* |
| **Actionable understanding**<br>A form of human-centred actionable knowledge that constitutes the capacity to interpret knowledge, assess its applicability in a given context, and select and justify appropriate actions accordingly. By our definition, actionable understanding requires logically necessary components: epistemic operations, which establish the relationship between a knowledge representation and the situation at hand, and intervention-oriented reasoning, which determines appropriate courses of action given that relationship. Without the first, there is no assessment of whether knowledge applies in the situation. Without the second, there is no action guidance. This is a conceptual claim about what the capacity of actionable understanding must involve to be coherent, not an empirical claim about human cognitive processes.<br>Actionable understanding may also depend on transformational operations that prepare or derive the representations required for epistemic interpretation and decision-making. These are enabling rather than intrinsic: they support the conditions under which actionable understanding can arise, but do not themselves constitute it.<br>For general statements (universal, prototypical, directive, and conditional directive), actionable understanding requires explicit applicability assessment, determining whether the referential and contextual applicability conditions under which a statement holds are satisfied in the specific situation at hand. In this sense, actionable understanding is not only a prerequisite for action but is itself refined and often achieved through the application of knowledge in concrete contexts. |

Understanding is refined not only through application but also through articulation. Explaining knowledge to others, instructing, and reasoning collaboratively force the externalisation of what was previously implicit and expose the gaps that abstract possession of knowledge conceals. In both directions the demand for actionability is what drives understanding rather than merely expressing it, whether the knowledge is made operational for oneself in a specific context or made communicable for another agent. By making the conditions of valid use explicit and representable, the framework supports both. It structures knowledge for context-sensitive action and organises it in a form that is communicable, assessable, and transferable across agents. In this sense actionable understanding is not only a prerequisite for action but an outcome of it, and an infrastructure that supports actionable understanding is at the same time an infrastructure for shared and cumulative knowledge.

This paper is not only a conceptual contribution, but also a governance-neutral one. A framework that represents where knowledge has been acted upon, where acting on it has failed, and where no grounding exists at all, is accountable to the communities whose places those records describe in a way that a framework concerned only with representing what is the case is not. The TripleA criteria extend FAIR and CLEAR, and in doing so they inherit, in a sharper form than either, the questions the CARE Principles raise. Section 6.7 sets out where the framework is exposed to them and where its own resources allow that exposure to be made visible rather than concealed.

Implementation and empirical evaluation are left to subsequent work, as is a fuller development of the framework's philosophical foundations. The triadic structure of the framework, in which knowledge is joined to world-directedness and operational specification, echoes Peirce's triadic conception of the sign rather than the Saussurean dyadic model that term-centric representation instantiates, and the convergence of Wittgenstein's account of rule-following, Hacking's distinction between representing and intervening, and the three operation classes suggests three traditions pointing toward the same representational architecture. Whether these parallels are structural or reflect deeper foundations is a question we leave open. What the present work establishes is narrower and, we hope, firmer. Bridging the knowledge-action gap requires representing not only what is known but the conditions and evidence under which it can be used, and the TripleA Principle specifies what a knowledge infrastructure must provide to do so.

# Glossary

**Actionability.** The extent to which a given representation can be used by an agent to perform epistemic, transformational, or intervention actions. Actionability depends on interpretability and the availability of operations that can process the representation as input. Actionability is not a Boolean property, but a multidimensional spectrum determined by:
(i) agent type (human or machine), reflecting differences in interpretative and operational capabilities; (ii) meaning-bearing representation level, distinguishing between non-propositional representations (terms) and propositional, meaning-bearing representations (statements); and (iii) statement generality (for propositional representations), distinguishing between assertional (instance-level) statements and universal or prototypical as well as directive and conditional directive (type-level) statements. These dimensions jointly determine the degree and type of actionability. Notably, statement generality is only defined for propositional representations.

**Actionable knowledge.** Knowledge that enables one or more classes of operations—epistemic, transformational, or intervention—depending on the agent, representation type, and context. In its full operational sense it concerns propositional representations (statements), since isolated terms do not carry propositional knowledge independently, though terms are themselves actionable at the level of recognition and designation. Actionable knowledge includes both human- and machine-actionable statements, and takes three forms depending on the class of operation it supports: epistemically actionable knowledge, transformationally actionable knowledge, and interventionally actionable knowledge, each defined below.

**Actionable understanding.** A form of human-centred actionable knowledge that constitutes the capacity to interpret knowledge, assess its applicability in a given context, and select and justify appropriate actions accordingly. By our definition, actionable understanding requires logically necessary components: epistemic operations, which establish the relationship between a knowledge representation and the situation at hand, and intervention-oriented reasoning, which determines appropriate courses of action given that relationship. Without the first, there is no assessment of whether knowledge applies in the situation. Without the second, there is no action guidance. This is a conceptual claim about what the capacity of actionable understanding must involve to be coherent, not an empirical claim about human cognitive processes.
Actionable understanding may also depend on transformational operations that prepare or derive the representations required for epistemic interpretation and decision-making. These are enabling rather than intrinsic: they support the conditions under which actionable understanding can arise, but do not themselves constitute it.
For general statements (universal, prototypical, directive, and conditional directive), actionable understanding requires explicit applicability assessment, determining whether the referential and contextual applicability conditions under which a statement holds are satisfied in the specific situation at hand. In this sense, actionable understanding is not only a prerequisite for action but is itself refined and often achieved through the application of knowledge in concrete contexts.

**Action Unit.** A semantic unit that integrates semantic content (i.e., knowledge) with input/output specifications, procedural knowledge, and explicit goals as mandatory components, and with applicability conditions and empirical evidence as optional components that become mandatory at the applicable and audited applicable action unit levels respectively. Action units extend plan specifications within the Semantic Units Framework by making applicability conditions and empirical evidence explicit as first-class, typed, and evaluable semantic unit components where available. At the base level, an action unit requires only input/output specifications, a plan specification, and an objective—encoding structural potential independently of any specific situation. Applicability conditions are added to form applicable action units, which are evaluated against contextual information provided at execution time; empirical evidence is added to form audited applicable action units. Action units represent knowledge in a structurally actionable form that supports epistemic, transformational, and intervention actions. They define how an action can, in principle, be performed, independent of whether it is applicable in a specific situation. Action units can also be granular and hierarchically composed, with a higher-order action unit consisting of ordered lower-order action units of potentially different types.

**Applicability.** The extent to which a given knowledge representation can be validly applied in a specific context. Applicability depends on the relationship between contextual information and the applicability conditions associated with the representation. Like actionability, applicability is not a Boolean property but graded and often uncertain. Applicability is also multi-dimensional, being assessed against a structured set of individual conditions, each of which may be satisfied, partially satisfied, or unsatisfied. The overall applicability profile is the structured set of these individual assessments rather than a single aggregate measure, and whether partial satisfaction of multiple conditions warrants action is a reflective judgment that depends on the relative importance of conditions and the costs of action versus inaction in the specific decision context. In many cases, it can only be assessed with varying degrees of confidence, depending on the completeness of contextual information, the validity of both explicit and tacit assumptions underlying the knowledge claim, and the availability of empirical evidence, reflecting ecological uncertainty and the challenges of real-world decision-making. Not all assumptions can be made fully explicit. Human applicability judgment is partly grounded in tacit knowledge and embodied expertise that resists complete propositional articulation. The framework makes explicitly representable applicability conditions visible and evaluable while acknowledging that human judgment will always involve a residue that the schema cannot fully capture. Applicability can be understood as the epistemic characterization of actionability in context. While actionability describes the structural potential for action, applicability determines whether such action can be reliably realized in a given context.
Applicability requirements depend on the type of operation: some operations are intrinsically applicable based on representational compatibility, whereas others require context-dependent evaluation of applicability conditions.

**Applicability Condition Unit.** A semantic unit that specifies the conditions under which a statement can be reliably applied. These compound units relate general knowledge to specific situations and can be operationalized as executable question units that evaluate contextual information. They provide the bridge between general knowledge and situational applicability.

**Applicable Action Unit.** An action unit whose applicability conditions have been evaluated against contextual information and found to be satisfied in a specific context, based on current knowledge. This evaluation may draw on general causal theories, expert judgment, and prior assumptions, but does not yet require documented empirical evidence of success in comparable contexts. An applicable action unit represents the judgment that the operation is theoretically appropriate for the context at hand, i.e., conditions are satisfied based on what is currently known, not on the basis of prior tested applications. At this level, the applicability conditions unit, which is optional at the base action unit level, becomes a mandatory component, and contextual information describing the specific situation must be provided as input for the applicability assessment.

**Auditability.** The extent to which a given knowledge representation supports assessment of the empirical grounding for successfully applying an operation in a specific context, through traceable documentation of where and why prior applications have succeeded or failed. Where actionability describes the structural potential for action and applicability the assessment of whether action can be reliably realized in a given context, auditability concerns the assessment of the evidential basis on which such applicability claims rest.

Like actionability and applicability, auditability is not a Boolean property but graded: a knowledge representation may carry no documented evidence, a single documented application, or extensive documentation spanning many comparable contexts including failure instances. Auditability is the capability that distinguishes audited applicable action units from merely applicable ones, and it is directly relevant to the iterative refinement of applicability conditions, since documented failure instances reveal the boundaries of reliable application and thereby refine the conditions under which an operation can be reliably applied. Auditability can be understood as the evidential characterization of applicability. While applicability assesses whether the known conditions for reliable application are satisfied, auditability assesses the empirical grounding of those conditions through traceable evidence of prior success and failure.

**Audited Applicable Action Unit.** An applicable action unit for which there is documented empirical evidence, including both success and failure instances, from prior applications in comparable contexts. Being audited goes beyond theoretical applicability as it grounds the claim empirically by documenting that the operation has been shown to work reliably in practice under conditions similar to those of the current context, and that its boundaries of valid application, including the contexts in which it has failed, are known. Failure documentation is as epistemically important as success documentation, since it refines and delimits the applicability conditions. The empirical evidence accumulated at this level thus constitutes the prerequisite for identifying reliable applicability conditions. It grounds the theoretical judgments encoded in applicable action units in documented operational outcomes, enabling progressive refinement of the conditions over time. At this level, the empirical evidence layer, which is optional at the base action unit and applicable action unit levels, becomes a mandatory component.

**Contextual Information.** Semantic units representing instance-level information about a concrete situation, including system states, environmental conditions, and observational data. Contextual information provides the situational input required to interpret and apply knowledge. It may be retrieved from the knowledge graph (e.g., via query-based question units) or provided by an agent in a specific decision context. Contextual information is provided at execution time as input to applicability assessment.

**Epistemically actionable knowledge.** A form of actionable knowledge that supports epistemic operations such as recognition, designation, description, and the evaluation of whether a statement applies to a given situation. For general statements (universal and prototypical as well as directive and conditional directive statements), epistemic actionability is realized through applicability assessment, i.e., determining whether the referential applicability conditions under which the statement holds are satisfied in a specific context. Epistemically actionable knowledge enables the question: *"Does this knowledge apply in this situation?"*

**Interventionally actionable knowledge.** A form of actionable knowledge that supports intervention operations aimed at modifying a system to achieve a specified goal. It requires the integration of semantic content, contextual information (assertional statements describing the actual state of the specific situation at hand), procedural knowledge, explicit goals, and contextual applicability conditions (the constraints that the situation must satisfy for the procedure to be reliably applied). Applicability assessment consists of evaluating whether the contextual information satisfies the contextual applicability conditions. Interventional actionability depends on prior applicability assessment. Interventionally actionable knowledge enables the question: *"What can be done in this situation to achieve a desired outcome?"*

**Semantic Unit.** A modular, self-contained, and referenceable representation of meaning, typically corresponding to a single statement (statement unit) or a structured collection of statements (compound unit). Semantic units constitute the foundational units of knowledge representation.

**Transformationally actionable knowledge.** A form of actionable knowledge that supports transformational operations on representations, such as data processing, aggregation, filtering, inference, or reformatting. It enables the transformation of representations into forms that support subsequent epistemic interpretation or intervention operations. Transformational actionability depends on the compatibility between input and output representations (formal applicability conditions), as well as the availability of operations (e.g., workflows, scripts, or graph-based transformations) that can be applied to them.
Unlike epistemically actionable knowledge, which relates representations to real-world situations, and interventionally actionable knowledge, which supports changes in real-world systems, transformationally actionable knowledge operates within the space of representations (information content entities) themselves. It enables the question: *"How can this representation be transformed into a form that supports interpretation*

*or action?"*

**Authors' Contributions**:
Lars Vogt - conceptualization, visualization, methodology, writing (original draft), writing (review and editing);
Robert Frühstückl - writing (review and editing);
Tim Alamenciak - writing (review and editing);

# References

1. IPBES. Summary for policymakers of the global assessment report on biodiversity and ecosystem services of the Intergovernmental Science-Policy Platform on Biodiversity and Ecosystem Service. Díaz S, Settele J, Brondízio ES, Ngo HT, Guèze M, Agard J, et al., editors. Bonn, Germany: IPBES secretariat; 2019. p. 56.

2. Hobbs RJ, Hallett LM, Ehrlich PR, Mooney HA. Intervention Ecology: Applying Ecological Science in the Twenty-first Century. BioScience. 2011 Jun;61(6):442–50. doi:10.1525/bio.2011.61.6.6

3. Buxton RT, Bennett JR, Reid AJ, Shulman C, Cooke SJ, Francis CM, et al. Key information needs to move from knowledge to action for biodiversity conservation in Canada. Biol Conserv. 2021 Apr;256:108983. doi:10.1016/j.biocon.2021.108983

4. Cooke SJ, Jeanson AL, Bishop I, Bryan BA, Chen C, Cvitanovic C, et al. On the theory-practice gap in the environmental realm: perspectives from and for diverse environmental professionals. Socio-Ecol Pract Res. 2021 Sep 1;3(3):243–55. doi:10.1007/s42532-021-00089-0

5. Buxton RT, Bennett JR, Reid AJ, Shulman C, Cooke SJ, Francis CM, et al. Key information needs to move from knowledge to action for biodiversity conservation in Canada. Biol Conserv. 2021;256:108983. doi:https://doi.org/10.1016/j.biocon.2021.108983

6. Cooke SJ, Rous AM, Donaldson LA, Taylor JJ, Rytwinski T, Prior KA, et al. Evidence-based restoration in the Anthropocene—from acting with purpose to acting for impact. Restor Ecol. 2018;26(2):201–5. doi:https://doi.org/10.1111/rec.12675

7. Roche DG, O'Dea RE, Kerr KA, Rytwinski T, Schuster R, Nguyen VM, et al. Closing the knowledge-action gap in conservation with open science. Conserv Biol. 2022;36(3). doi:10.1111/cobi.13835

8. Nonaka I, Takeuchi H. The Knowledge-Creating Company: How Japanese Companies Create the Dynamics of Innovation [Internet]. Oxford University Press; 1995. Available from:

https://doi.org/10.1093/oso/9780195092691.001.0001 doi:10.1093/oso/9780195092691.001.0001

9. Hoerner L, Schamberger M, Bodendorf F. Using Tacit Expert Knowledge to Support Shop-floor Operators Through a Knowledge-based Assistance System. Comput Support Coop Work CSCW. 2023 Mar 1;32(1):55–91. doi:10.1007/s10606-022-09445-4

10. Hamberg K. Gender Bias in Medicine. Womens Health. 2008;4(3):237–43. doi:10.2217/17455057.4.3.237

11. Criado Pérez C. Invisible Women: Exposing data bias in a world designed for men. Great Britain: Chatto & Windus; 2019. 411 p.

12. Bower SD, Brownscombe JW, Birnie-Gauvin K, Ford MI, Moraga AD, Pusiak RJP, et al. Making Tough Choices: Picking the Appropriate Conservation Decision-Making Tool. Conserv Lett. 2018;11(2):e12418. doi:https://doi.org/10.1111/conl.12418

13. Alamenciak T, Gornish E, Murphy SD. Dimensions of effective volunteer restoration techniques in North America. Restor Ecol. 2025;33(4):e70028. doi:https://doi.org/10.1111/rec.70028

14. Wilkinson MD, Dumontier M, Aalbersberg IJ, Appleton G, Axton M, Baak A, et al. The FAIR Guiding Principles for scientific data management and stewardship. Sci Data. 2016 Dec;3(1):160018. doi:10.1038/sdata.2016.18

15. Vogt L. The CLEAR Principle: organizing data and metadata into semantically meaningful types of FAIR Digital Objects to increase their human explorability and cognitive interoperability. J Biomed. 2024;16(18):1–26. doi:10.1186/s13326-025-00340-7

16. Arp R, Smith B, Spear AD. Building Ontologies with Basic Formal Ontology [Internet]. Cambridge, Massachusetts: The MIT Press; 2015. 248 p. Available from: http://mitpress.universitypressscholarship.com/view/10.7551/mitpress/9780262527811.001.0001/upso-9780262527811 doi:10.7551/mitpress/9780262527811.001.0001

17. Gangemi A, Guarino N, Masolo C, Oltramari A, Schneider L. Sweetening ontologies with DOLCE. In: Gómez-Perez A, Benjamins VR, editors. European Knowledge Aquisition Workshop (EKAW-2002), Siguenza, Spain. Springer; 2002. p. 166–81.

18. Hogan A, Blomqvist E, Cochez M, D'amato C, de Melo G, Gutierrez C, et al. Knowledge Graphs. ACM Comput Surv. 2021;54(4):1–37. doi:10.1145/3447772

19. Guizzardi G. Ontology, Ontologies and the "I" of FAIR. Data Intell. 2020 Jan;2(1–2):181–91. doi:10.1162/dint_a_00040

20. Dickson I, Butchart SHM, Catalano A, Gibbons D, Jones JPG, Lee-Brooks K, et al. Introducing a common taxonomy to support learning from failure in conservation. Conserv Biol. 2023;37(1):e13967. doi:https://doi.org/10.1111/cobi.13967

21. Frühstückl R. Spatiotemporal variability and extrapolation from ecological experiments. Biol Philos. 2025 Apr;40(2):9. doi:10.1007/s10539-025-09981-5

22. Vogt L, Kuhn T, Hoehndorf R. Semantic units: organizing knowledge graphs into semantically meaningful units of representation. J Biomed Semant. 2024 May;15(7):1–18. doi:10.1186/s13326-024-00310-5

23. Vogt L, Mons B. The Grammar of FAIR: A Granular Architecture of Semantic Units for FAIR Semantics, Inspired by Biology and Linguistics [Internet]. 2025. Available from: https://arxiv.org/abs/2509.26434v1 doi:10.48550/arXiv.2509.26434

24. Vogt L. The Semantic Ladder: A Framework for Progressive Formalization of Natural Language Content for Knowledge Graphs and AI Systems [Internet]. arXiv; 2026 [cited 2026 Mar 30]. Available from: http://arxiv.org/abs/2603.22136 doi:10.48550/arXiv.2603.22136

25. Vogt L. The Semantic Units Framework, a technology-agnostic representational approach to FAIR and CLEAR knowledge infrastructures. Sci Data. 2026 Jun;13(936):936. doi:10.1038/s41597-026-07588-3

26. Vogt L, Strömert P, Matentzoglu N, Karam N, Konrad M, Prinz M, et al. Suggestions for extending the FAIR Principles based on a linguistic perspective on semantic interoperability. Sci Data. 2025 Apr;12(1):688. doi:10.1038/s41597-025-05011-x

27. Vogt L, Mikó I, Bartolomaeus T. Anatomy and the type concept in biology show that ontologies must be adapted to the diagnostic needs of research. J Biomed Semant. 2022 Dec;13(18):27. doi:10.1186/s13326-022-00268-2

28. Hacking I. Representing and Intervening: Introductory Topics in the Philosophy of Natural Science. Cambridge: Cambridge University Press; 1983.

29. Weiland C, Islam S, Broder D, Anders I, Wittenburg P. FDO Machine Actionability - Version 2.1 - FDO Forum Proposed Recommendation 19 August 2022 [Internet]. 2022. p. 10. Report August. Available from: https://docs.google.com/document/d/1hbCRJvMTmEmpPcYb4_x6dv1OWrBtKUUW5CEXB2gqsRo/edit#

30. Marconi D. On the Structure of Lexical Competence. Proc Aristot Soc. 1995 Jun 1;95(1):131–50. doi:10.1093/aristotelian/95.1.131

31. Searle JR. Speech acts: an essay in the philosophy of language. Cambridge University Press; 1969. 203 p.

32. Searle JR. A Taxonomy of Illocutionary Acts. Minn Stud Philos Sci. 1975;07:344–69.

33. Schulz S, Jansen L. Formal ontologies in biomedical knowledge representation. IMIA Yearb Med Inform 2013. 2013 Jan;8(1):132–46. PubMed PMID: 23974561.

34. Vogt L, Farfar KE, Karanth P, Konrad M, Oelen A, Prinz M, et al. Rosetta Statements: simplifying FAIR knowledge graph construction with a user-centred approach. Database J Biol Databases Curation. 2026;2026:33. doi:10.1093/database/baag030

35. Bandrowski A, Brinkman R, Brochhausen M, Brush MH, Bug B, Chibucos MC, et al. The Ontology for Biomedical Investigations. PLoS ONE. 2016;11(4):1–19. doi:10.1371/journal.pone.0154556 PubMed PMID: 27128319.

36. Vogt L, König-Ries B, Alamenciak T, Brian JI, Arnillas CA, Korell L, et al. A Framework for FAIR and CLEAR Ecological Data and Knowledge: Semantic Units for Synthesis and Causal Modelling [Internet]. arXiv; 2025 [cited 2026 Mar 30]. Available from: http://arxiv.org/abs/2508.08959 doi:10.48550/arXiv.2508.08959

37. Woodward J. Making Things Happen: A Theory of Causal Explanation. New York: Oxford University Press; 2003.

38. Illari P, Russo F. Causality: Philosophical Theory meets Scientific Practice. Oxford, UK: Oxford University Press; 2014. 328 p.

39. Carroll SR, Garba I, Figueroa-Rodríguez OL, Holbrook J, Lovett R, Materechera S, et al. The CARE Principles for Indigenous Data Governance. Data Sci J. 2020 Nov 4;19:43. doi:10.5334/dsj-2020-043

40. Jennings L, Anderson T, Martinez A, Sterling R, Chavez DD, Garba I, et al. Applying the 'CARE Principles for Indigenous Data Governance' to ecology and biodiversity research. Nat Ecol Evol. 2023 Aug 9;7(10):1547–51. doi:10.1038/s41559-023-02161-2

41. Parker WS. Model Evaluation: An Adequacy-for-Purpose View. Philos Sci. 2020;87(3):457–77. doi:10.1086/708691

42. Kodikara KAS, Mukherjee N, Jayatissa LP, Dahdouh-Guebas F, Koedam N. Have mangrove restoration projects worked? An in-depth study in Sri Lanka. Restor Ecol. 2017;25(5):705–16. doi:https://doi.org/10.1111/rec.12492

43. Sequeira AMM, Bouchet PJ, Yates KL, Mengersen K, Caley MJ. Transferring biodiversity models for conservation: Opportunities and challenges. Methods Ecol Evol. 2018;9(5):1250–64. doi:https://doi.org/10.1111/2041-210X.12998

44. Yates KL, Bouchet PJ, Caley MJ, Mengersen K, Randin CF, Parnell S, et al. Outstanding Challenges in the Transferability of Ecological Models. Trends Ecol Evol. 2018;33(10):790–802. doi:https://doi.org/10.1016/j.tree.2018.08.001

45. Lommen STE, Jongejans E, Leitsch-Vitalos M, Tokarska-Guzik B, Zalai M, Müller-Schärer H, et al. Time to cut: population models reveal how to mow invasive common ragweed cost-effectively. NeoBiota. 2018;39:53–78. doi:10.3897/neobiota.39.23398

46. Vitalos M, Karrer G. Dispersal of Ambrosia artemisiifolia seeds along roads:the contribution of traffic and mowing machines. In: Biological Invasions: Towards a Synthesis. p. 53–60. (Neobiota; no. 8).

47. Milakovic I, Fiedler K, Karrer G. Management of roadside populations of invasive Ambrosia artemisiifolia by mowing. Weed Res. 2014;54(3):256–64. doi:https://doi.org/10.1111/wre.12074

48. Polanyi M. The Tacit Dimension. New York, NY, USA: Doubledaxy; 1966. (Anchor books).

49. Hart HLA. The Concept of Law (3rd edn) [Internet]. Oxford: Oxford University Press; 2012. (Clarendon Law Series). Available from: https://www.oxfordlawtrove.com/view/10.1093/he/9780199644704.001.0001/he-9780199644704 doi:10.1093/he/9780199644704.001.0001

50. Kelsen H. Pure Theory of Law - Translation from the Second (Revised and Enlarged) German Edition by MAX KNIGHT [Internet]. 1st edn. University of California Press; 1967. 366 p. Available from: https://www.jstor.org/stable/jj.13167921 doi:https://doi.org/10.2307/jj.13167921

51. Wittgenstein L. On Certainty. Anscombe GEM, Wright GH, editors. Oxford: Basil Blackwell; 1969. 93 p.

52. Ehrlinger L, Wöß W. Towards a Definition of Knowledge Graphs. In: Joint Proceedings of the Posters and Demos Track of the 12th International Conference on Semantic Systems — SEMANTiCS2016 and the 1st International Workshop on Semantic Change & Evolving Semantics (SuCCESS'16) [Internet]. Leipzig, Germany: CEUR Workshop Proceedings; 2016. Available from: http://ceur-ws.org/Vol-1695/paper4.pdf

53. Cyganiak R, Lanthaler M, Wood D. RDF 1.1 Concepts and Abstract Syntax; W3C Recommendation 25 February 2014 [Internet]. 2014. Available from: https://www.w3.org/TR/rdf11-concepts/

54. Frey J, Müller K, Hellmann S, Rahm E, Vidal ME. Evaluation of metadata representations in RDF stores. Semantic Web. 2019;10(2):205–29. doi:10.3233/SW-180307

55. Groth P, Gibson A, Velterop J. The Anatomy of a Nano-publication. Inf Serv Use. 2010;30(1–2):51–6. doi:https://www.doi.org/10.3233/ISU-2010-0613

56. Heath T, Bizer C. Linked Data: Evolving the Web into a Global Data Space. Springer Cham; 2011. XIII, 122. (Synthesis Lectures on Data, Semantics, and Knowledge). doi:10.1007/978-3-031-79432-2

57. Hartig O. Foundations of RDF* and SPARQL*. In: CEUR Workshop Proceedings 1912 [Internet]. Aachen, Germany: CEUR-WS.org; 2017. Available from: https://ceur-ws.org/Vol-1912/paper12.pdf

58. Mill JS. A System of Logic, Ratiocinative and Inductive: Being a Connected View of the Principles of Evidence, and the Methods of Scientific Investigation. Cambridge University Press; 2011. (Cambridge Library Collection - Philosophy).

59. Mackie JL. Causes and Conditions. Am Philos Q. 1965;2(4):245–64.

60. Mackie JL. The Cement of the Universe: A Study of Causation [Internet]. Oxford University Press; 1980. Available from: https://doi.org/10.1093/0198246420.001.0001 doi:10.1093/0198246420.001.0001

61. Waters CK. Causes That Make a Difference. J Philos. 2007;104(11):551–79. doi:10.5840/jphil2007104111

62. Frühstückl R. Interactive causes and causal selection. Philos Sci. 2026;1–32. doi:10.1017/psa.2025.10189

63. Earman J, Roberts JT, Smith S. Ceteris Paribus Lost. In: Earman J, Glymour C, Mitchell S, editors. Ceterus Paribus Laws [Internet]. Dordrecht: Springer Netherlands; 2002. p. 5–25. Available from: https://doi.org/10.1007/978-94-017-1009-1_1 doi:10.1007/978-94-017-1009-1_1

64. Reutlinger A, Schurz G, Hüttemann A. Ceteris Paribus Laws. In: Zalta E, editor. The Stanford Encyclopedia of Philosophy. 2011.

65. Fodor JA. Psychosemantics: The Problem of Meaning in the Philosophy of Mind [Internet]. The MIT Press; 1987. Available from: https://doi.org/10.7551/mitpress/5684.001.0001 doi:10.7551/mitpress/5684.001.0001

66. Pietroski P, Rey G. When Other Things Aren't Equal: Saving Ceteris Paribus Laws from Vacuity. Br J Philos Sci. 1995;46(1):81–110. doi:10.1093/bjps/46.1.81

67. Schurz G. Ceteris Paribus Laws: Classification and Deconstruction. In: Earman J, Glymour C, Mitchell S, editors. Ceterus Paribus Laws [Internet]. Dordrecht: Springer Netherlands; 2002. p. 75–96. Available from: https://doi.org/10.1007/978-94-017-1009-1_4 doi:10.1007/978-94-017-1009-1_4

68. Woodward J. There is No Such Thing as a Ceteris Paribus Law. Erkenntnis. 2002 Nov 1;57(3):303–28. doi:10.1023/A:1021578127039

69. Lange M. Who's Afraid of Ceteris-Paribus Laws? Or: How I Learned to Stop Worrying and Love Them. Erkenntnis. 2002 Nov 1;57(3):407–23. doi:10.1023/A:1021546731582

70. Heger T, Aguilar C, Bartram I, Braga RR, Dietl GP, Enders M, et al. The hierarchy-of-hypotheses approach: a synthesis method for enhancing theory development in ecology and evolution. BioScience. 2021;71(4):337–49. doi:10.1093/biosci/biaa130

71. Schurz G. The Hierarchy-of-Hypotheses Approach from a Philosophy of Science Perspective. BioScience. 2021 Apr;71(4):350–6. doi:10.1093/biosci/biaa097

72. Peirce CS. Reason and the Logic of Things - The Cambridge Conference Lectures of 1898. Ketner KL, editor. Cambridge, Massachusetts: Harvard University Press; 1993. 312 p.

73. Lebo T, Sahoo S, McGuinness D. PROV-O: The PROV Ontology; W3C Recommendation 30 April 2013 [Internet]. 2013. Available from: https://www.w3.org/TR/prov-o/

74. Garijo D, Gil Y. Augmenting PROV with Plans in P-PLAN: Scientific Processes as Linked Data. In: LISC@ISWC [Internet]. 2012. Available from: https://api.semanticscholar.org/CorpusID:5826456

75. Soiland-Reyes S, Sefton P, Crosas M, Castro LJ, Coppens F, Fernández JM, et al. Packaging research artefacts with RO-Crate. Data Sci. 2022;5(2):97–138. doi:10.3233/DS-210053

76. Crusoe MR, Abeln S, Iosup A, Amstutz P, Chilton J, Tijanić N, et al. Methods included: standardizing computational reuse and portability with the Common Workflow Language. Commun ACM. 2022 May;65(6):54–63. doi:10.1145/3486897

77. Genesereth MR, Nilsson NJ. Logical foundations of artificial intelligence. Los Altos, California: Morgan Kaufmann Publishers Inc.; 1987. 405 p.

78. Brooks RA. A Robust Layered Control System for a Mobile Robot. IEEE J Robot Autom. 1986 Mar;2(1):14–23.

79. Ciccio CD, Marrella A, Russo A. Knowledge-Intensive Processes: Characteristics, Requirements and Analysis of Contemporary Approaches. J Data Semant. 2015;4:29–57.

80. Nezhad HRM, Swenson KD. Adaptive Case Management: Overview and Research Challenges. 2013 IEEE 15th Conf Bus Inform. 2013;264–9.

81. Pereira HM, Ferrier S, Walters M, Geller GN, Jongman RHG, Scholes RJ, et al. Essential Biodiversity Variables. Science. 2013;339(6117):277–8. doi:10.1126/science.1229931

82. Arnillas CA, Carscadden K. When indices disagree: facing conceptual and practical challenges. Trends Ecol Evol. 2024 Mar 20. doi:10.1016/j.tree.2024.02.001

83. Matentzoglu N, Flack J, Graybeal J, Harris NL, Hegde HB, Hoyt CT, et al. A Simple Standard for Ontological Mappings 2022: Updates of data model and outlook. CEUR Workshop Proc. 2022;3324:61–6.

84. Wittgenstein L. Philosophical Investigations. Translated by GEM Anscombe, PMS Hacker, J Schulte. Revised 4th edition. Oxford: Wiley-Blackwell; 2009. 1–246 p.

# Acknowledgements

The authors acknowledge the use of an AI-based language model to assist with language editing, phrasing, and manuscript structuring. The AI system was not used to generate scientific content, and all ideas, concepts, analyses, and conclusions presented in this work are those of the authors.